\documentclass[letterpaper,twocolumn,10pt]{article}
\usepackage{zhanggroup}

\usepackage{tikz}
\usepackage{amsmath}
\usepackage{amssymb}
\usepackage{multirow}
\usepackage{color}
\usepackage{xcolor}
\usepackage{url}
\usepackage{caption}
\usepackage{subcaption}
\usepackage{float}
\usepackage{balance}
\usepackage{graphicx}
\usepackage{textcomp}
\usepackage{booktabs}
\usepackage{colortbl}
\usepackage[ruled,linesnumbered]{algorithm2e}

\usepackage[absolute]{textpos}

\newcommand{\mypara}[1]{\smallskip\noindent\textbf{#1}}

\newcommand{\TargetModel}{\mathcal{T}}
\newcommand{\ShadowModel}{\mathcal{S}}
\newcommand{\DataPoint}{x}
\newcommand{\Dataset}{\mathcal{D}}

\newcommand{\TargetTrain}{\mathcal{D}_{\textit{target}}^{\textit{train}}}
\newcommand{\TargetSet}{\mathcal{D}_{\textit{target}}}
\newcommand{\TargetTest}{\mathcal{D}_{\textit{target}}^{\textit{test}}}
\newcommand{\ShadowTrain}{\mathcal{D}_{\textit{shadow}}^{\textit{train}}}
\newcommand{\ShadowSet}{\mathcal{D}_{\textit{shadow}}}
\newcommand{\ShadowTest}{\mathcal{D}_{\textit{shadow}}^{\textit{test}}}

\begin{document}

\begin{textblock}{15}(1.9,1)
To Appear in 2022 ACM SIGSAC Conference on Computer and Communications Security, November 2022
\end{textblock}

\title{\Large \bf Auditing Membership Leakages of Multi-Exit Networks}

\author{
Zheng Li\textsuperscript{1} \ \ \ Yiyong Liu\textsuperscript{1} \ \ \ Xinlei He\textsuperscript{1}
\\
\ \ \ Ning Yu\textsuperscript{2} \ \ \ Michael Backes\textsuperscript{1} \ \ \ Yang Zhang\textsuperscript{1}
\\
\\
\textsuperscript{1}\textit{CISPA Helmholtz Center for Information Security} \ \ \ \textsuperscript{2}\textit{Salesforce Research}
}
\date{}
\maketitle

\begin{abstract}
Relying on the fact that not all inputs require the same amount of computation to yield a confident prediction, multi-exit networks are gaining attention as a prominent approach for pushing the limits of efficient deployment. 
Multi-exit networks endow a backbone model with early exits, allowing to obtain predictions at intermediate layers of the model and thus save computation time and/or energy.
However, current various designs of multi-exit networks are only considered to achieve the best trade-off between resource usage efficiency and prediction accuracy, the privacy risks stemming from
them have never been explored. 
This prompts the need for a comprehensive investigation of privacy risks in multi-exit networks.

In this paper, we perform the first privacy analysis of multi-exit networks through the lens of membership leakages. 
In particular, we first leverage the existing attack methodologies to quantify the multi-exit networks' vulnerability to membership leakages. 
Our experimental results show that multi-exit networks are less vulnerable to membership leakages and the exit (number and depth) attached to the backbone model is highly correlated with the attack performance.
Furthermore, we propose a hybrid attack that exploits the exit information to improve the performance of existing attacks. 
We evaluate membership leakage threat caused by our hybrid attack under three different adversarial setups, ultimately arriving at a model-free and data-free adversary.
These results clearly demonstrate that our hybrid attacks are very broadly applicable, thereby the corresponding risks are much more severe than shown by existing membership inference attacks.
We further present a defense mechanism called \textit{TimeGuard} specifically for multi-exit networks and show that \textit{TimeGuard} mitigates the newly proposed attacks perfectly.\footnote{Our code is available at \url{https://github.com/zhenglisec/Multi-Exit-Privacy}.}
\end{abstract}

\section{Introduction}
\label{sec:intro}

Machine Learning (ML) has established itself as a cornerstone for a wide range of critical applications, such as image classification and face recognition.
To achieve better performance, large ML models with increasing complexity are proposed. 
The improvement in performance stems from the fact that the deeper ML model fixes the errors of the shallower one.
This means some samples that are already correctly classified or recognized by the shallow ML model do not require additional complexity. 
However, such progression to deeper ML models has dramatically increased the latency and energy required for feedforward inference, which further contradicts the research activity in Green AI~\cite{SDSE20,GreenAICloud} that aims to decrease AI's environmental footprint and increase its inclusivity.
This reality has motivated research on input-adaptive mechanisms, i.e., multi-exit networks, which is an emerging direction for fast inference and energy-efficient computing.

The multi-exit model consists of a backbone model (i.e., a large vanilla model) and multiple exits (i.e., lightweight classifiers) attached to the backbone model at different depths.
The backbone model is used for feature extraction and the lightweight classifiers allow data samples to be predicted and to exit at an early layer of the model based on tunable early-exit criteria. 
Multi-exit architecture can be applied to many critical applications as it can effectively reduce computational costs.
For example, exiting early means low latency, which is crucial for operating under real-time constraints in robotics applications, such as self-driving cars. 
Furthermore, exiting early can improve energy efficiency, which directly influences battery life and heat release, especially on mobile devices.

\subsection{Our Contributions}

\mypara{Motivation.} 
Multi-exit networks, despite their low latency and high energy efficiency, also rely on large-scale data to train themselves, as the way vanilla ML models are trained.
In many cases, the data contains sensitive and private information of individuals, such as shopping preferences, social relationships, and health status.
Various recent studies have shown that vanilla ML models, represented by image classifiers, are vulnerable to privacy attacks~\cite{NSH19,LF20,SSSS17,SZHBFB19,LLR21,SM21,LZ21}. 
One major attack in this domain is membership inference: An adversary aims to infer whether a data sample is part of a target ML model’s training dataset.

However, current various designs of multi-exit networks are only considered to achieve the best trade-off between resource usage efficiency and prediction accuracy, the privacy risks stemming from them have never been explored. 
This prompts the need for a comprehensive investigation of privacy risks in multi-exit networks, such as the vulnerability of multi-export networks to privacy attacks, the reasons inherent in this vulnerability, the factors that affect attack performance, and whether or how these factors can be exploited to improve or reduce attack performance.

In this paper, we take the first step to audit the privacy risks of multi-exit networks through the lens of membership inference.
More specifically, we focus on machine learning classification, which is the most common machine learning task, and conduct experiments with 3 types of membership inference attacks, 6 benchmark datasets, and 8 model architectures.

\mypara{Main Findings.} 
We first leverage the existing attack methodologies (gradient-based, score-based, and label-only) to audit the multi-exit networks' vulnerability through membership inference attacks. We conduct extensive experiments and the empirical results demonstrate that multi-exit models are less vulnerable to membership inference attacks than vanilla ML models. 
For instance, considering the score-based attacks, we achieve an attack success rate of 0.5413 on the multi-exit model trained on CIFAR-10 with the backbone model being ResNet-56, while the result is 0.7122 on the corresponding vanilla ResNet-56.
Furthermore, we delve more deeply into the reasons for the lower vulnerability and reveal that the reason behind this is that the multi-exit models are less likely to be overfitted.

We also find that the number of exits is negatively correlated with the attack performance, i.e., multi-exit models with more exits are less vulnerable to membership inference.
Besides, a more interesting observation is that considering a certain multi-exit model, exit depth is positively correlated with attack performance, i.e., exits attached to the backbone model at deeper locations are more vulnerable to membership inference.
These observations are due to the fact that different depths of exits in the backbone model actually imply different capacity models, and that deeper exits imply higher capacity models, which are more likely to be overfitted by memorizing properties of the training set.

\mypara{Hybrid Attack.}
The above findings render us a new factor to improve the attack performance. 
More concretely, we propose a novel hybrid attack against multi-exit networks that exploit the exit information as new knowledge of the adversary.
The hybrid attack’s methodology can be divided into two stages, each of which actually represents one type of attack against ML models:

\begin{itemize}
    \item Hyperparameter stealing: the adversary's goal is to steal the hyperparameters, i.e., the number of exits and the exit depth of a given multi-exit network designed by the model owner.
    \item Enhanced membership inference: the adversary then exploits the stolen exit information as new knowledge to launch more powerful member inference attacks.
\end{itemize}

In particular, we study three different adversaries for obtaining exit information by starting with some strong assumptions, and gradually relaxing these assumptions in order to show that far more broadly applicable attack scenarios are possible.
Our investigation shows that indeed, our proposed hybrid attack can achieve better attack performance by exploiting extra exit information, compared to original membership inference attacks.

\smallskip
\noindent\textsl{Adversary 1.} For the first adversary, we assume they have direct access to the exit information, i.e., exit depth, as well as train a shadow model of the same
architecture (especially the exit placements) as the target model.
Further, the adversary trains the shadow models on a shadow dataset that comes from the same distribution as the target dataset. 
The assumption of same-architecture and same-distribution also holds for almost all existing membership inference attacks~\cite{NSH19,LF20,SSSS17,SZHBFB19,LLR21,SM21,LZ21}.

We start by querying the target model using a large number of data samples to determine the number of exits attached to the model.
Then we propose different methods (e.g., one-hot encoding) based on the attack models adopted by existing attacks to exploit this exit information.
Extensive experimental evaluation shows that extra exit information indeed leaks more membership information about training data.
For example, our hybrid attack achieves an attack success rate of 0.7681 on a multi-exit WideResNet-32 trained on CIFAR-100, while the result of the original attack is 0.6799.

\smallskip
\noindent\textsl{Adversary 2.} For this adversary, we relax the assumption that they have direct access to exit information and keep the assumption of the same architecture and same distribution unchanged.
This is a more challenging scenario compared to the previous one.

In this scenario, we propose \textit{time-based hyperparameter stealing} to obtain the exit information.
Concretely, we feed a set of samples to the target multi-exit model and record the inference time of these samples.
We then propose a simple but effective unsupervised method to cluster the samples based on different inference times.
Thus, the number of clusters implies the number of exits, and the index of cluster implies the exit depth.

The intuition is that the goal of multi-exit models is to reduce the computational costs by allowing data samples to be predicted and to exit at an early point.
Therefore, the inference time for data samples inevitably varies with the depth of the exit, i.e., data samples leaving deeper exit points imply longer inference times.
Thus, we can determine the exit depths by observing the magnitude of inference time.
Experimental results show that our hybrid attack achieves a strong performance as our \textit{time-based hyperparameter stealing} can achieve almost 100\% prediction accuracy of exit depths.

\smallskip
\noindent\textsl{Adversary 3.} 
This adversary works without any knowledge about the target models and target datasets, that is, the adversary can only construct a shadow model that is different from the target model or a dataset from a different distribution from the target dataset.
Meanwhile, the different architectures between the shadow model and the target model will inevitably lead to different exit placements between them.
Encouragingly, our hybrid attack still has better attack performance than the original attacks, suggesting that the extra exit information has a broader range of applicable attack scenarios.

Finally, we propose a simple but effective defense mechanism called \textit{TimeGuard}, which postpones giving the prediction, rather than giving them immediately.
Our in-depth analysis shows that \textit{TimeGuard} can reduce attack performance to a lower bound and maintain high efficiency, i.e., achieve the best trade-off between privacy and efficiency.

Abstractly, our contributions can be summarized as:
\begin{itemize}
    \item We take the first step to audit the privacy risks of multi-exit networks through the lens of membership inference attacks.
    \item Our empirical evaluation shows that the multi-exit networks are less vulnerable to member inference, and the exit information is highly correlated with the attack performance.
    \item We propose a hybrid attack that exploits the exit information to improve the attack performance of membership inference.
    \item We evaluate membership leakage threat caused by hybrid attack under three different adversarial setups, ultimately arriving at a model-free and data-free adversary, which further enlarges the scope of the hybrid attack.
    \item We propose \textit{TimeGuard} to mitigate privacy risks stemming from our attack and empirically evaluate its effectiveness.
\end{itemize}

\section{Preliminaries}
\label{sec:preli}

\subsection{Membership Leakages in Machine Learning Models}

Membership leakages in ML models emerge when an adversary aims to identify whether a data sample was used to train a certain model or not.
It can raise severe privacy risks as the membership can reveal an individual’s private information.
For example, identifying an individual’s participation in a hospital’s health analytic training set reveals that this individual was once a patient in that hospital.

To evaluate the vulnerability of a given ML model to membership leakages, membership inference attacks are used as an auditing tool to quantify the private information that a model leaks about the individual data samples in its training set.
In this work, we focus on auditing the membership leakages of multi-exit networks.

\begin{figure}[t]
    \centering
    \includegraphics[width=0.9\linewidth]{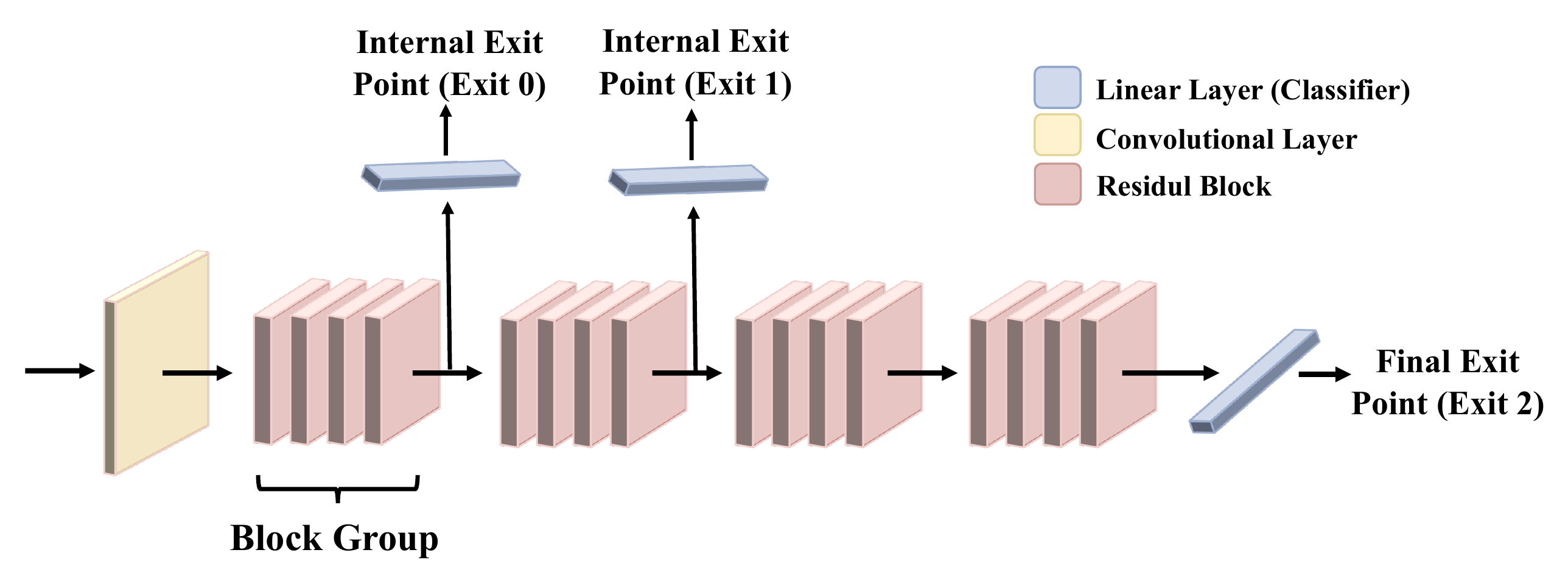}
    \caption{An illustration of multi-exit network with 3 exits inserted, including 2 internal exit points and 1 final exit point.}
    \label{fig:multi-exit}
\end{figure}

\subsection{Multi-Exit Networks}\label{sec:multi-exit}

Multi-exit networks save computation by making input-specific decisions about bypassing the remaining layers, once the model becomes
confident. 
More concretely, a multi-exit network applies multiple lightweight classifiers on a vanilla ML model to allow the inference to preemptively finish at one of the exit points when the network is sufficiently confident with a predefined stopping criterion. 
See \autoref{fig:multi-exit} for an illustration of the design of multi-exit networks.
In this paper, we focus on the pioneer and meanwhile the most representative design of multi-exit networks.

\mypara{Backbone Initialisation.}
As aforementioned, multi-exit networks modify the vanilla ML model by adding multiple lightweight classifiers at certain placements throughout the network.
Here, vanilla ML models are also referred to as backbone models.
A backbone model can be any regular machine learning model architecture.
For example, in this work, we focus on the image classification task, so such backbone models can be convolutional networks applicable to vision, such as VGG~\cite{SZ15}, ResNet~\cite{HZRS16}, and MobileNet~\cite{SHZZC18}.

\mypara{Exit Placement.}
For simplicity, exit placements are restricted to be at the output of individual network blocks, following an approximately equidistant workload distribution.

\mypara{Multi-Exit Network Training.}
Given a training dataset, a multi-exit model is optimized by minimizing the loss function of all training samples and exit points.
The training process consists of two steps: the feedforward pass and the backward pass. 
In the former, a data sample is passed through the model, including both final exit point and internal exit points, the output from the network at all exit points is recorded, and the loss of the network is then calculated. 
In backward propagation, the loss is passed back through the network and the model's weights are updated using gradient descent. 

\mypara{Early-Exit Criteria.}
Given a data sample, it will leave at one of the exit points when the network is sufficiently confident with a predefined stopping criterion.
To quantify confidence, we use the estimated probability of the sample belonging to the predicted class.
We deem a prediction confident if this probability exceeds the threshold $\tau$.
The threshold facilitates on-the-fly adjustment of the early exits based on the resource availability and the performance requirements. 
Following most previous works~\cite{TMK16,HCLWMW18,KHD19,PL19,HVD15,RBKCGB15}, the principle of threshold selection is to guarantee the same or similar classification performance as vanilla models while gaining a lower computational cost.
See Appendix \autoref{fig:choose_threshold} for how to select a threshold.

\mypara{Application Scenarios.}
As introduced in \autoref{sec:intro}, with the benefits of low latency and energy efficiency, multi-exit architectures can be used in many real-world applications.
In the industry, IT companies leverage multi-exit networks to accelerate forward inference. 
For instance, Intel~\cite{Multi_Intel} has developed multi-exit architectures to reduce the computational cost of DNNs. 
Microsoft~\cite{ZXGMXW20} and Huawei~\cite{HHSJCL20} conducted research on reducing the computational complexity of BERT for bringing language models to IoT devices. 
Besides, there is a growing body of applications deploying multi-exit architectures to IoT scenarios and real-time systems~\cite{KHGRMMT17,LZC18,ZWTD19,HBWL19,JSZYZSH19}. 
In academia, there has been substantial research focusing on minimizing the computational and energy requirements of multi-exit networks for efficient inference~\cite{TMK16,HCLWMW18,KHD19,PL19,HVD15,RBKCGB15,WL21,HCWW20,HKMD21,DQZLLL21,LZWZDJ20,XTLYL20,LVALL20,KVLL21,WSHXNBWL20}.
In short, multi-exit networks are gaining significant attention and rapid development in both industry and academia.

\section{Quantifying Membership Leakage Risks}
\label{sec:quantify}

In this section, we quantify the privacy risks of multi-exit networks through the lens of membership inference attacks.
We start by defining the threat model.
Then, we describe the attack methodology.
Finally, we present the evaluation results.
Note that our goal here is not to propose a novel membership inference attack, instead, we aim to quantify the membership leakages of multi-exit networks. 
Therefore, we follow the existing attacks and their threat models.

\subsection{Threat Model}

Here, we outline the threat models considered in this paper.
There are three existing categories of scenarios, i.e., white-box scenario, black-box scenario, and label-only scenario.

Given a target model, we assume the adversary has an auxiliary dataset (namely shadow dataset) that comes from the same distribution as the target model’s training set.
The shadow dataset is used to train a shadow model, the goal of which is to mimic the behavior of the target model to perform the attack. 
Furthermore, we assume the shadow model has the same architecture as the target model following previous works~\cite{NSH19,LF20,SSSS17,SZHBFB19,LLR21,SM21,LZ21}.
In particular, the exit placements of the shadow multi-exit model are also the same as that of the target multi-exit model.

\subsection{Attack Methodology}

We leverage existing membership inference attacks, which are designed for vanilla ML models, to multi-exit models.
More specifically, for three different scenarios, we consider three representative attacks, namely gradient-based attacks~\cite{NSH19,LF20} in the white-box scenario, score-based attacks~\cite{SSSS17,SZHBFB19,LLR21,SM21} in the black-box scenario, and label-only attacks~\cite{LZ21,CTCP21} in the label-only scenario.

\mypara{Gradient-based Attacks.}
In gradient-based attacks~\cite{NSH19,LF20}, the adversary obtains all adversarial knowledge and has full access to the target model.
This means for any data sample $\DataPoint$, the adversary not only obtains the prediction (score and label), but also knows the intermediate computations (features and gradients) of $\DataPoint$ on the target model.
Given a shadow dataset $\ShadowSet$, the adversary first splits it into two disjoint sets, i.e., shadow training set $\ShadowTrain$ and shadow testing set $\ShadowTest$.
Then the adversary queries the shadow model $\ShadowModel$ on each data sample $\DataPoint$ from $\ShadowTrain$, and computes the prediction score, the feature of the second to the last layer, the loss in a forward pass, and the gradient of the loss with respect to the last layer's parameters in the backward pass.
These computations, in addition to the one-hot encoding of the true label, are concatenated into a flat vector and labeled as a member if $\DataPoint$ is in the shadow training set $\ShadowTrain$, otherwise labeled as a non-member.
In this way, the adversary can derive all data samples of $\ShadowSet$ as an attack training data set.
With the attack training dataset, the adversary then trains the attack model, which is a binary classifier. 
Once the attack model is trained, the adversary can perform the attack to query the target model $\TargetModel$ to differentiate members and nonmembers of the target dataset $\TargetSet$.

\mypara{Score-based Attacks.}
Score-based attacks~\cite{SSSS17,SZHBFB19,LLR21,SM21} need to train the shadow model as well. 
Unlike gradient-based attacks, score-based attacks do not require intermediate features or gradients of the target model, but only access to the output scores of the model.
The adversary also derives the attack training dataset by querying the shadow model using the shadow training dataset (labeled as members) and the shadow test dataset (labeled as non-members).
The adversary can then use the attack training set to construct an attack model.

\mypara{Label-only Attacks.}
Label-only attacks~\cite{LZ21,CTCP21} consider a more restricted scenario where the target model only exposes the predicted label instead of intermediate features or gradients, or even output scores.
Thus, label-only attacks solely rely on the target model's predicated label as their attack model’s input.
Similar to previous attacks, this attack requires the adversary to train a shadow model.
The adversary queries the target model on a data sample and perturbs it to change the model’s predicted labels.
Then, the adversary measures the magnitude of the perturbation and considers the data samples as members if their magnitude is greater than a predefined threshold, which can be derived by perturbing the shadow dataset on the shadow model.

\subsection{Experimental Settings}
\label{setup_quantify}

\mypara{Datasets.} 
We consider six benchmark datasets of different tasks, sizes and complexity to conduct our experiments. 
Concretely, we adopt three computer vision tasks, namely CIFAR-10~\cite{CIFAR}, CIFAR-100~\cite{CIFAR}, TinyImageNet~\cite{TinyImageNet}, and three non-computer vision tasks, namely Purchases~\cite{Purchases}, Locations~\cite{Locations} and Texas~\cite{TexasHealthCare}.
In particular, the latter three datasets are privacy-sensitive: Purchases relate to shopping preferences, Locations relate to social connections, and Texas relate to health status.
Details of all six datasets can be found in Appendix \autoref{appendix:datasets}.


\begin{figure*}[!t]
\centering
\begin{subfigure}{0.99\columnwidth}
\includegraphics[width=\columnwidth]{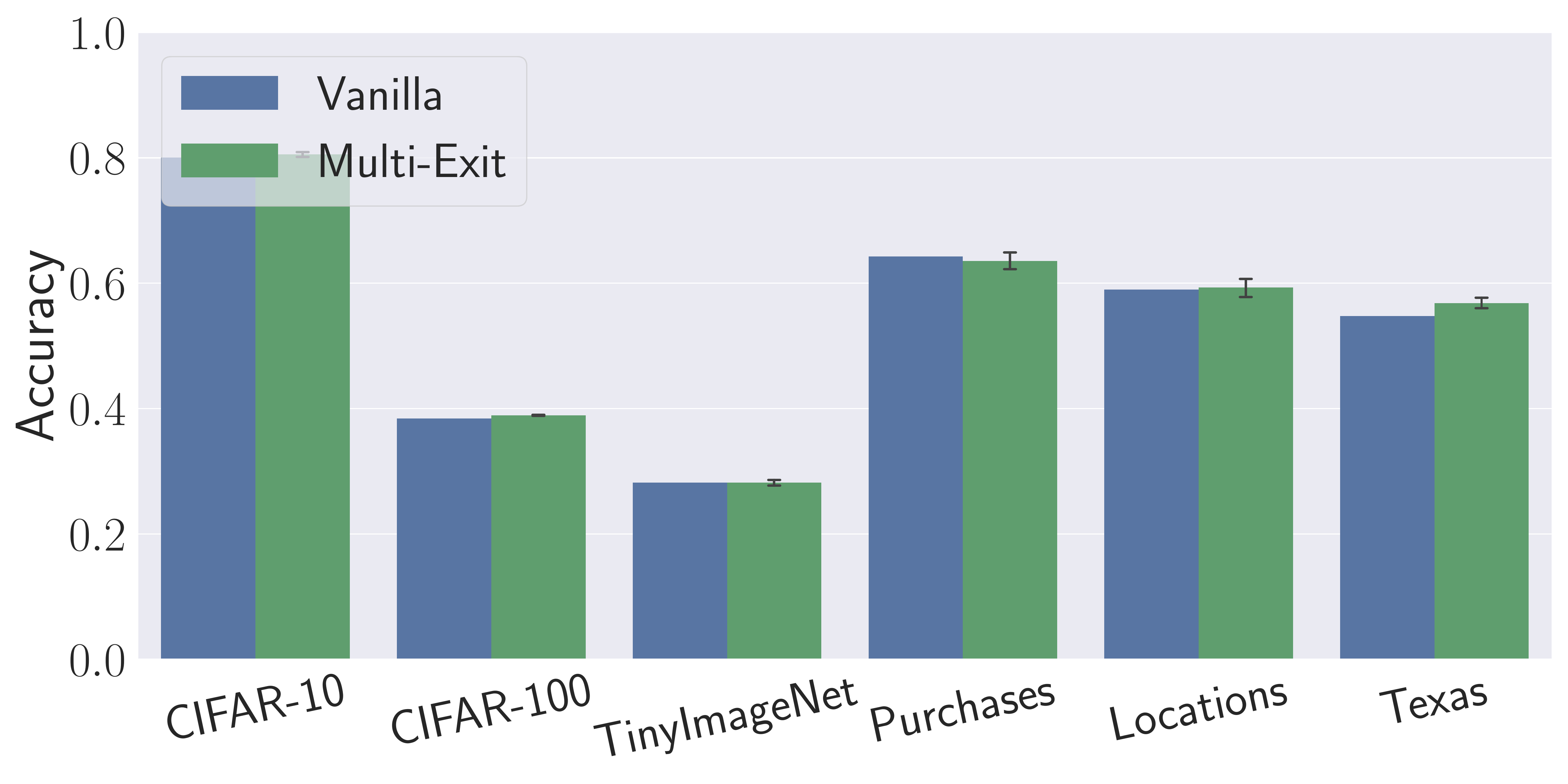}
\caption{Classification Accuracy}
\end{subfigure}
\begin{subfigure}{0.99\columnwidth}
\includegraphics[width=\columnwidth]{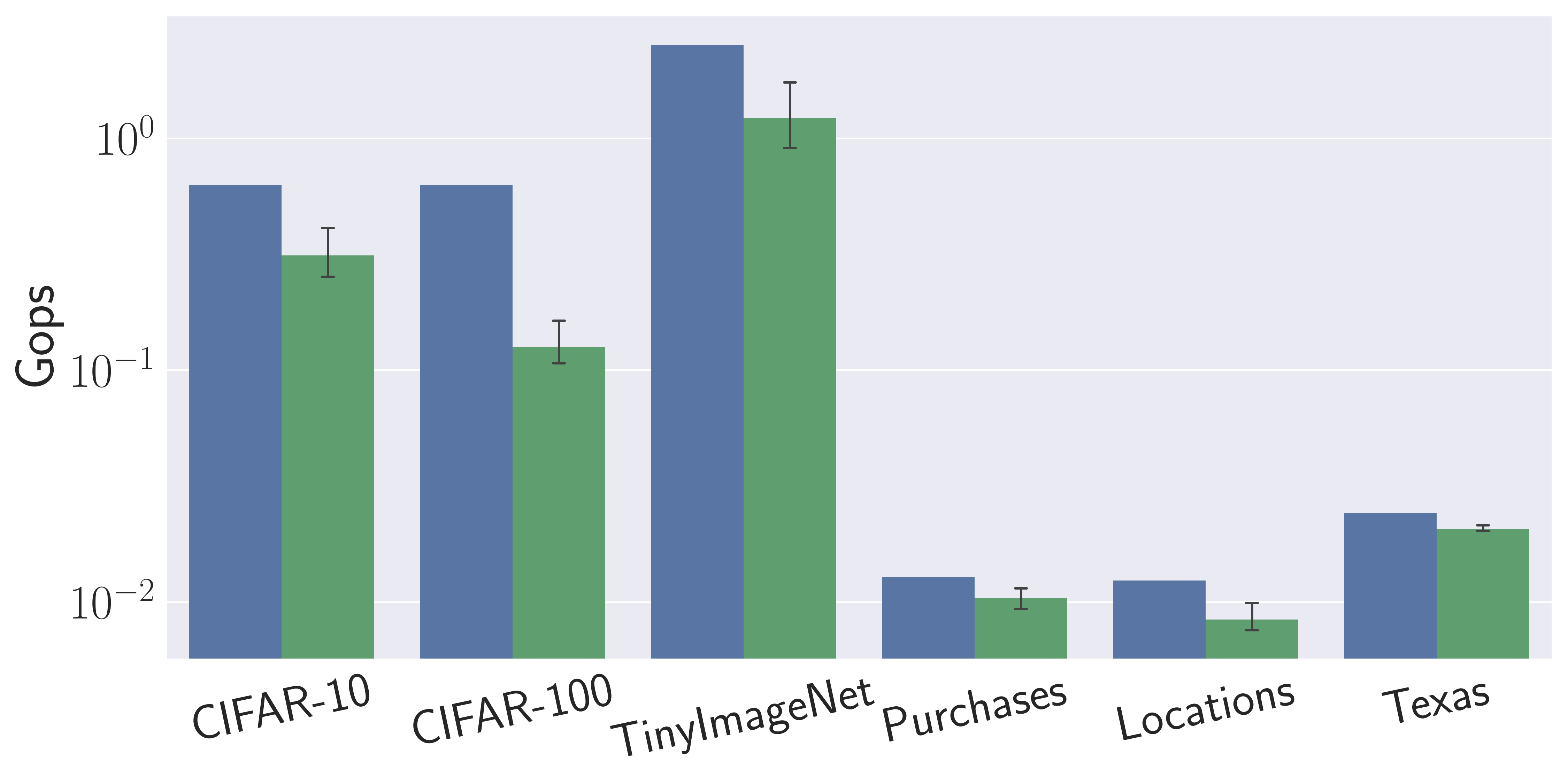}
\caption{Computational Cost}
\end{subfigure}
\caption{The performance of original classification tasks and computational costs for both vanilla models and multi-exit models. computer vision tasks are on VGG-16, and non-computer vision tasks are on FCN-18-1.}
\label{fig:classification_performance}
\end{figure*}

\mypara{Datasets Configuration.} 
For a given dataset $\Dataset$, we randomly split it into four disjoint equal parts: $\TargetTrain$, $\TargetTest$, $\ShadowTrain$, and $\ShadowTest$. 
We use $\TargetTrain$ to train the target model $\TargetModel$ and treat it as the members of the target model. 
We treat $\TargetTest$ as the non-members of the target model.
Similarly, we use $\ShadowTrain$ to train the shadow model $\ShadowModel$ and treat it as the members of the shadow model.
We again treat $\ShadowTest$ as the non-members of the shadow model.
We feed all $\ShadowTrain$ and $\ShadowTest$ to the shadow model to create an attack training dataset to train the attack models.

\mypara{Attack Model.}
Here we establish three types of attack models and each type for one attack.
\begin{itemize}
    \item \mypara{Gradient-based.} This attack has five inputs for the attack model, like the one used by Nasr et al.~\cite{NSH19}, including the target sample’s prediction score, the feature of second to last layer, classification loss, gradients of the last layer's parameters, and one-hot encoding of its true label. Each input is fed into a different MLP (2 or 3 layers) and the resulted embeddings are concatenated together as one vector to a 4-layer MLP.
    \item \mypara{Score-based.} The score-based attack utilizes the predicted score as input to the attack model, which is constructed as a 4-layer MLP with one input component.
    \item \mypara{Label-only.} Here, the attack model is not a specific MLP, but a decision function that measures the magnitude of the perturbation and considers data samples as members if their magnitude is larger than a predefined threshold, which can be derived by perturbing the shadow dataset on shadow model. 
\end{itemize}

\mypara{Target Model (Multi-Exit Model).}
For computer vision tasks, we adopt four popular architectures as the backbone to construct multi-exit models, including VGG-16~\cite{SZ15}, ResNet-56~\cite{HZRS16}, MobileNetV2~\cite{SHZZC18}, and WideResNet-32~\cite{ZK16}.
For non-computer vision tasks, we designed four 18-layer fully connected networks (FCN-18) with different numbers of hidden neural units (1024, 2048, 3072, 4096), named FCN-18-1/2/3/4 throughout the paper.
See Appendix \autoref{appendix:fcn-18} for more detailed about these architectures.
For the exit placement, we follow the principle of Kaya et al.~\cite{KHD19} by attaching an additional lightweight classifier (2- or 3-layer MLP) as an exit, i.e., exit placements are restricted to be at the output of individual network blocks, following an approximately equidistant workload distribution.
In particular, for each backbone model, we construct 5 different target models with the number of exits varying from 2 to 6.
Note that here we consider the backbone model's own classifier as the final exit point and count it in the total number of exits.
For early-exit threshold $\tau$ $(0 \le \tau \le 1)$, we manually search for suitable $\tau$ value (among 0 to 1 in 0.05 steps) that achieve the same or similar classification performance as vanilla (backbone) models while gaining lower computational cost.
To evaluate computational cost, we calculate the number of mathematical operations (denoted as \textit{ops}) in the feedforward pass process by averaging over 10,000 images.
The early-exit threshold we set for multi-exit models can be found in Appendix \autoref{table:end_threshold_cv} and \autoref{table:end_threshold_non_cv}.


\begin{figure*}[!t]
\centering
\begin{subfigure}{0.33\linewidth}
\includegraphics[width=\linewidth]{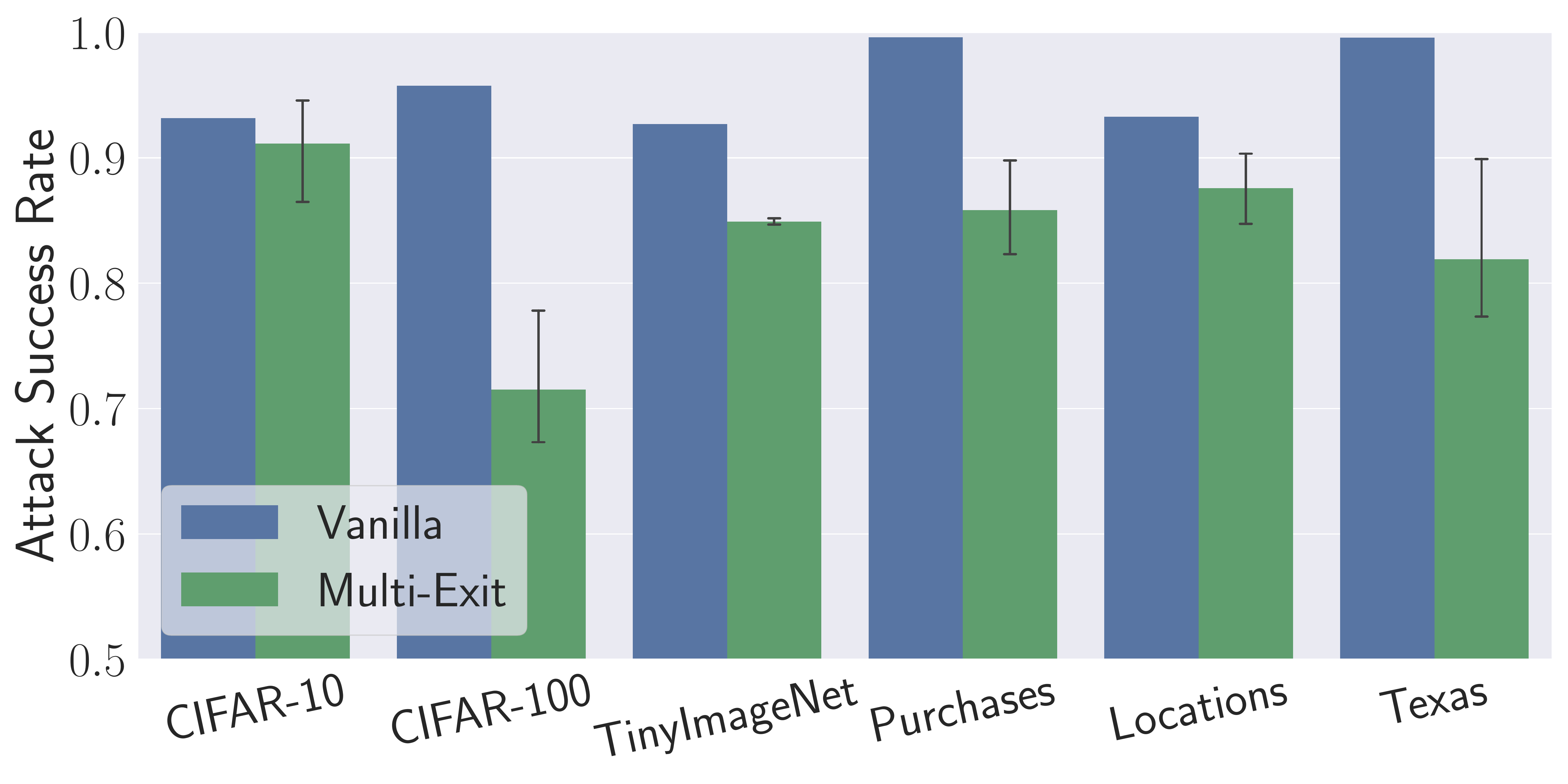}
\caption{Gradient-based}
\end{subfigure}
\begin{subfigure}{0.33\linewidth}
\includegraphics[width=\linewidth]{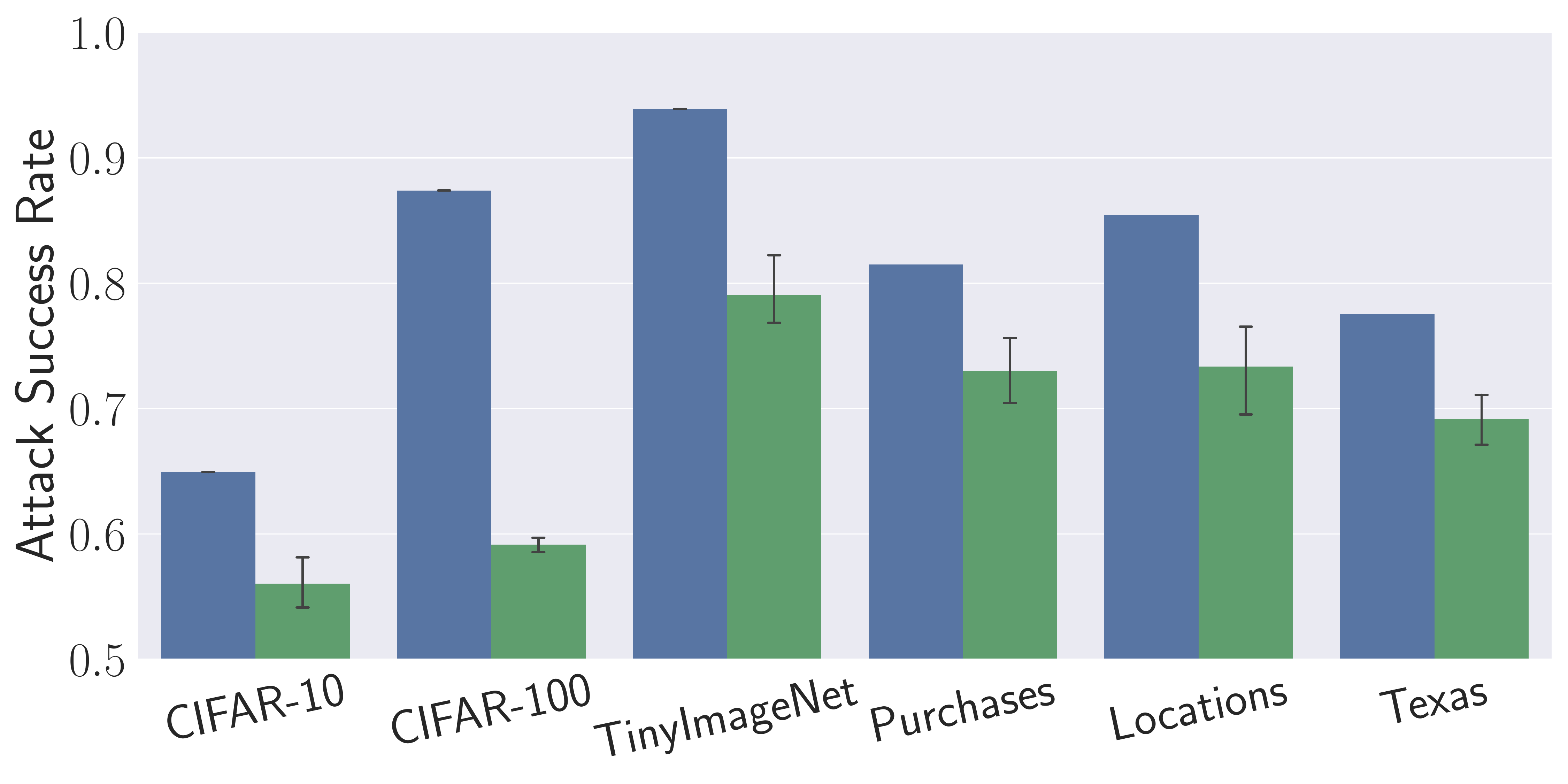}
\caption{Score-based}
\end{subfigure}
\begin{subfigure}{0.33\linewidth}
\includegraphics[width=\linewidth]{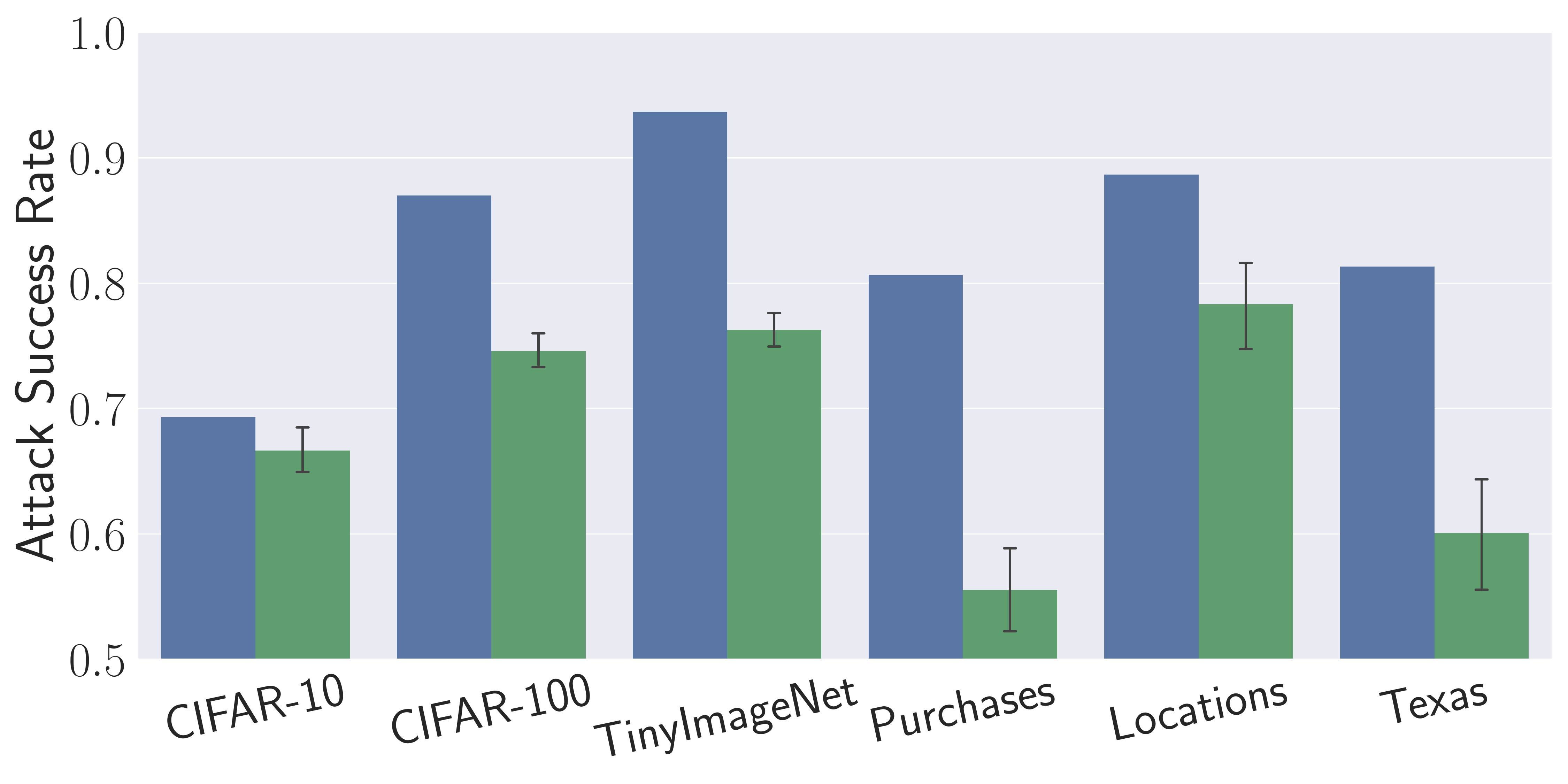}
\caption{Label-only}
\end{subfigure}
\caption{The attack performance of original membership inference attacks again vanilla and multi-exit models. Computer vision tasks are on VGG-16, and non-computer vision tasks are on FCN-18-1.}
\label{fig:attacASR}
\end{figure*}



\begin{figure}[!t]
\centering
\begin{subfigure}{0.49\columnwidth}
\includegraphics[width=\columnwidth]{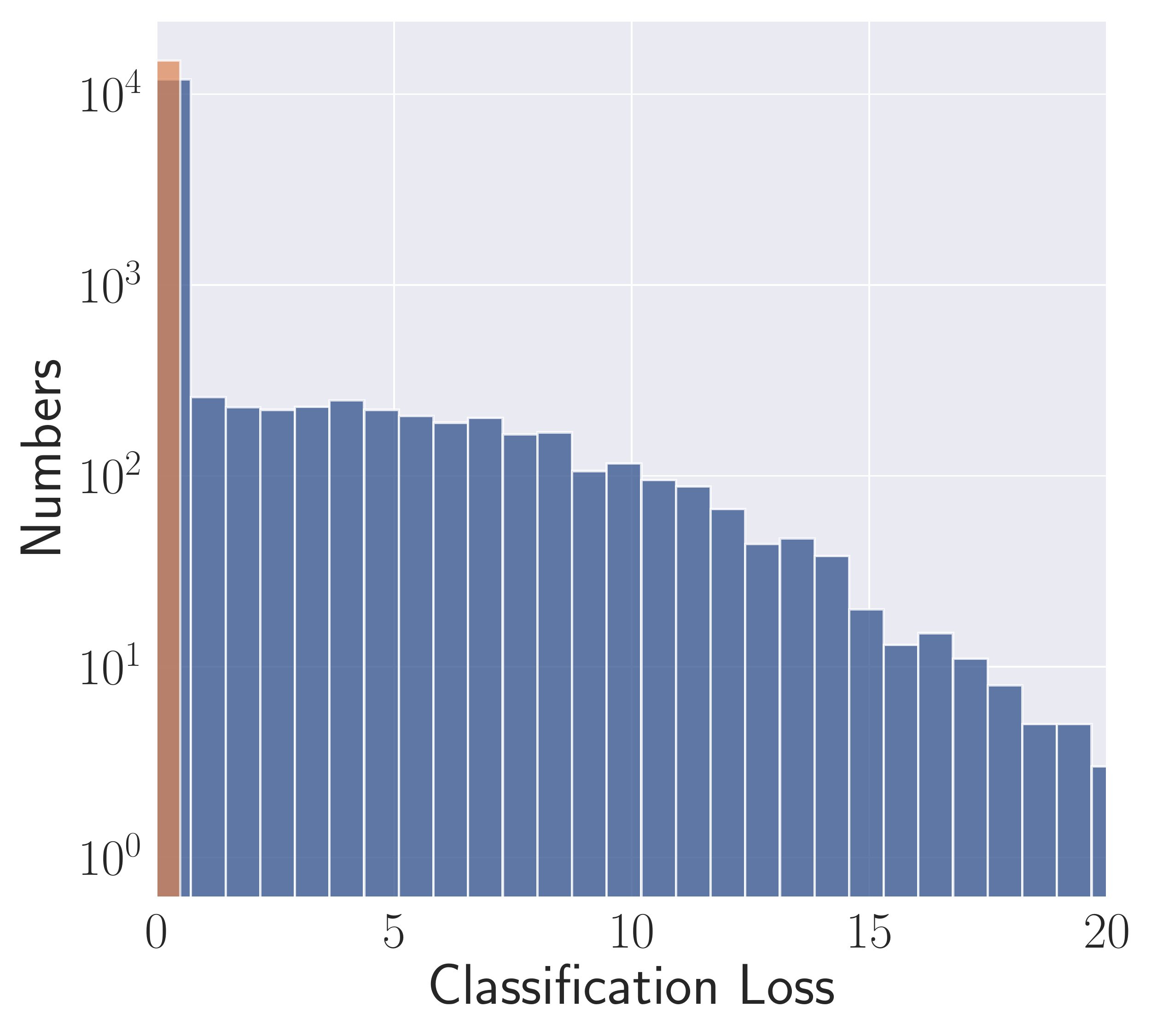}
\caption{Vanilla}
\end{subfigure}
\begin{subfigure}{0.49\columnwidth}
\includegraphics[width=\columnwidth]{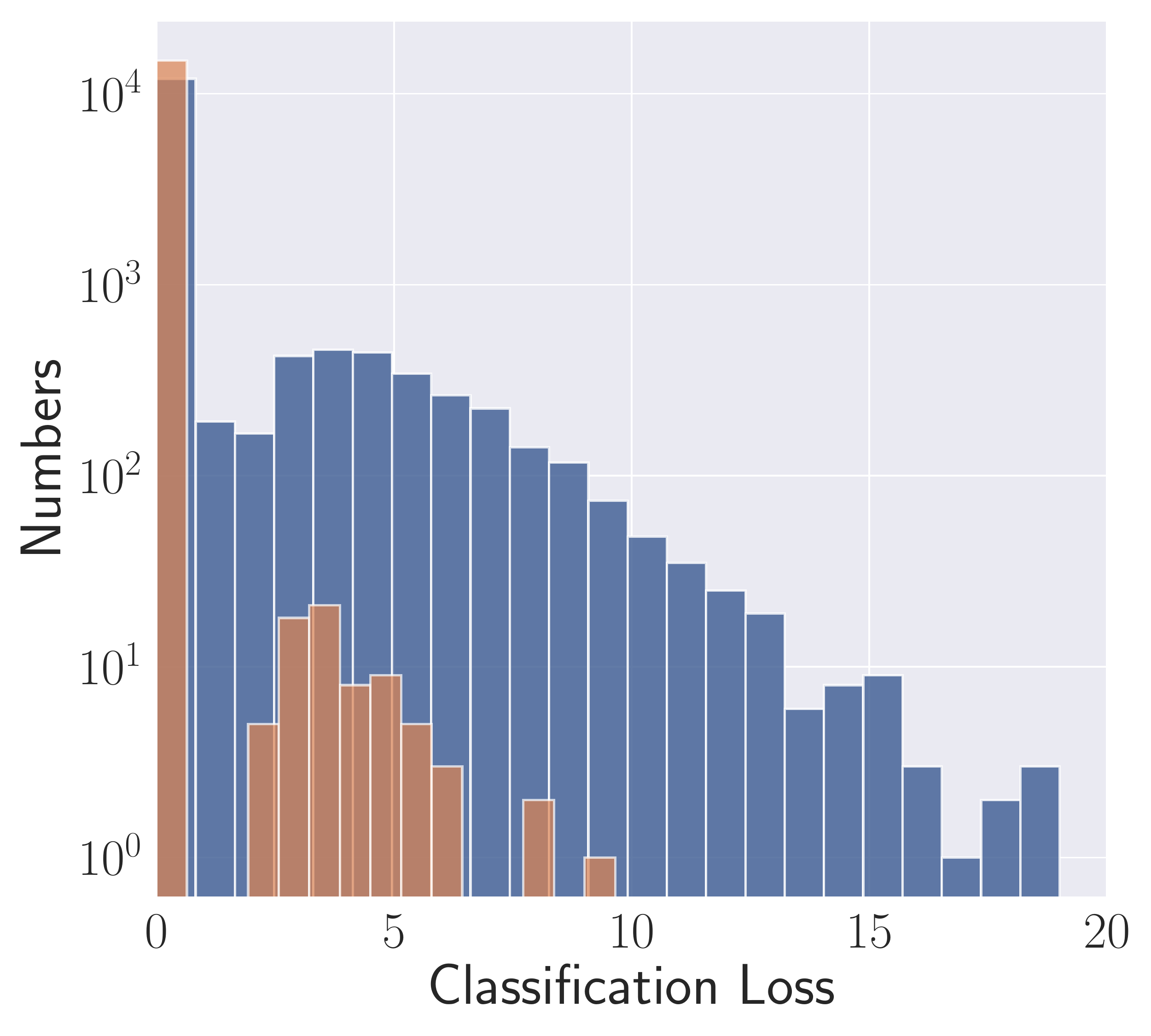}
\caption{Multi-Exit}
\end{subfigure}
\caption{The distribution of loss with respect to original classification tasks for member and non-member samples between the vanilla VGG-16 and the 4-Exit VGG-16 on CIFAR-10.}
\label{fig:vgg_loss_distribution_overall}
\end{figure}

\mypara{Baseline (Vanilla Model).}
To fully understand the membership leakages of multi-exit models, we further use the vanilla model as the baseline model.
We train eight models from scratch for all datasets, including both computer vision and non-computer vision models.
In all cases, including vanilla and multi-exit models, we adopt cross-entropy as the loss function and Adam as the optimizer, and train them for 100 epochs.
Our code is implemented in Python 3.8 and PyTorch 1.8.1 and runs on an NVIDIA HGX-A100 server with Ubuntu 18.04.

\mypara{Metric.}
Following previous work, we adopt the accuracy, i.e., attack success rate (denoted as \texttt{ASR}) thorough the paper, as the attack model’s training and testing datasets are both balanced with respect to membership distribution.
Note that we average the performance of different multi-exit models with the number of exits varying from 2 to 6 and report the mean and standard deviation.
Besides, our evaluation adopts different datasets, architectures, and attack methods, which inevitably lead to a wide variety of results.

\subsection{Results}

\mypara{Classification Accuracy and Computational Cost.} 
We first show the performance of vanilla and multi-exit models on their original classification tasks and computational costs in \autoref{fig:classification_performance}.
See more results in Appendix \autoref{fig:appendix_classification_performance}.
We observe that the multi-exit model performs at least on par with the vanilla model on the classification task, but is much better in terms of computational cost.
For instance, the multi-exit VGG-16 trained on CIFAR-10 achieves 80.558\% accuracy, which is better than 80.04\% accuracy of vanilla VGG-16.
As for the computational cost, the multi-exit VGG-16 achieves 0.3125 \texttt{Gops} while the vanilla model achieves 0.6283 \texttt{Gops}.

\mypara{Attack \texttt{ASR} Score.} 
Regarding membership inference against vanilla and multi-exit models, we report \texttt{ASR} score on all datasets and model architectures in \autoref{fig:attacASR}.
See more results in Appendix \autoref{fig:appendix_attackASR}.
We can observe that all the multi-exit models have lower \texttt{ASR} than the vanilla models.
For example, score-based \texttt{ASR} on vanilla VGG-16 trained on CIFAR-100 is 0.8738, while the mean \texttt{ASR} on multi-exit VGG-16 is only 0.5914. 
Label-only \texttt{ASR} on vanilla FCN-18-1 trained on Locations is 0.8866, while the mean \texttt{ASR} on multi-exit VGG-16 is 0.7831.
However, these results may lead to premature claims of privacy.
\autoref{sec:hybrid} presents that the membership leakage risks stemming from our hybrid attack are much more severe than shown by existing attacks.

\begin{figure}[!t]
\centering
\begin{subfigure}{0.49\columnwidth}
\includegraphics[width=\columnwidth]{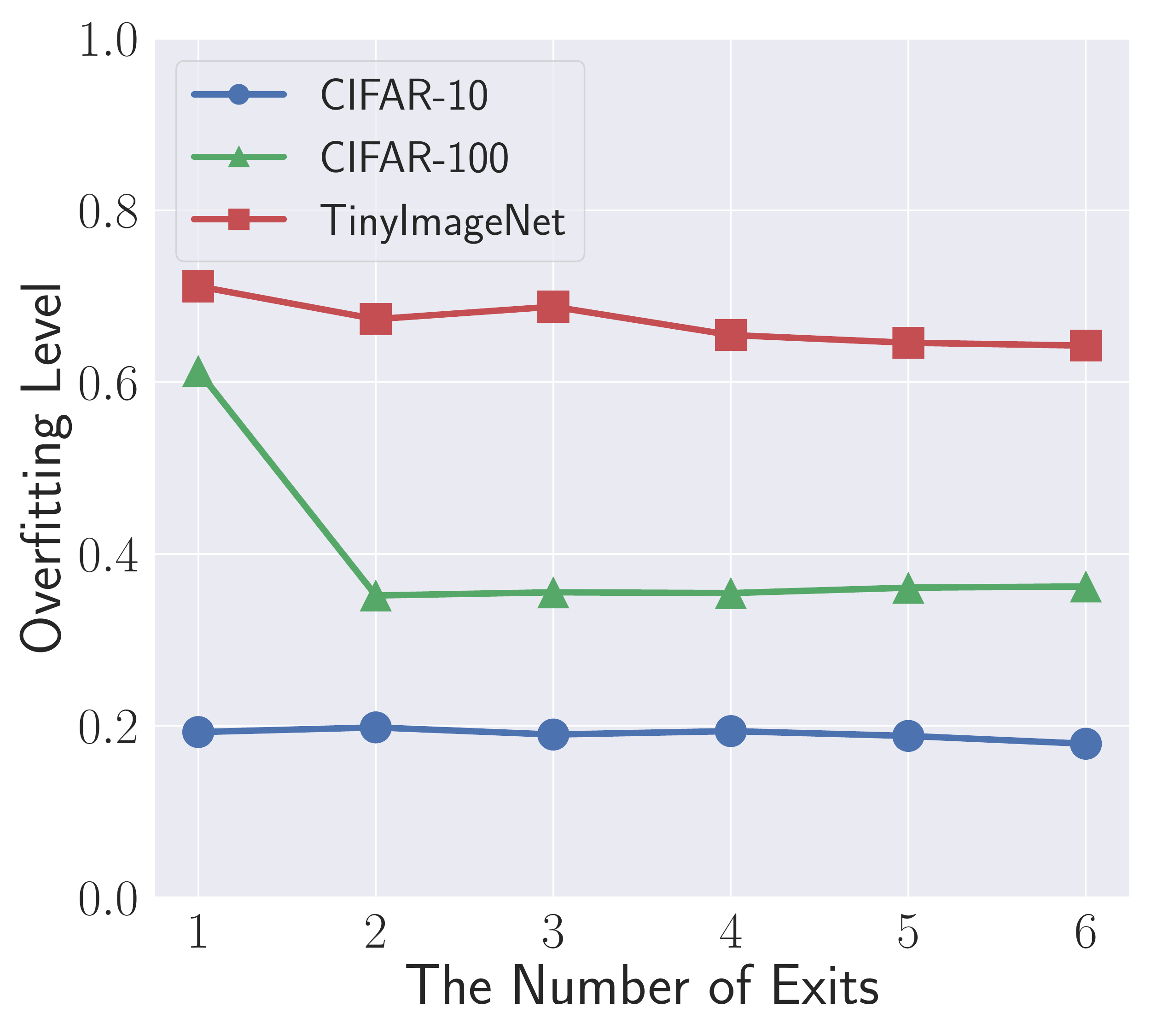}
\caption{VGG-16}
\label{fig:overfitting_a}
\end{subfigure}
\begin{subfigure}{0.49\columnwidth}
\includegraphics[width=\columnwidth]{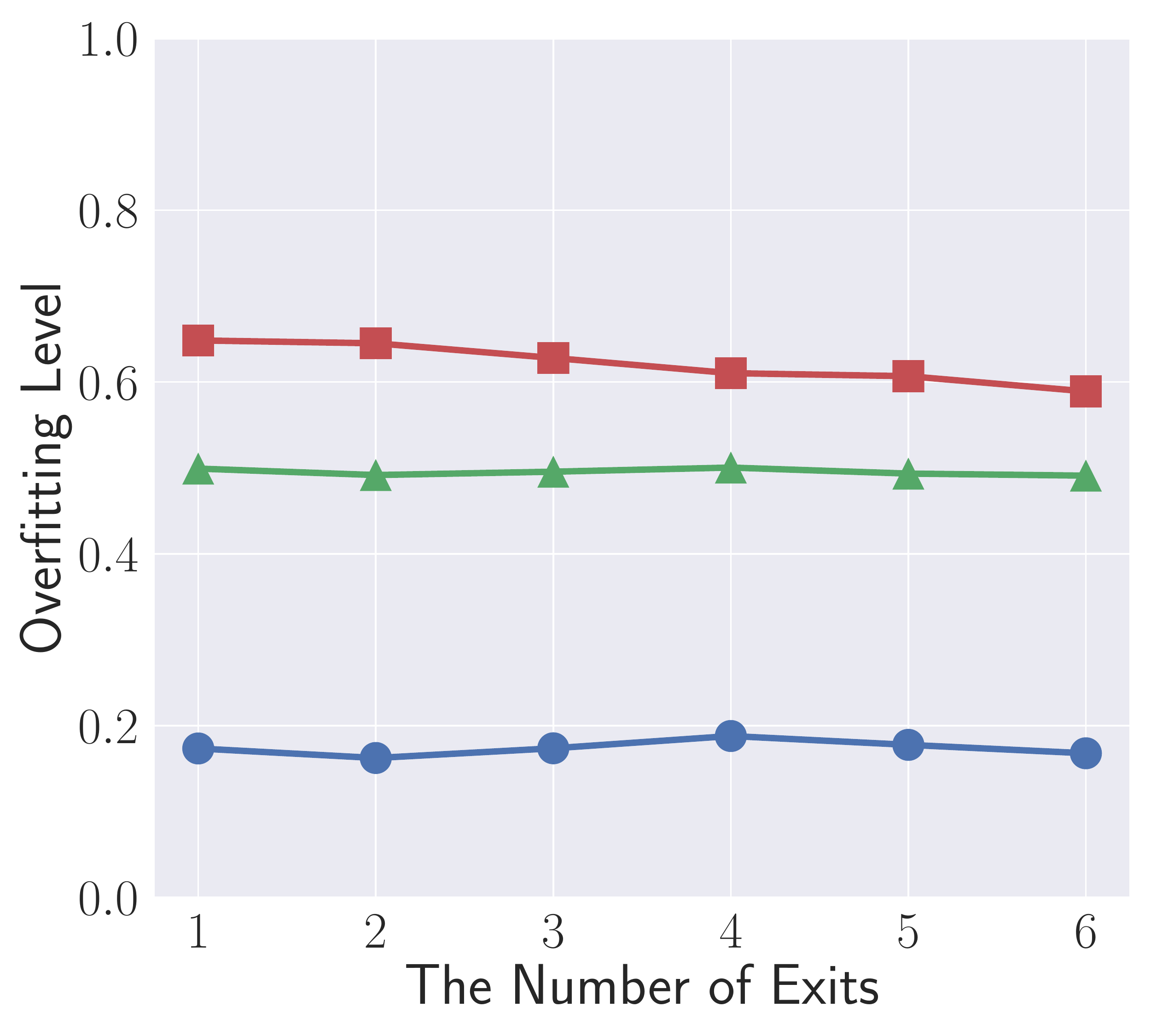}
\caption{WideResNet-32}
\label{fig:overfitting_b}
\end{subfigure}
\caption{Comparison of overfitting levels between vanilla and multi-exit model. 
    Note that 1 exit represents the vanilla model and 2-6 exits represent different multi-exit models.}
\label{fig:overfitting}
\end{figure}

\mypara{Overfitting Level.}
Here, we delve more deeply into the reasons for the less vulnerability of multi-exit models.
As almost all previous works~\cite{SZHBFB19,SSSS17,LZ21,SM21,LLR21,HZ21,YGFJ18,NSH18} claim that the overfitting level is the main factor contributing to the vulnerability of the model to membership inference, i.e., a lower overfitting level leads to less vulnerability to membership inference.
Here, we also relate this to the different overfitting levels of ML models.
The overfitting level of a given model is measured by calculating the difference between its training accuracy and testing accuracy, i.e., subtracting testing accuracy from training accuracy, which is adopted by previous works.
In \autoref{fig:overfitting}, however, we see that the overfitting level of multi-exit models remains almost the same compared to the vanilla model, especially in VGG-16 and WideResNet-32 trained CIFAR-10 dataset.
More results on other datasets can be found in \autoref{fig:overfitting_rest}.


This observation which is contradictory to the previous conclusion inspires us to rethink the relationship between overfitting levels and vulnerability to membership inference.
More precisely, we argue that the current calculation, i.e., subtracting test accuracy from training accuracy, is not the best way to characterize overfitting level, which leads to no strong correlation between overfitting level and vulnerability to membership inference, at least for multi-exit models.


\begin{figure*}[!t]
\centering
\begin{subfigure}{0.5\columnwidth}
\includegraphics[width=\columnwidth]{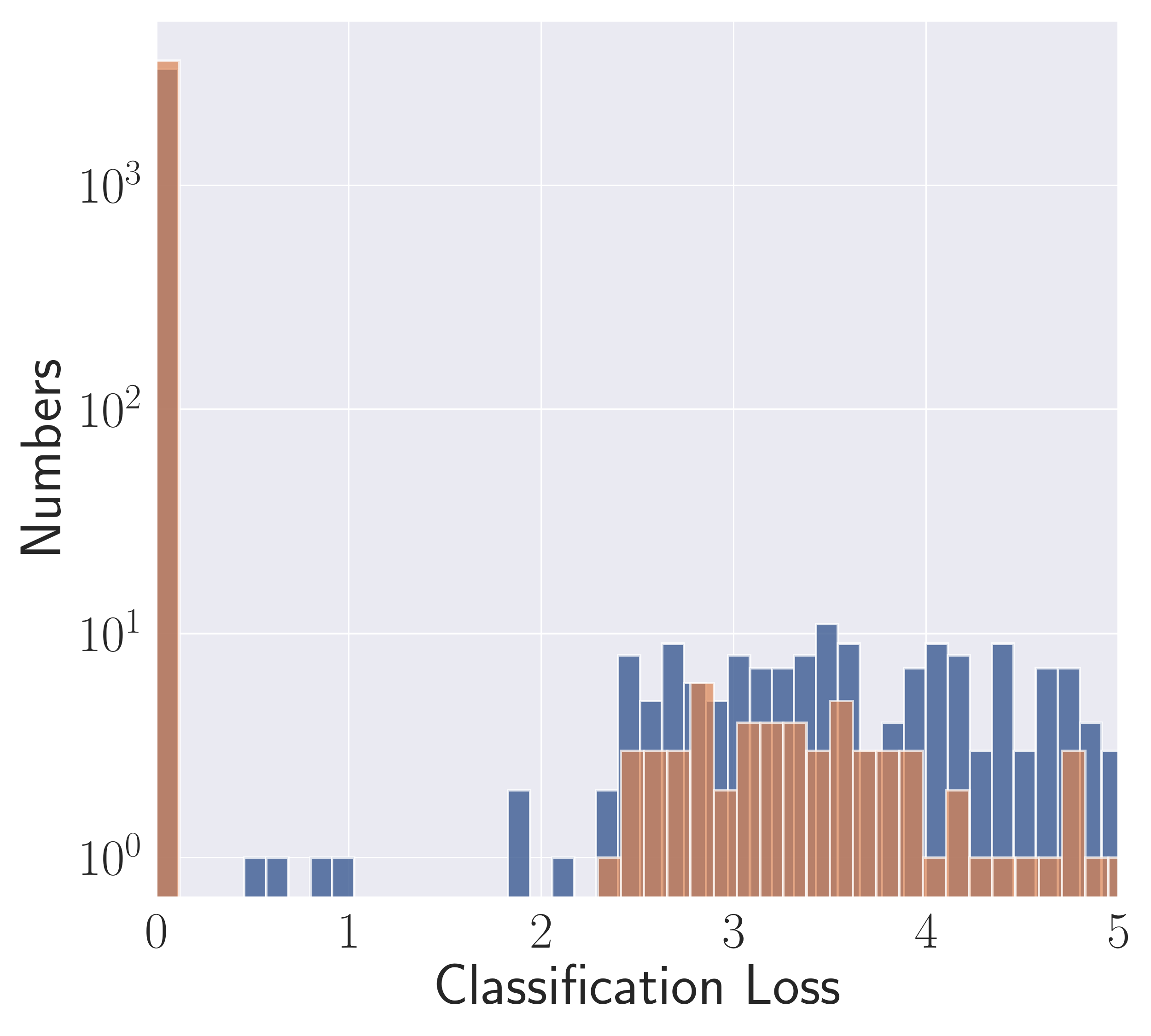}
\caption{Exit \#0}
\end{subfigure}
\begin{subfigure}{0.5\columnwidth}
\includegraphics[width=\columnwidth]{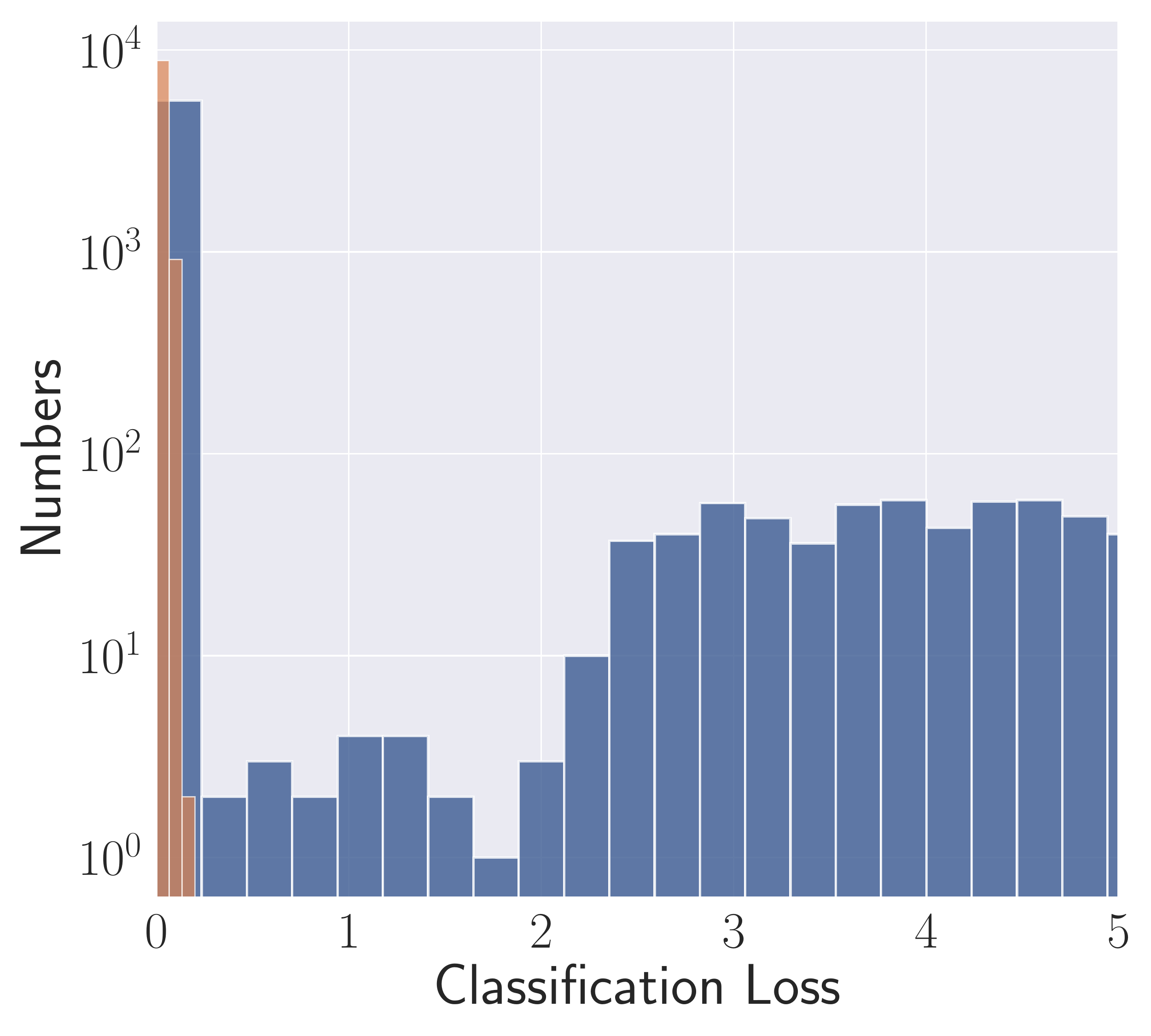}
\caption{Exit \#1}
\end{subfigure}
\begin{subfigure}{0.5\columnwidth}
\includegraphics[width=\columnwidth]{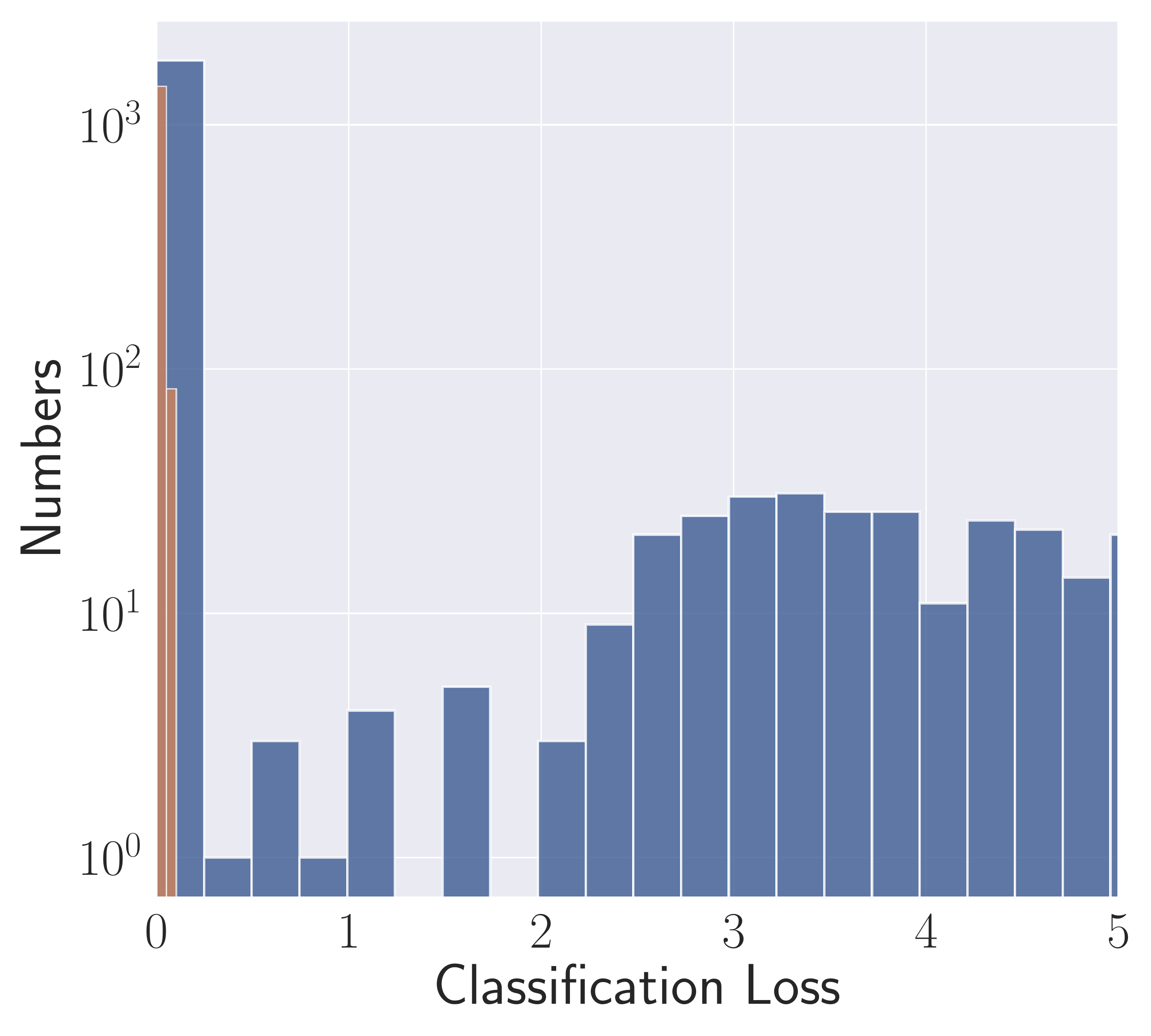}
\caption{Exit \#2}
\end{subfigure}
\begin{subfigure}{0.5\columnwidth}
\includegraphics[width=\columnwidth]{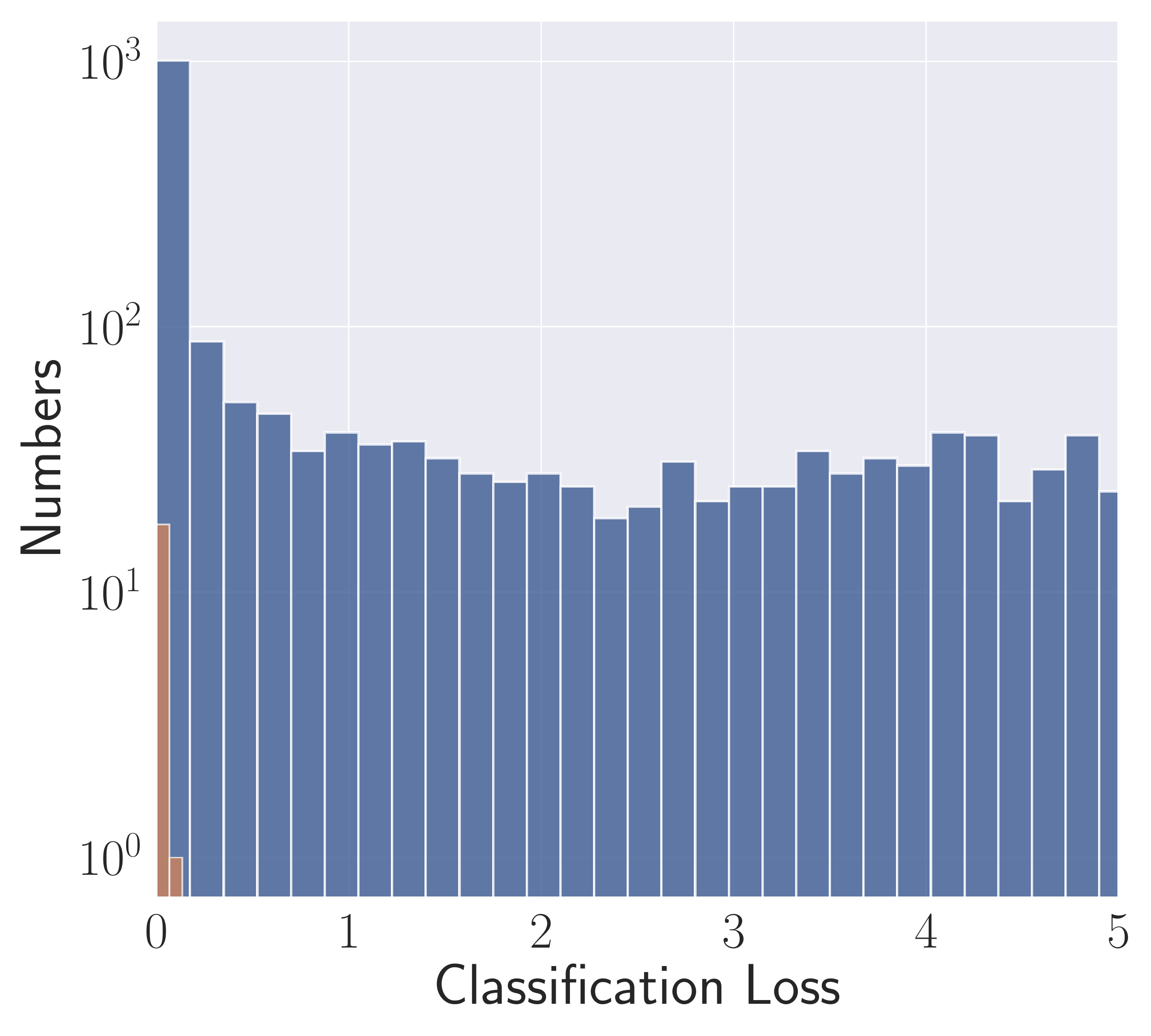}
\caption{Exit \#3}
\end{subfigure}
\caption{The distribution of loss with respect to original classification tasks for member and non-member samples of 4-Exit VGG-16 on CIFAR-10.}
\label{fig:vgg_loss_distribution_exit_taken}
\end{figure*}

\mypara{Loss Distribution.}
To find a more appropriate way to characterize the overfitting level and also to further investigate why the multi-exit model is less vulnerable to membership inference, we analyze the loss distribution between members and non-members in both vanilla and multi-exit models.
Due to space limitations, we only show the results of VGG-16 trained on the CIFAR-10 dataset in \autoref{fig:vgg_loss_distribution_overall}.
See Appendix \autoref{fig:vgg_TN_loss_distribution} for more results. 
A clear trend is that compared to the vanilla VGG-16, the multi-exit VGG-16 has a much lower divergence between the classification loss (cross-entropy) for members and non-members, especially the  classification loss of members becomes larger.
Note that in \autoref{fig:overfitting_a}, the overfitting level calculated by subtracting test accuracy from training accuracy is almost the same between vanilla and multi-exit VGG-16 trained on CIFAR-10.

\begin{figure}[!t]
\centering
\begin{subfigure}{0.49\columnwidth}
\includegraphics[width=\columnwidth]{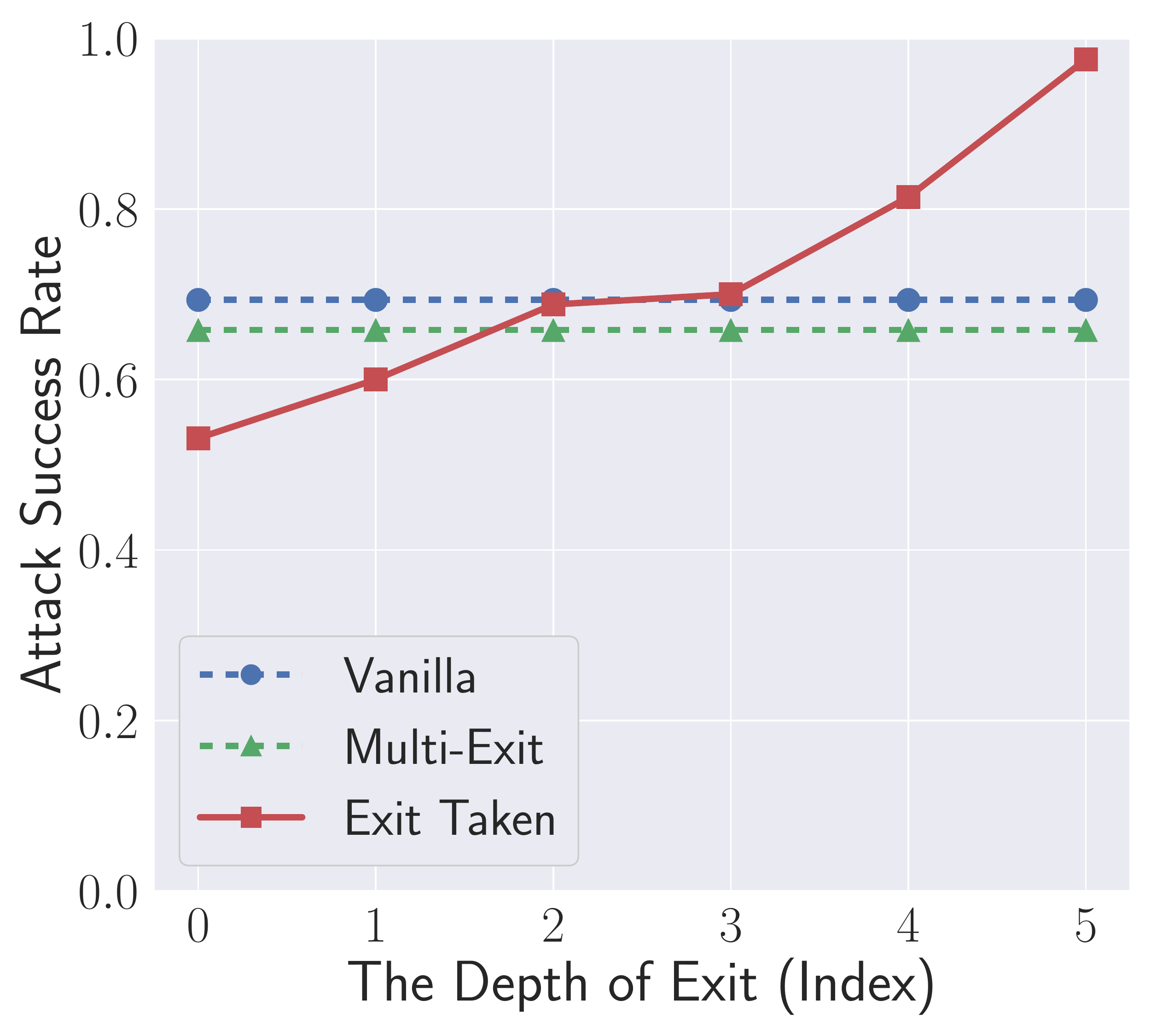}
\caption{CIFAR-10, VGG-16}
\end{subfigure}
\begin{subfigure}{0.49\columnwidth}
\includegraphics[width=\columnwidth]{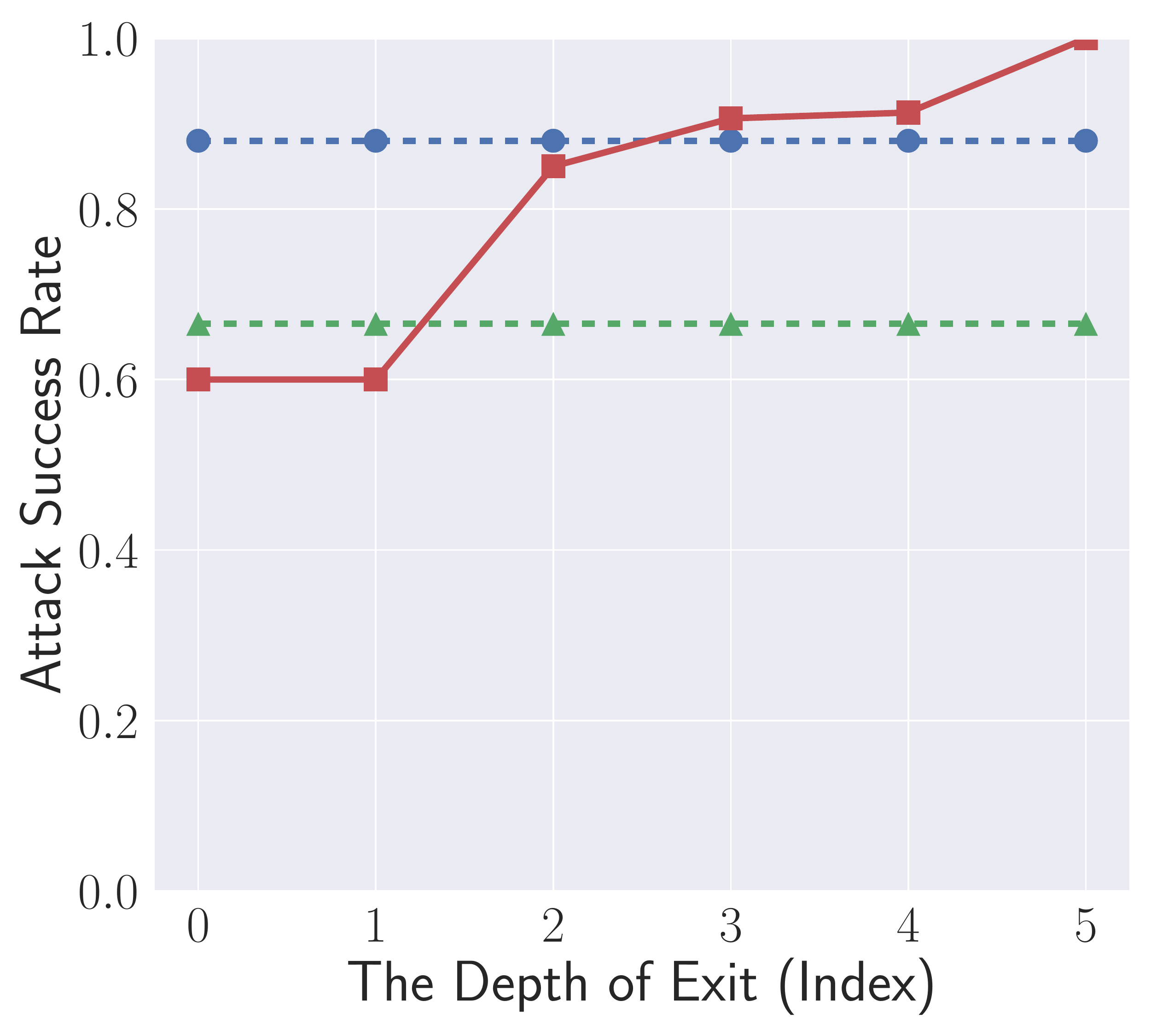}
\caption{CIFAR-100, ResNet-56}
\end{subfigure}
\caption{The attack performance of original label-only attack against vanilla model and 6-exit model with one specific exit (i.e., “exit taken”).}
\label{fig:ASR_exit_taken}
\end{figure}
\begin{figure}[!t]
\centering
\begin{subfigure}{0.49\columnwidth}
\includegraphics[width=\columnwidth]{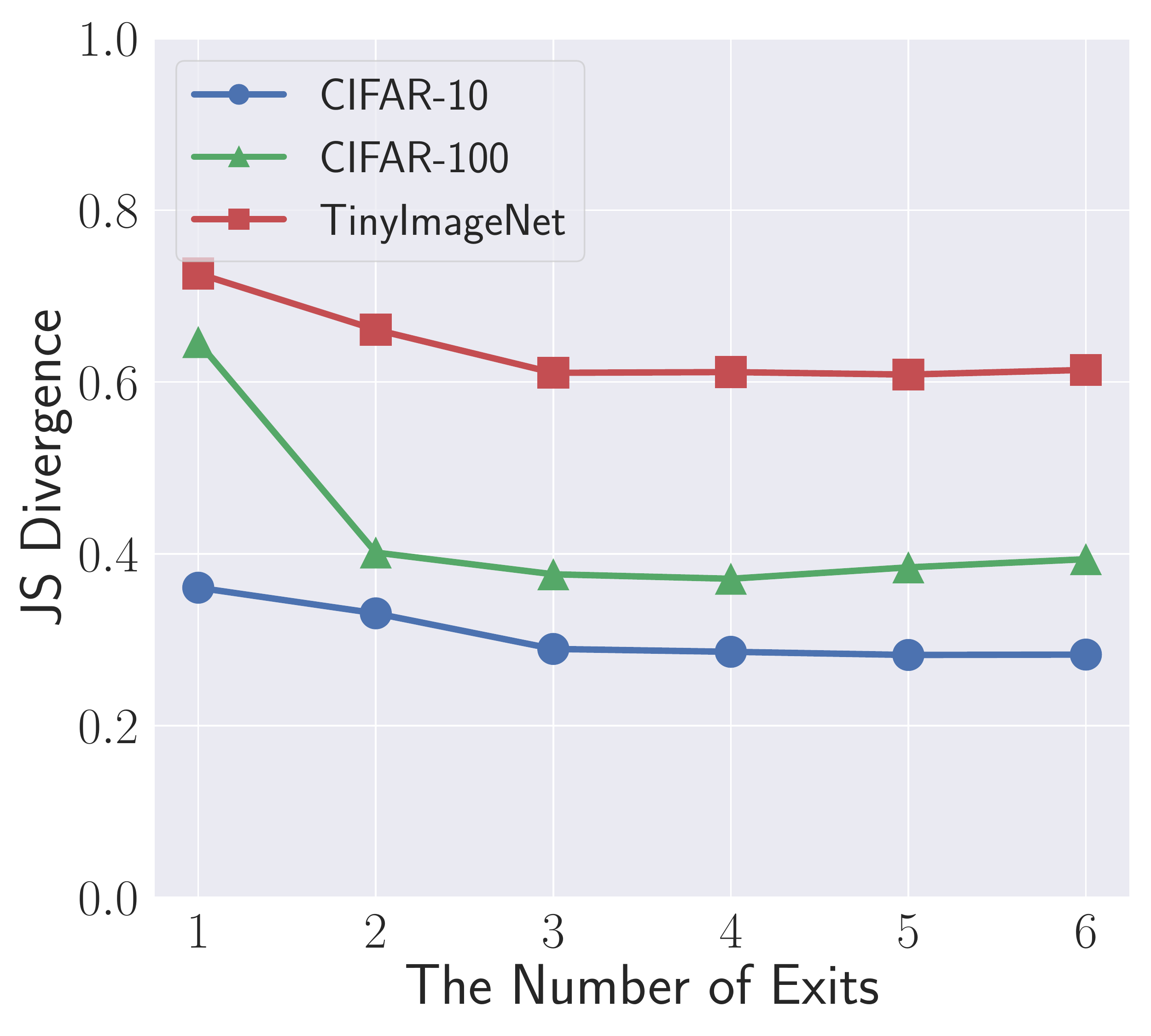}
\caption{VGG-16}
\end{subfigure}
\begin{subfigure}{0.49\columnwidth}
\includegraphics[width=\columnwidth]{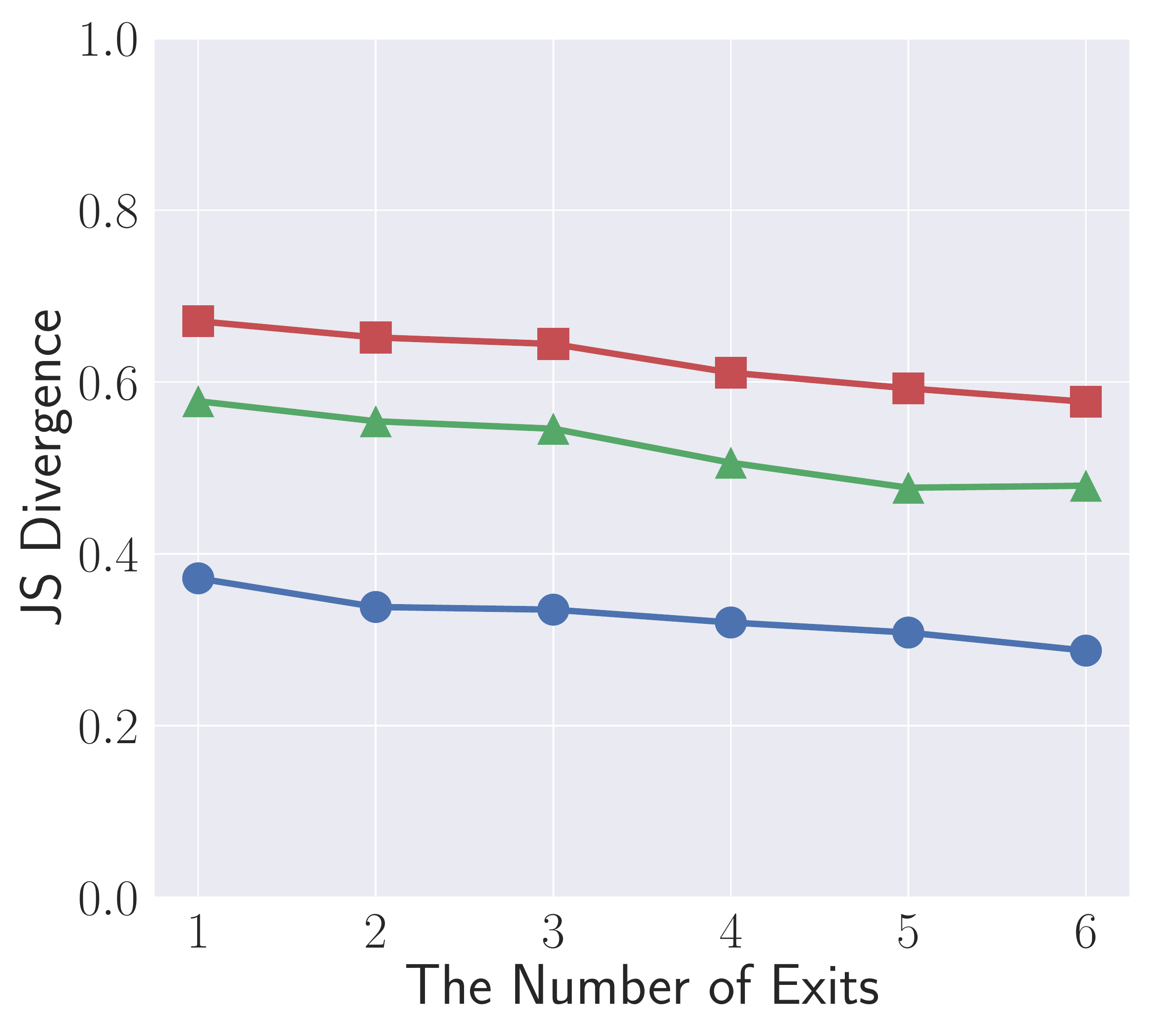}
\caption{WideResNet-32}
\end{subfigure}
\caption{Comparison of \texttt{JS} divergence between vanilla and multi-exit model. 
    Note that 1 exit represents the vanilla model and 2-6 exits represent different multi-exit models.}
\label{fig:stand_JS}
\end{figure}

We further report the loss distribution and attack performance by ``exit taken'', i.e., specific exit depth of one certain multi-exit model.
\autoref{fig:vgg_loss_distribution_exit_taken} shows the the loss distribution for members and non-members by ``exit taken'' for 4-exit VGG-16.
We can see that the divergence between the loss distribution for members and non-members is becoming larger by exit depth.
In other words, the early exits are less overfitting while the latter exits are much more overfitting.
We further report the attack \textit{ASR} score of original label-only attack against vanilla, multi-exit, and multi-exit with one specific exit (i.e., “exit taken”) in \autoref{fig:ASR_exit_taken}.
We can find that latter exits are much more vulnerable to membership leakages, which is consistent with the trend in loss distribution of “exit taken”.

Based on the above observation, we believe that calculating the divergence between the loss distribution of members and non-members can better characterize the overfitting level.
More concretely, we leverage Jensen-Shannon (denoted as \texttt{JS}) divergence, a widely used metric, to measure the distance of two probability distributions~\cite{GPMXWOCB14}.
In \autoref{fig:stand_JS}, we display \texttt{JS} divergence between the classification loss of members and non-members with respect to the number of exits.
We can see that the \texttt{JS} divergence of multi-exit models is clearly lower than that of vanilla models.
These results show that \texttt{JS} divergence is indeed a better way to characterize the overfitting level, compared to subtracting test accuracy from training accuracy.


\begin{figure}[!t]
\centering
\begin{subfigure}{0.49\columnwidth}
\includegraphics[width=\columnwidth]{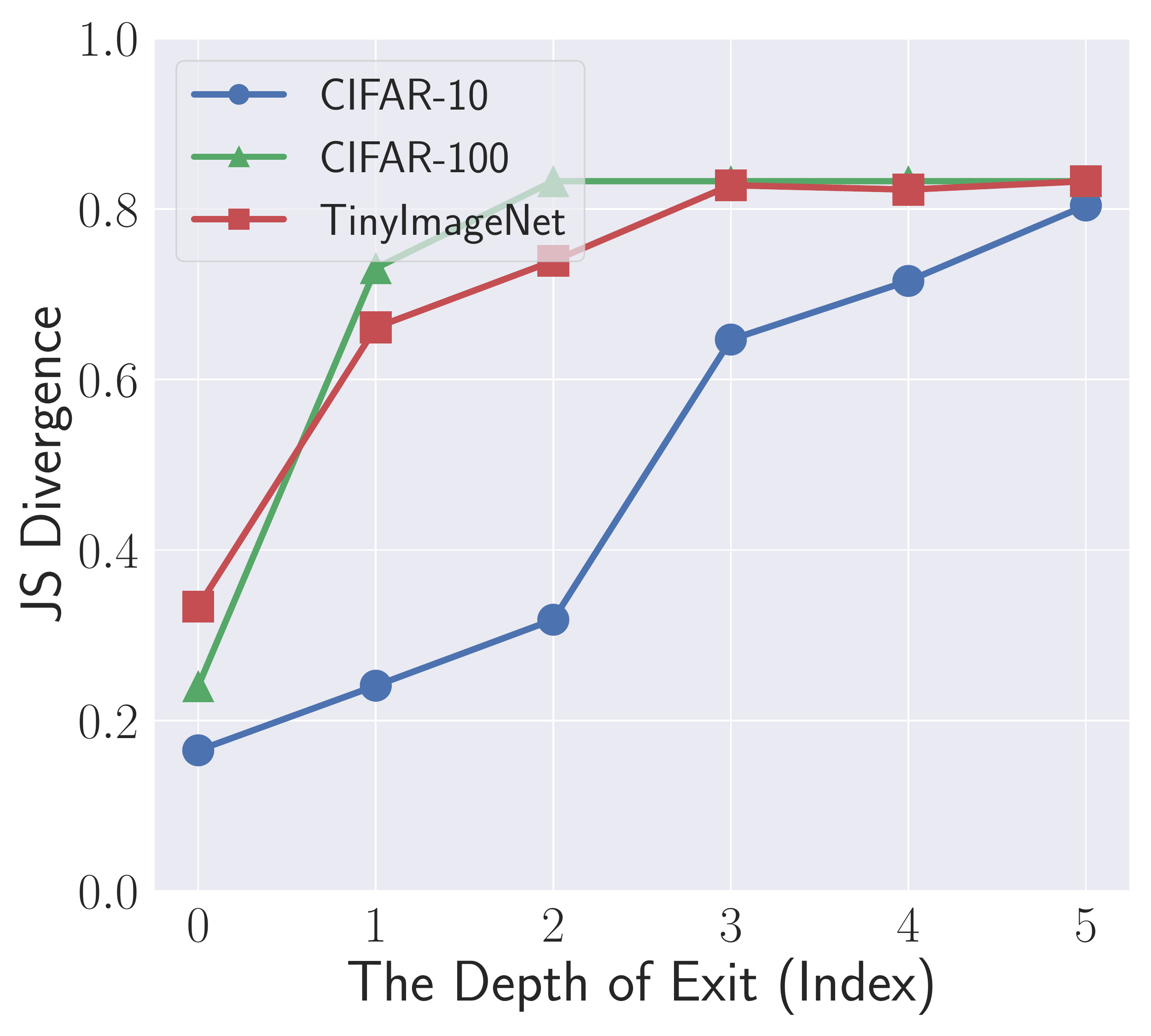}
\caption{VGG-16 (6 Exits)}
\end{subfigure}
\begin{subfigure}{0.49\columnwidth}
\includegraphics[width=\columnwidth]{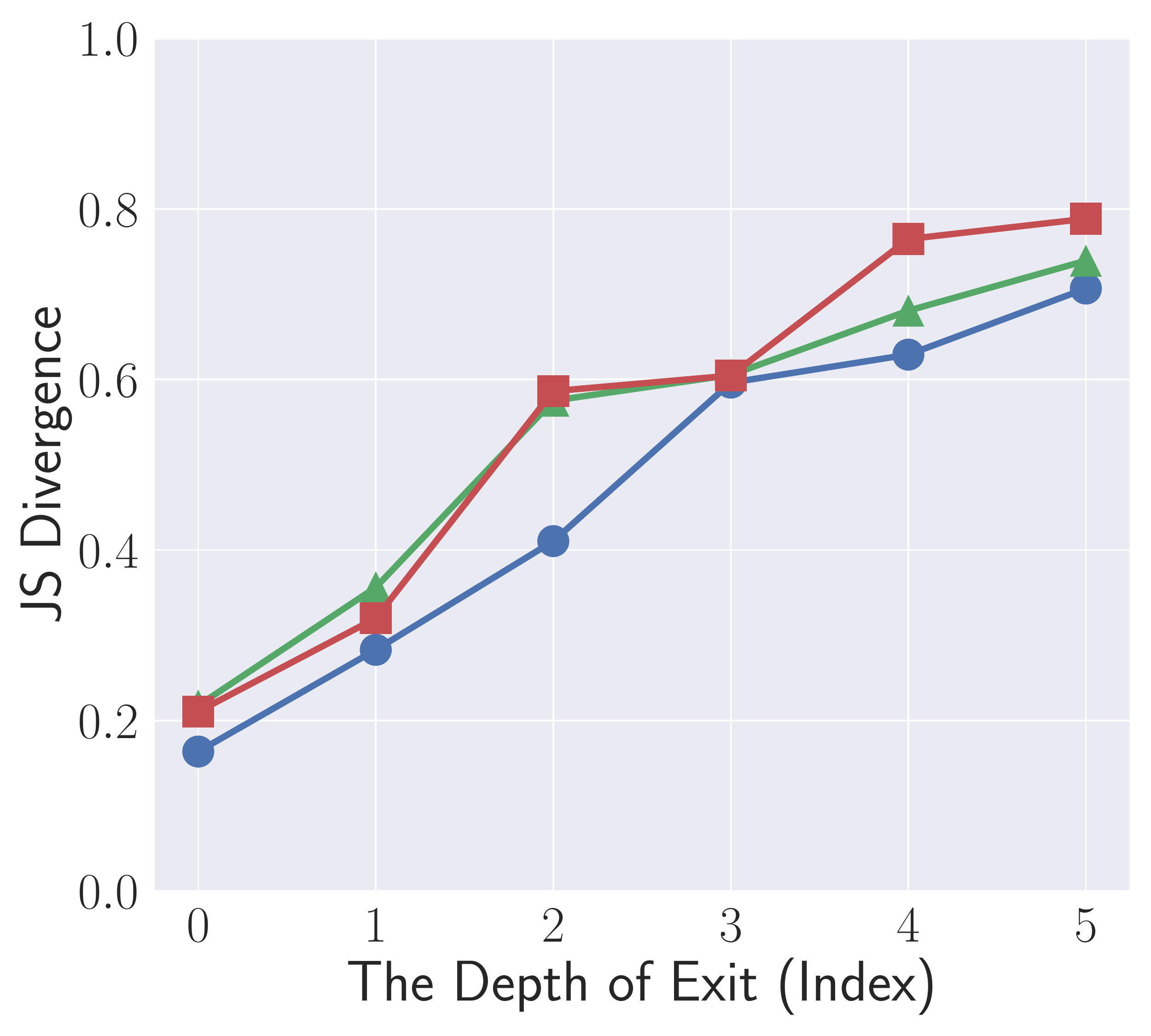}
\caption{WideResNet-32 (6 Exits)}
\end{subfigure}
\caption{The \texttt{JS} divergence of classification loss with respect to the depth of exits. 
    The x-axis represents the depth of exit. The y-axis represents the \texttt{JS} divergence.}
\label{fig:JS_6_exits}
\end{figure}

\mypara{Effects of the Number of Exits.} 
We further investigate the effects of the number of exits attached to the backbone models.
More interestingly, in \autoref{fig:stand_JS}, we can also find the model with more number of exits leads to lower divergence.
This indicates that the number of exits is negatively correlated to the vulnerability to membership leakages.
The reason is that more exits attached to the backbone model mean that more data samples leave the earlier exit points than the final exit points, which makes the model less likely to be overfitted.

\mypara{Effects of the Depth of Exits.} 
Here we investigate the effects of the depth of exits attached to the backbone models.
Given a backbone model with 6 exits, we use the exit index (from 0 to 5) to represent the depth of exits.
We calculate the \texttt{JS} divergence for members and non-members of each exit point separately. 
As shown in \autoref{fig:JS_6_exits}, we can see that the \texttt{JS} divergence increases with the depth of exit.
These results indicate that the depth of exits is positively correlated to the vulnerability to membership leakages, i.e., the samples leaving from deeper exit points were easier to distinguish between members and non-members.
The reason for this observation is that deeper exit points imply higher capacity models, which are more likely to be overfitted to the training set.

\section{Hybrid Attack}
\label{sec:hybrid}

After quantifying membership leakages of multi-exit models, we conclude that the multi-exit models are less vulnerable to membership leakages and, more interestingly, we find that exit information is highly correlated with attack performance.
The latter motivates us to present a new research question: \textit{Can extra exit information (number and depth) of the multi-exit model leak more membership information about the training set?}.
Before answering the above research question, more importantly, we need to answer these two-step questions first:
\begin{itemize}
    \item how to obtain the exit information of target multi-exit models, especially in black-box and label-only scenarios.
    \item how to leverage the exit information to improve existing membership inference attacks.
\end{itemize}

In the next section, we propose a novel hybrid attack that first steals exit information and then exploits the exit information as new knowledge for the adversary.
In particular, we study three different scenarios for hybrid attacks by starting with some strong assumptions and gradually relaxing them to show that far more broadly applicable attack scenarios are possible.


\begin{figure*}[!t]
\centering
\begin{subfigure}{0.33\linewidth}
\includegraphics[width=\linewidth]{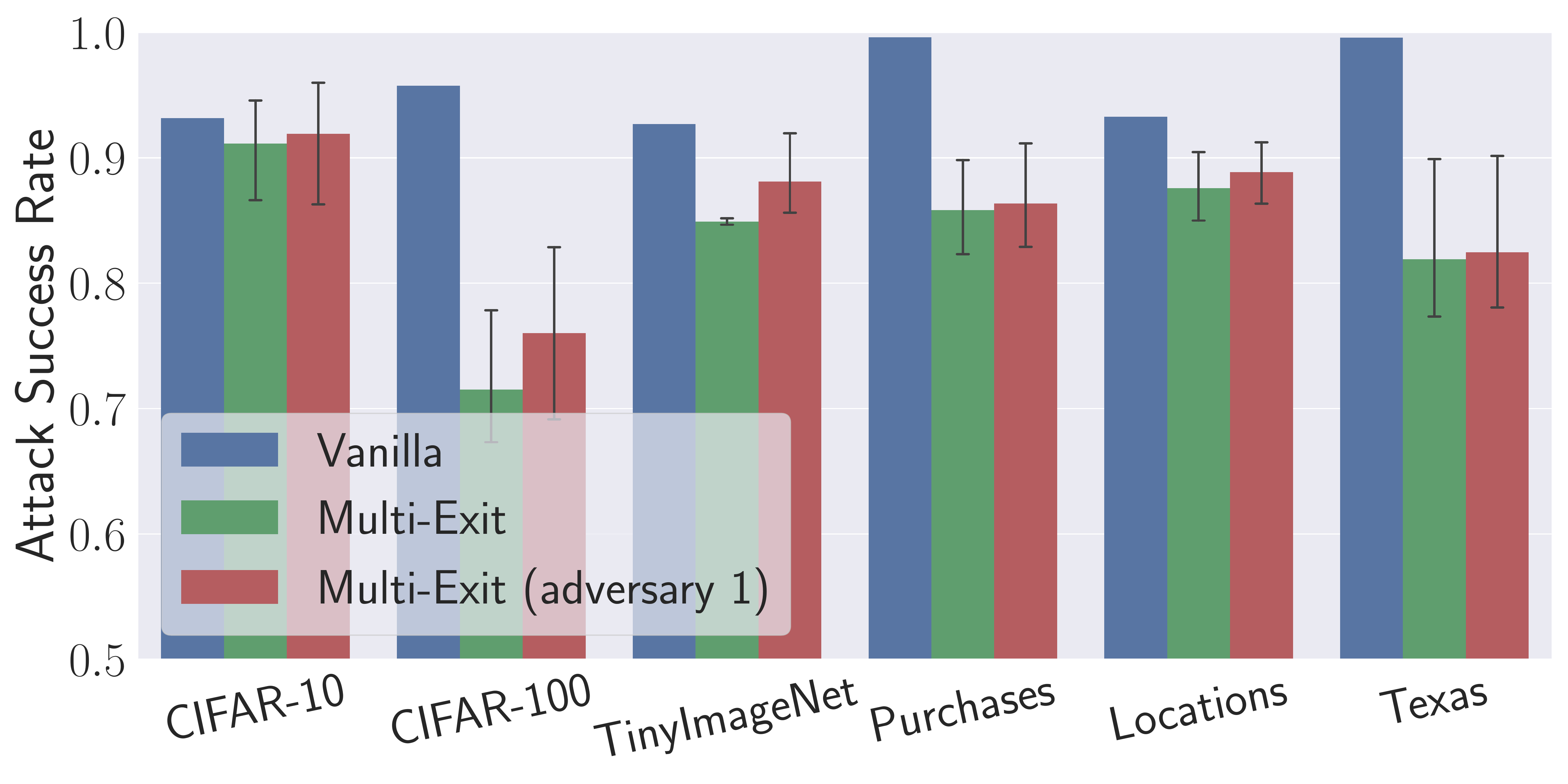}
\caption{Gradient-based}
\end{subfigure}
\begin{subfigure}{0.33\linewidth}
\includegraphics[width=\linewidth]{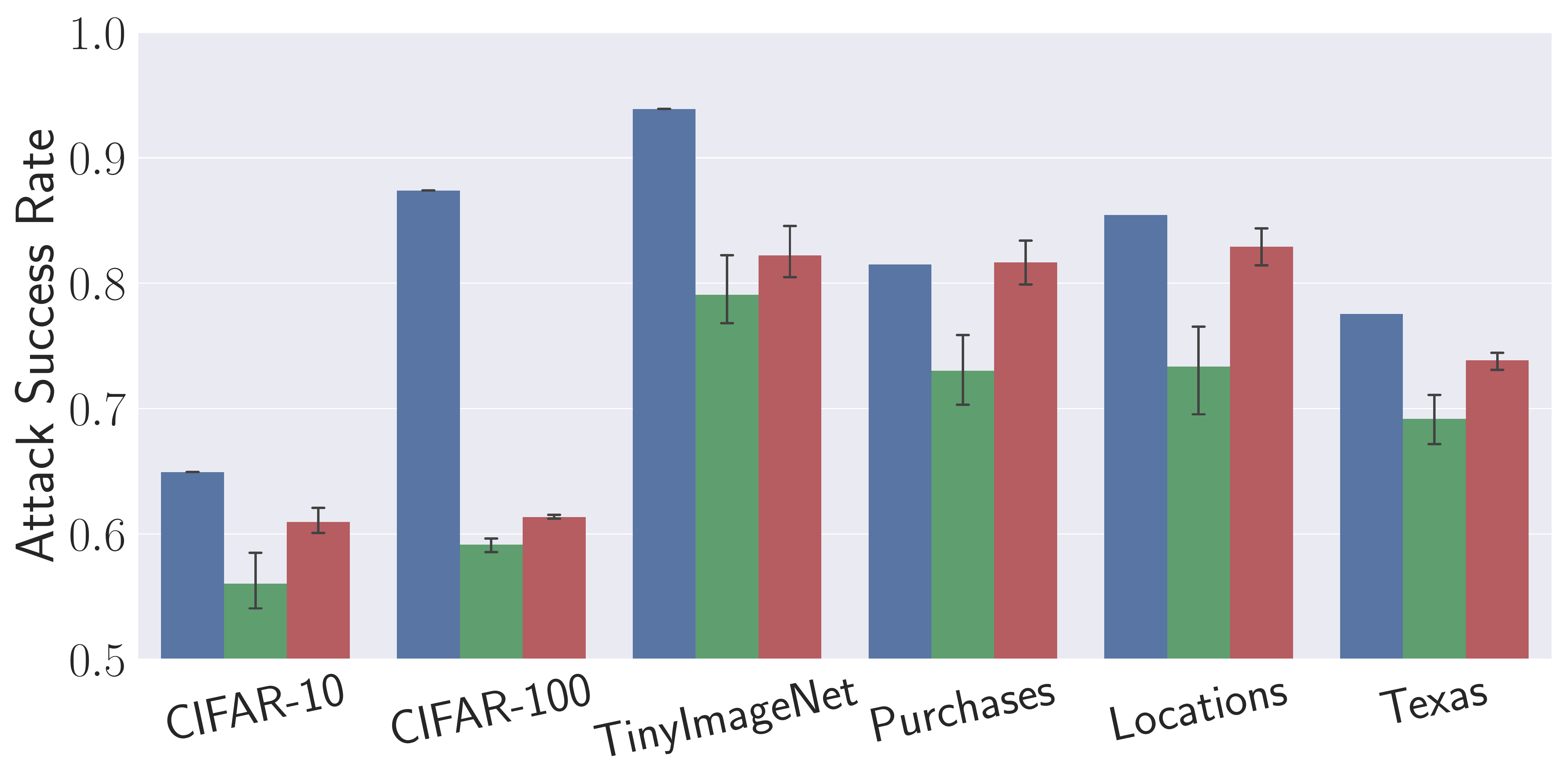}
\caption{Score-based}
\end{subfigure}
\begin{subfigure}{0.33\linewidth}
\includegraphics[width=\linewidth]{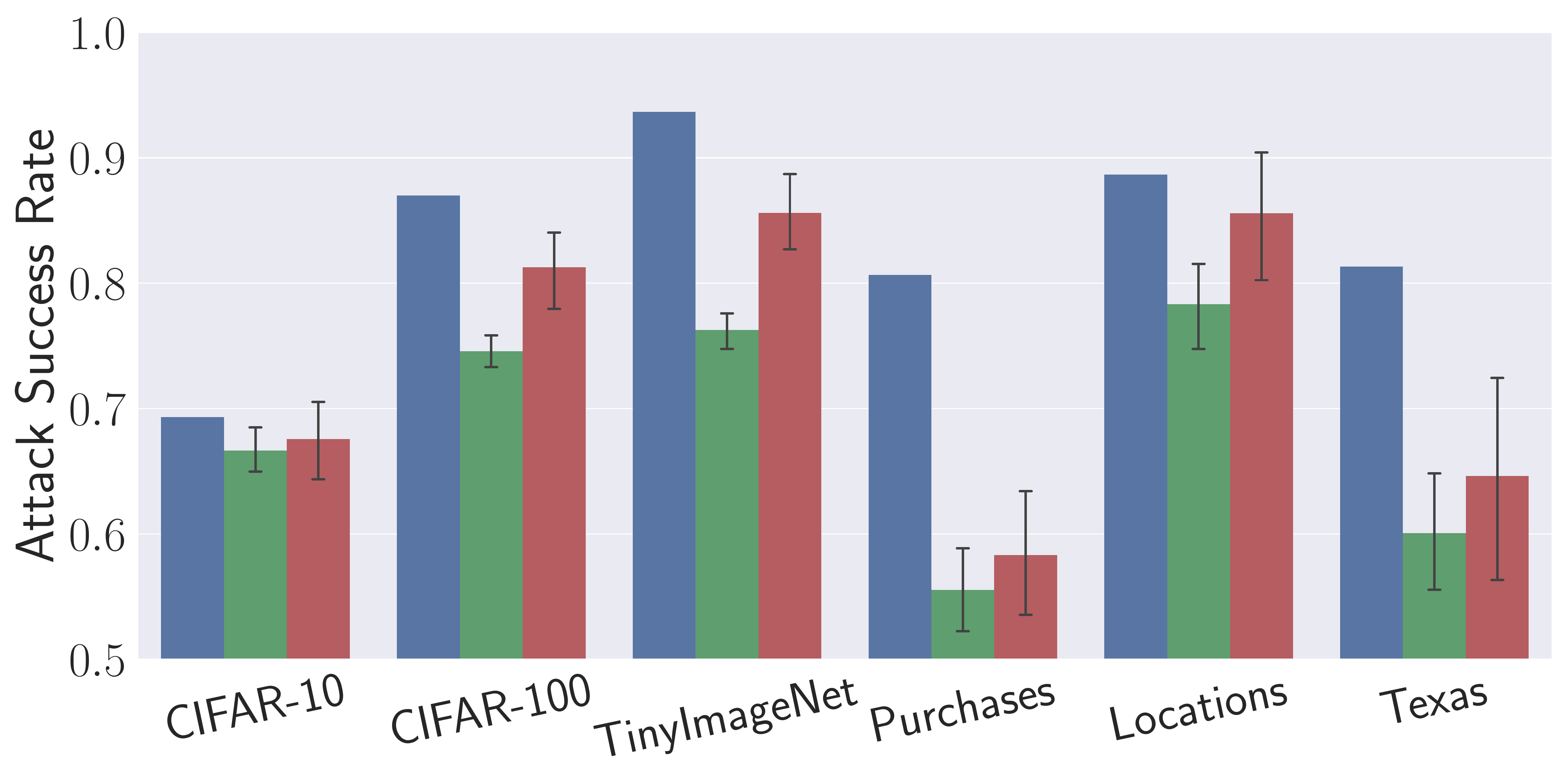}
\caption{Label-only}
\end{subfigure}
\caption{The attack performance of different membership inference attacks on all datasets. 
    The blue and green bars indicate the original attack on the vanilla and multi-exit models, while the red bar indicates our hybrid attack on the multi-exit model. 
    Computer vision tasks are on VGG-16, and non-computer vision tasks are on FCN-18-1.}
\label{fig:adversary_1}
\end{figure*}

\subsection{Adversary 1}
\label{Adersary_1}

In this section, we describe our first adversary considered for leveraging exit information to mount membership inference attacks.
For this adversary, we mainly make a strong assumption about the adversary's knowledge.
In consequence, the research question of whether extra exit information will leak more membership information can be investigated in a more effective and lower-cost way.
In the following, we start by defining the threat model, then describe the adversary’s attack methodology.
In the end, we present a comprehensive experimental evaluation.

\mypara{Threat Model.}
We assume that the adversary has direct access to exit information, i.e., exit depth.
More concretely, given a data sample and a 6-exit model, the model outputs not only predictions (score or label) but also exit information, e.g., predictions from the first exit point (\textit{exit 0}) or the sixth exit point (\textit{exit 5}).
Note that, here we directly consider the exit index as the exit depth.
\textit{exit 0} means the shortest path from the entry point to the first exit, while \textit{exit 5} means the longest path from the entry point to the final exit.

In addition, we make the same assumptions for other settings, such as data knowledge, training knowledge, model knowledge, and output knowledge.
For example, in the gradient-based attack, we keep the assumption unchanged that the adversary has access to the intermediate computations of the target model.

\mypara{Methodology.} 
The methodology is organized into two stages: hyperparameter stealing and enhanced membership inference.

\smallskip
\noindent\textit{Hyperparameter Stealing.} 
The adversary first queries the target model using a large number of data samples, which can come from the shadow dataset or random data samples collected from the Internet.
They then count all exit indexes and sort them from smallest to largest.
Thus, the largest index implies the number of exits attached to the backbone model.

\smallskip
\noindent\textit{Enhanced Membership Inference.}
According to the two different types of attack models used in existing attacks, we propose different methods for each attack model to exploit the exit information.
\begin{itemize}
\item \mypara{MLP Attack Model.} In gradient-based and score-based attacks using MLP as the attack model, given the exit information (number and depth), the adversary first converts it to a one-hot encoding, which is the same as the one-hot encoding of the true label used in the gradient-based attack.
They then provide the one-hot encoding of the exit information and other existing information to the attack model.
\item \mypara{Decision Function.} In the original label-only attack, the adversary measures the magnitude of the perturbation and treats the data samples as members if their magnitude is larger than a predefined threshold.
Here, instead of performing the above operation directly on all data samples, the adversary first separates the data samples according to their exit depths and then performs the above operation to distinguish members and non-members of each exit depth. 
The thresholds are also derived in this way on the shadow model. 
Note that label-only attack proceeds directly to the second stage without the hyperparameter stealing stage, because the adversary does not need to generate one-hot encoding based on the exit depth and number.
\end{itemize}
In addition, all other attack steps are the same as those used in original attacks, such as shadow model training and attack training dataset building.

\mypara{Experimental Setup.}
For the attack models in the gradient-based and score-based attacks, the adversary has six inputs and two inputs, respectively, where the extra one is the one-hot encoding of the exit depths.
Thus the new attack model has one more input component.
For the evaluation metric, we again use the attack success rate (denoted as \texttt{ASR}).
Note that in label-only attacks, we average \texttt{ASR} scores across all exit depths, as \texttt{ASR} is independent of exit depths. Besides that, we use the same experimental setup as presented in \autoref{setup_quantify}, such as the datasets, multi-exit model structures and training settings.

\mypara{Results.}
\autoref{fig:adversary_1} depicts the performance of original attacks and our hybrid attacks (see Appendix \autoref{fig:appendix_attackASR} for more results).
Note that, we also average the performance of multi-exit models with the number of exits varying from 2 to 6 and report the mean and standard deviation. 
Encouragingly, we can observe that hybrid attacks achieve clearly higher \texttt{ASR} score than original attacks, regardless of datasets, architectures and attack types.
These results convincingly demonstrate that extra exit information of the multi-exit model leaks more membership information about training set, compared to original information only.

More interestingly, we also find that compared with gradient-based attacks, the extra exit information used in score-based and label-only attacks can significantly improve the performance of the original attacks.
Recall that gradient-based attacks are applicable in white-box scenarios.
This indicates that the original gradient-based attack has already exploited almost all the information and thus can achieve the attack performance close to the upper bound.
Therefore, in gradient-based attacks, extra exit information can not lead to much higher attack performance gains.
In contrast, we can observe that in label-only attacks, extra exit information leads to much higher attack performance gains.


\begin{figure}[!t]
\centering
\begin{subfigure}{0.49\columnwidth}
\includegraphics[width=\columnwidth]{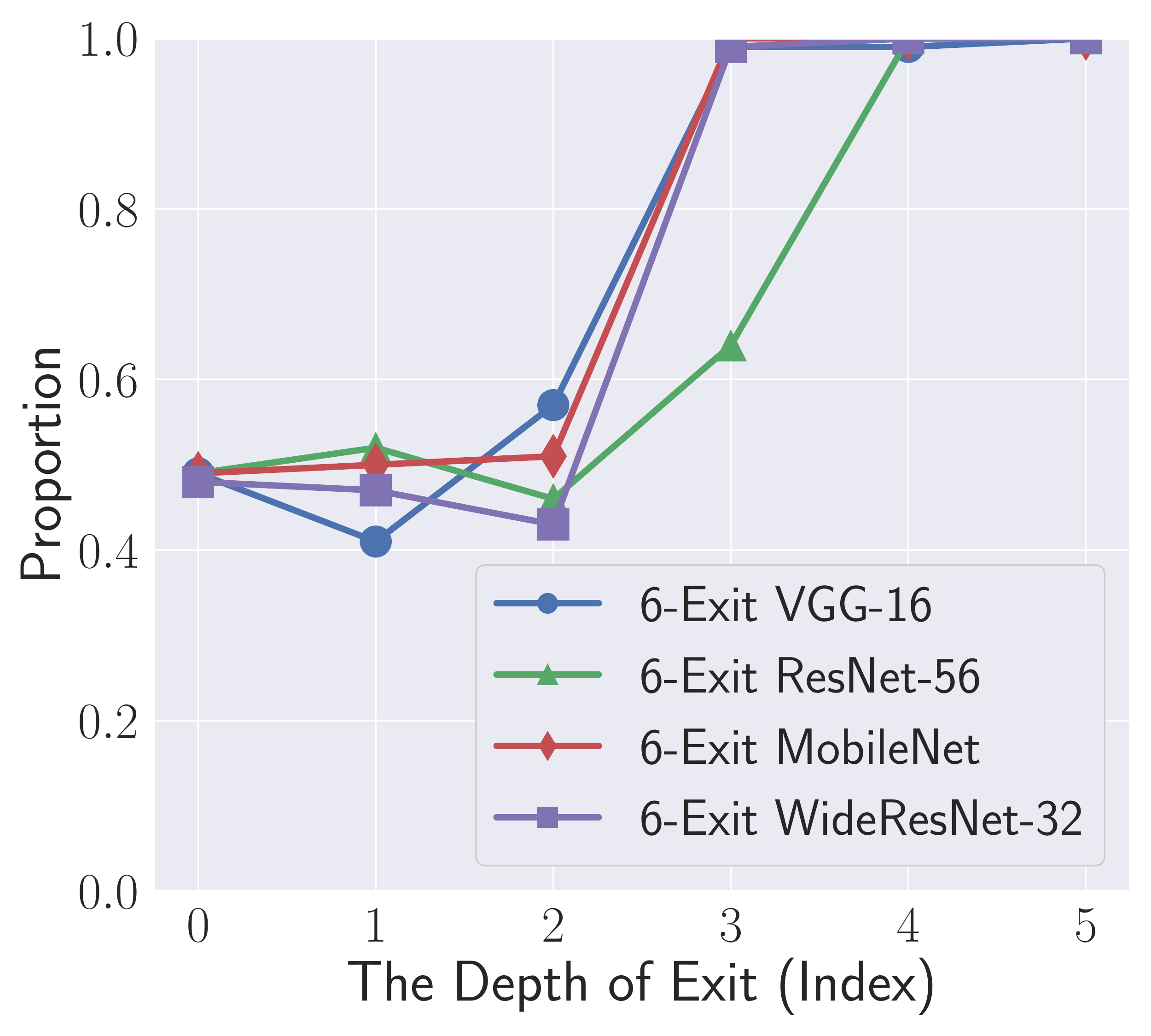}
\caption{CIFAR-10}
\end{subfigure}
\begin{subfigure}{0.49\columnwidth}
\includegraphics[width=\columnwidth]{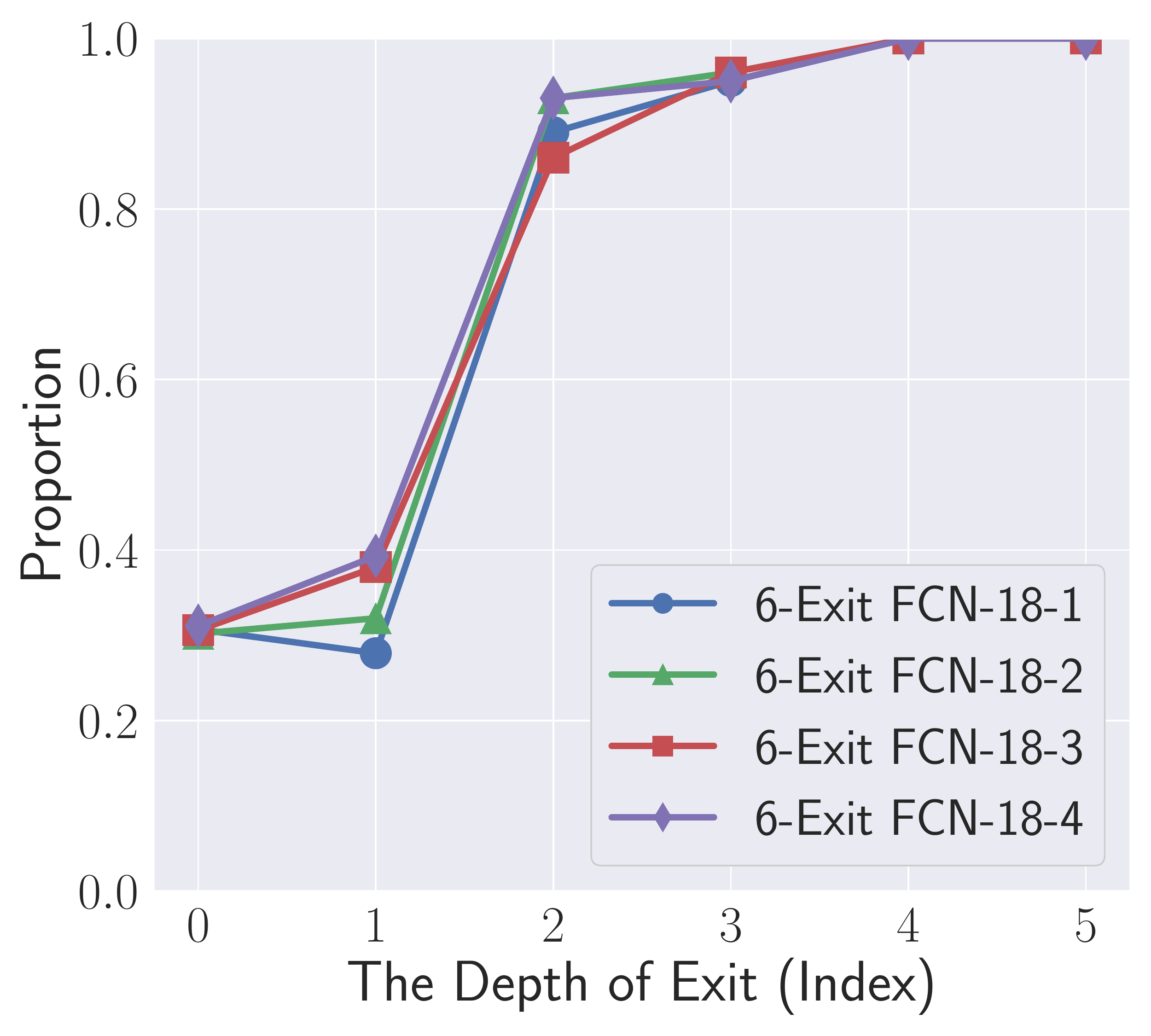}
\caption{Purchases}
\end{subfigure}
\caption{Proportion of non-members in all samples leaving at each exit.}
\label{fig:ratio}
\end{figure}


\begin{figure*}[!t]
\centering
\begin{subfigure}{0.99\columnwidth}
\includegraphics[width=\columnwidth]{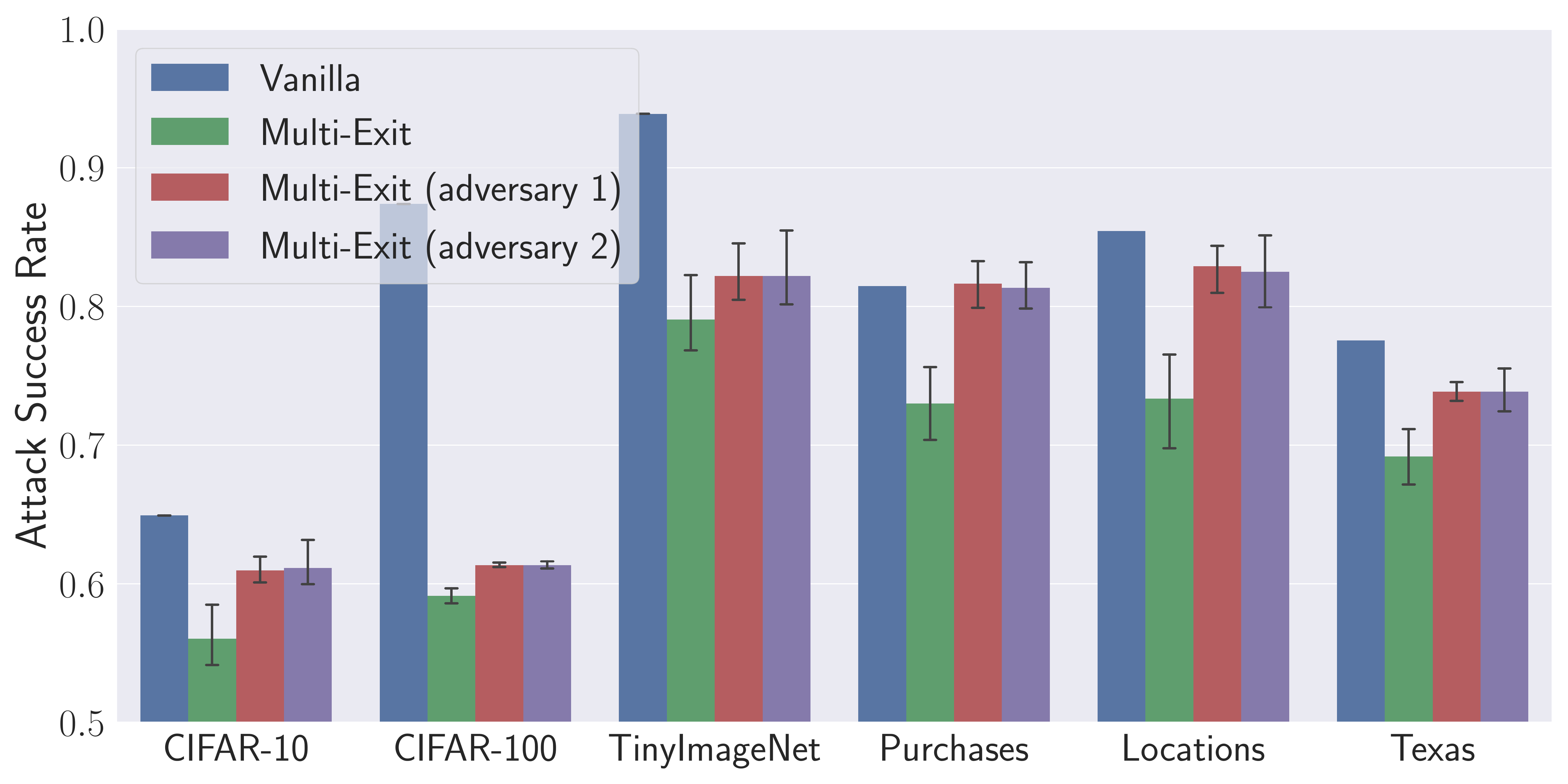}
\caption{Score-based}
\end{subfigure}
\begin{subfigure}{0.99\columnwidth}
\includegraphics[width=\columnwidth]{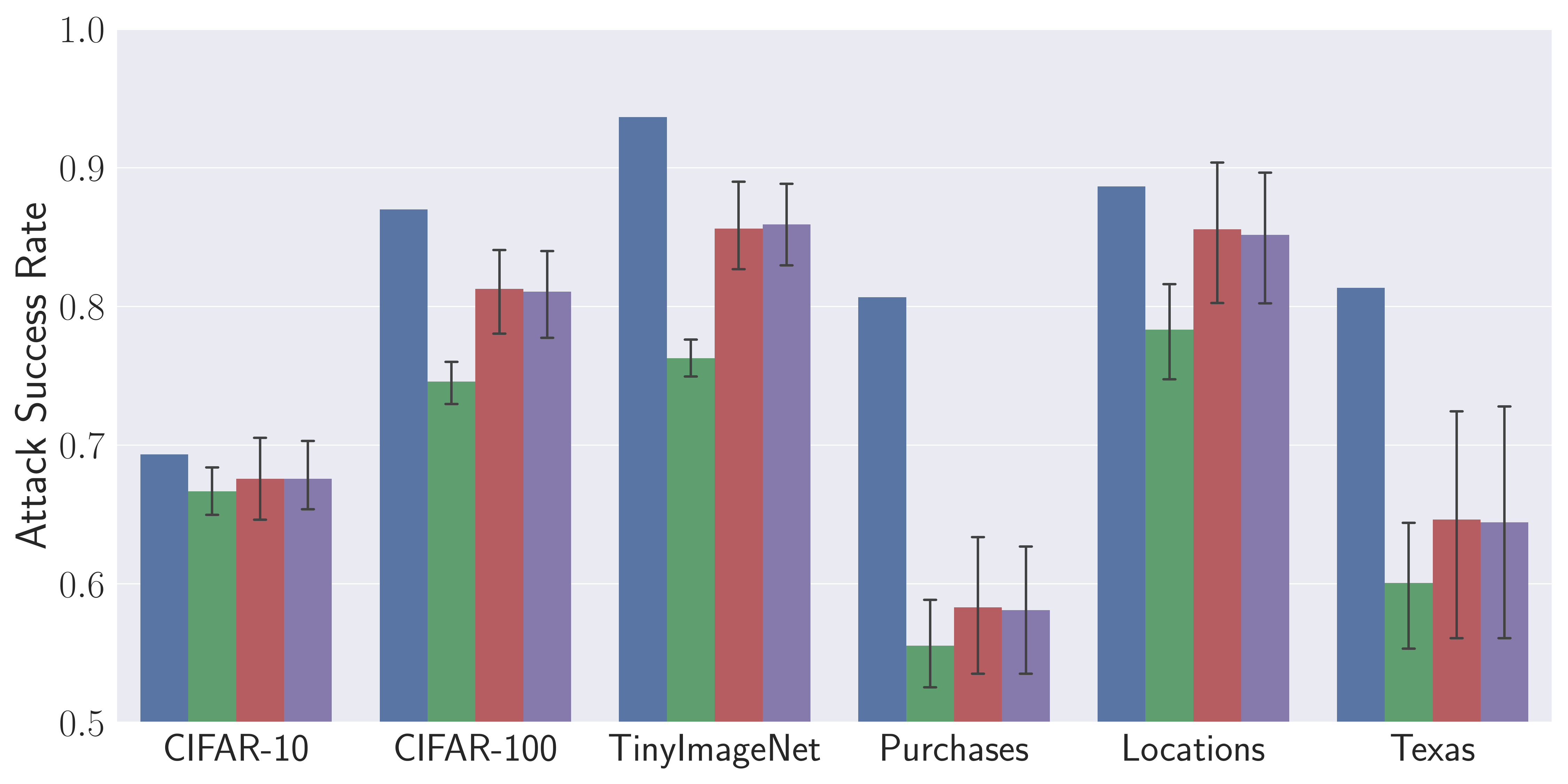}
\caption{Label-only}
\end{subfigure}
\caption{The attack performance of different membership inference attacks again vanilla and multi-exit models. 
    The blue and green bars indicate the original attack on the vanilla and multi-exit models, while the red and purple bars indicate our hybrid attack on the multi-exit model. 
    Computer vision tasks are on VGG-16, and non-computer vision tasks are on FCN-18-1.}
\label{fig:adversary_2}
\end{figure*}

The above results fully demonstrate the efficacy of our hybrid attack. Here, we delve more deeply into the reasons for the success.
Our insight is that the exit depth is a critical indicator for membership inference. 
\autoref{fig:ratio} shows the proportion of non-members in all samples leaving at each exit.
See Appendix \autoref{fig:appendix_ratio} for more results.
We can see that members tend to exit early, while non-members tend to exit late.
In other words, the later exit depth itself indicates that the samples leaving here are likely to be non-members, in contrast to early exits where the samples are likely to be members.
Recall that \autoref{fig:JS_6_exits} shows that the JS divergence of late exits is much larger than early exits, which further contributes to our hybrid attack.
Such separability of members/non-members in terms of exit depth guarantees the efficacy of our hybrid attack.

\subsection{Adversary 2}
\label{Adversary_2}

In this section, we relax the assumption that the adversary has direct access to exit information.
We start by explaining the threat model, then describe the adversary's attack methodology. 
In the end, we present a comprehensive experimental evaluation.

\mypara{Threat Model.}
Different from the threat model in \autoref{Adersary_1}, we remove the assumption that the adversary has direct access to exit information, i.e., exit depth.
This largely reduces the attack capabilities of the adversary.
Given a data sample, the multi-exit model gives a prediction that includes only the score or label and does not include any exit information.
This is a more realistic but also more challenging scenario.
Note that we only focus on score-based and label-only attacks, as in this scenario it is unlikely the adversary can obtain gradients or features of target models.

\begin{figure}[!t]
\centering
\begin{subfigure}{0.49\columnwidth}
\includegraphics[width=\columnwidth]{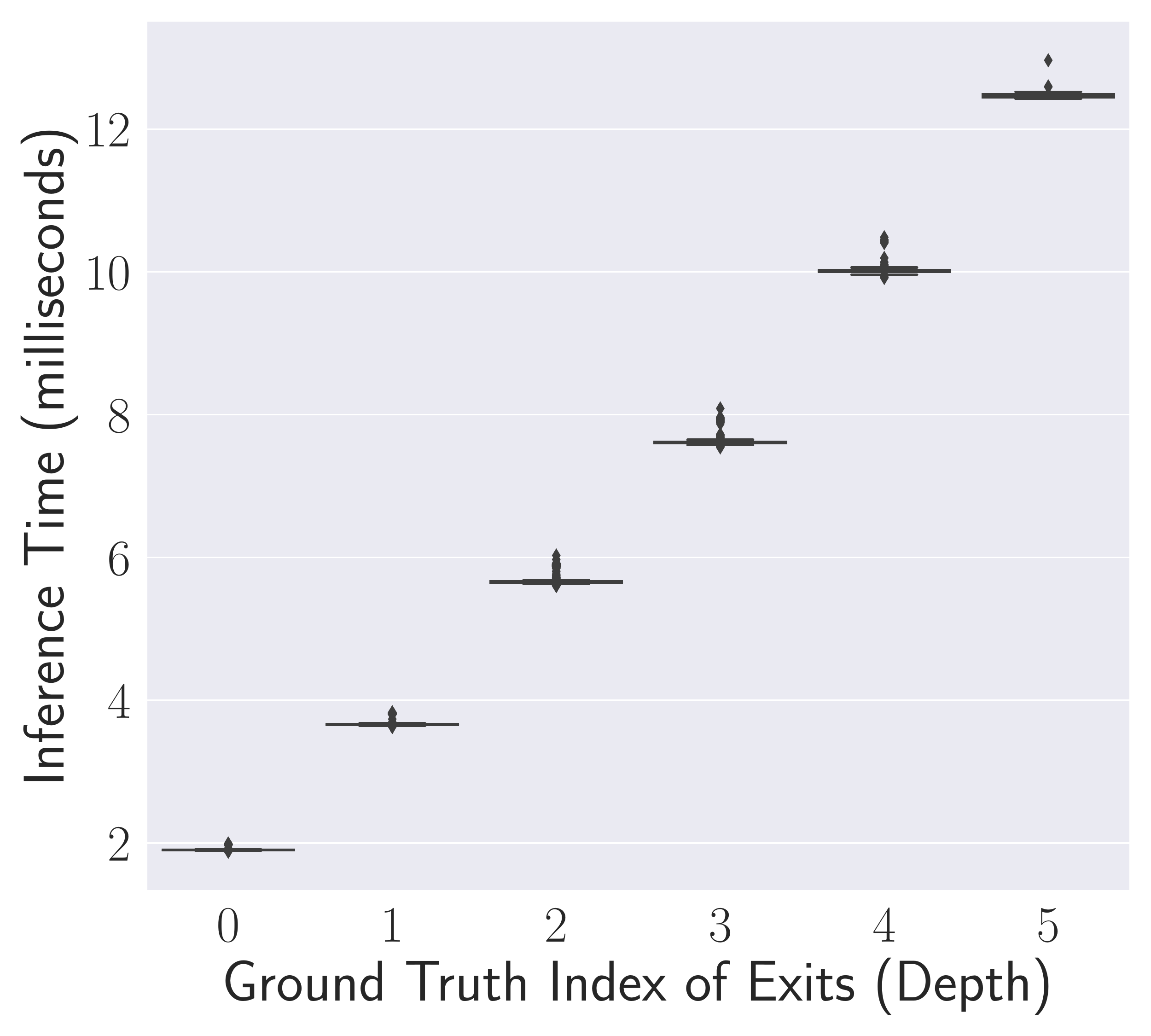}
\caption{6-Exit ResNet-56}
\label{fig:time_wrt_truth}
\end{subfigure}
\begin{subfigure}{0.49\columnwidth}
\includegraphics[width=\columnwidth]{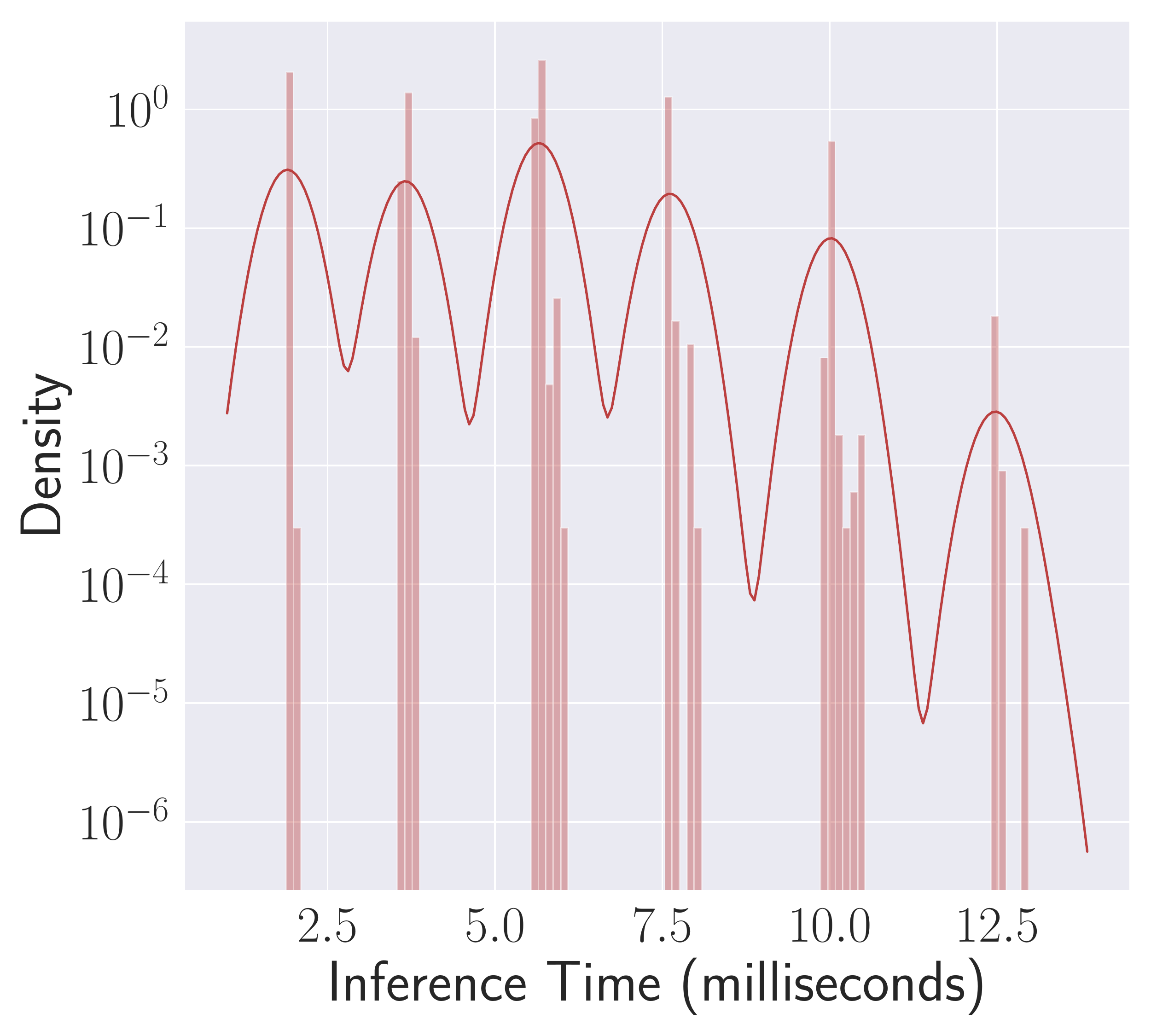}
\caption{6-Exit ResNet-56}
\label{fig:KDE}
\end{subfigure}
\caption{The inference time with respect to ground truth index of exit (a), and the density estimation by KDE based on inference time (b). 
    They are both obtained from the same model, i.e., 6-exit ResNet-56 trained on CIFAR-100.}
\label{fig:inference_time}
\end{figure}

\mypara{Methodology.}
Recall that the goal of multi-exit models is to reduce the computational costs by allowing data samples to be predicted and to exit at an early layer.
Therefore, the inference time for data samples inevitably varies with the depth of the exit, i.e., data samples leaving deeper exit points imply longer inference times, as shown in \autoref{fig:time_wrt_truth}.
This renders us a new perspective to determine the exit information, i.e., the magnitude of inference time actually represents the different exit depths. 
We refer to this method as \textit{time-based hyperparameter stealing}.

\begin{table}[!t]
\definecolor{mygray}{gray}{0.9}
\centering
\caption{The prediction accuracy of exit depths when we run 4 models simultaneously at a time, each on a single GPU. 
We averaged the performance with the number of exits varying from 2 to 6 and report the mean and stand deviations.}
\label{table:pred_exit_1perGPU}
\setlength{\tabcolsep}{5pt}
\scalebox{0.8}{
\begin{tabular}
{c|c|c|c }
\toprule
Target Model&CIFAR-10&CIFAR-100&TinyImageNet\\
\midrule
VGG& 0.9998$\pm$2e-4& 0.9999$\pm$1e-5&  0.9999$\pm$1e-4\\
ResNet&  0.9999$\pm$2e-5& 1.0$\pm$0.0&  1.0$\pm$0.0\\
MobileNet&  0.9998$\pm$8e-5& 1.0$\pm$0.0&  0.9998$\pm$2e-4\\
WideResNet&  0.9996$\pm$2e-7& 0.9999$\pm$1e-5&  0.9999$\pm$1e-5\\
\bottomrule
\end{tabular}
}
\end{table}

The adversary first queries the target multi-exit model using a large number of data samples and records the inference time of these samples.
These query samples can come from the shadow dataset, or random data collected from the Internet or any source.
The adversary then sorts all recorded inference time as a one-dimensional array. 
Note that a longer inference time indicates a deeper exit point.
Thus the adversary can partition this one-dimensional array into different clusters.
Here we leverage Kernel Density Estimation (\textit{KDE})\cite{R56}, an unsupervised statistical method for clustering one-dimensional data.
\autoref{fig:KDE} shows a set of records of inference time, and we can see that \textit{KDE} fits these records with a smoothed line.
Then, several minima of the smoothed line can be used to partition them into different clusters.
Thus the number of clusters means the number of exits attached to the target model, and the index of each cluster means the exit depth.
The reason why we adopt \textit{KDE} is that we want to cluster one-dimensional arrays (i.e., recorded time), for which \textit{KDE} is well suited, while other popular techniques such as K-means~\cite{L82}, kNN~\cite{NDG03} and DBSCAN~\cite{EKSX96} are multidimensional clustering algorithms.

\mypara{Experimental Setup.}
We use all the same setup as presented in \autoref{setup_quantify}, such as the attack model design and evaluation metric. 
All experiments are conducted on an NVIDIA HGX-A100 server with 4-GPU deployed.
We run 4 models simultaneously at a time, each on a single GPU.
Practically, in order to get a stable inference time, we calculate the inference time by averaging the time of each sample 10 times.

\mypara{Results.} 
First, we report the prediction accuracy of exit depth overall datasets and model architectures in \autoref{table:pred_exit_1perGPU}.
As we can see, our proposed \textit{time-based hyperparameter stealing} can achieve almost 100\% accuracy.
This indicates that the magnitude of inference time indeed can represent the exit depth, i.e., a longer inference time represents a deeper exit point and vice versa.
Consequently, we can observe that our two adversaries achieve very similar performance for all datasets and model architectures in \autoref{fig:adversary_2}.
More results can be found in Appendix \autoref{fig:appendix_attackASR}.
These results clearly demonstrate our hybrid attacks are very broadly applicable.

Recall that we run 4 target models simultaneously at a time, each on a single GPU. 
We further investigate whether a more complex system environment affects attack performance.
To this end, we increase the number of models running simultaneously from 4 to 16, and every 4 on a single GPU.
We summarize the prediction performance of exit depth and attack performance in  \autoref{table:pred_exit_4perGPU} and \autoref{fig:more_process}, respectively.
More results can be found in Appendix \autoref{fig:appendix_more_process}.
We can see that the prediction performance of the exit depth has become lower and the variance has become larger.
Encouragingly, \autoref{fig:more_process} shows that our hybrid attack still achieves better attack performance than the original attack, albeit a bit worse than other adversaries.
Moreover, we believe that programs or models are more likely to run in simple or clean environments, at least not as complex as ours, especially in some real-time constrained applications, such as self-driving cars and robotic surgery.
These results show that our attacks achieve significant performance even in such complex environments, further implying that more broadly applicable attack scenarios are possible.

\begin{table}[!t]
\definecolor{mygray}{gray}{0.9}
\centering
\caption{The prediction accuracy of exit depths when we run 16 models simultaneously at a time, every 4 on a single GPU. 
We averaged the performance with the number of exits varying from 2 to 6 and report the mean and stand deviations.}
\label{table:pred_exit_4perGPU}
\setlength{\tabcolsep}{5pt}
\scalebox{0.8}{
\begin{tabular}
{c|c|c|c}
\toprule
Target Model&CIFAR-10&CIFAR-100&TinyImageNet\\
\midrule
VGG& 0.8027$\pm$0.1485& 0.9968$\pm$0.0019&  0.7167$\pm$0.0799\\
ResNet&  0.8966$\pm$0.2158& 0.9118$\pm$0.1972&  0.8279$\pm$0.1859\\
MobileNet&  0.7062$\pm$0.4015& 0.9008$\pm$0.1351&  0.8059$\pm$0.1336\\
WideResNet&  0.7500$\pm$0.3561& 0.9622$\pm$0.0840&  0.9441$\pm$0.1105\\
\bottomrule
\end{tabular}
}
\end{table}


\begin{figure}[!t]
\centering
\begin{subfigure}{0.49\columnwidth}
\includegraphics[width=\columnwidth]{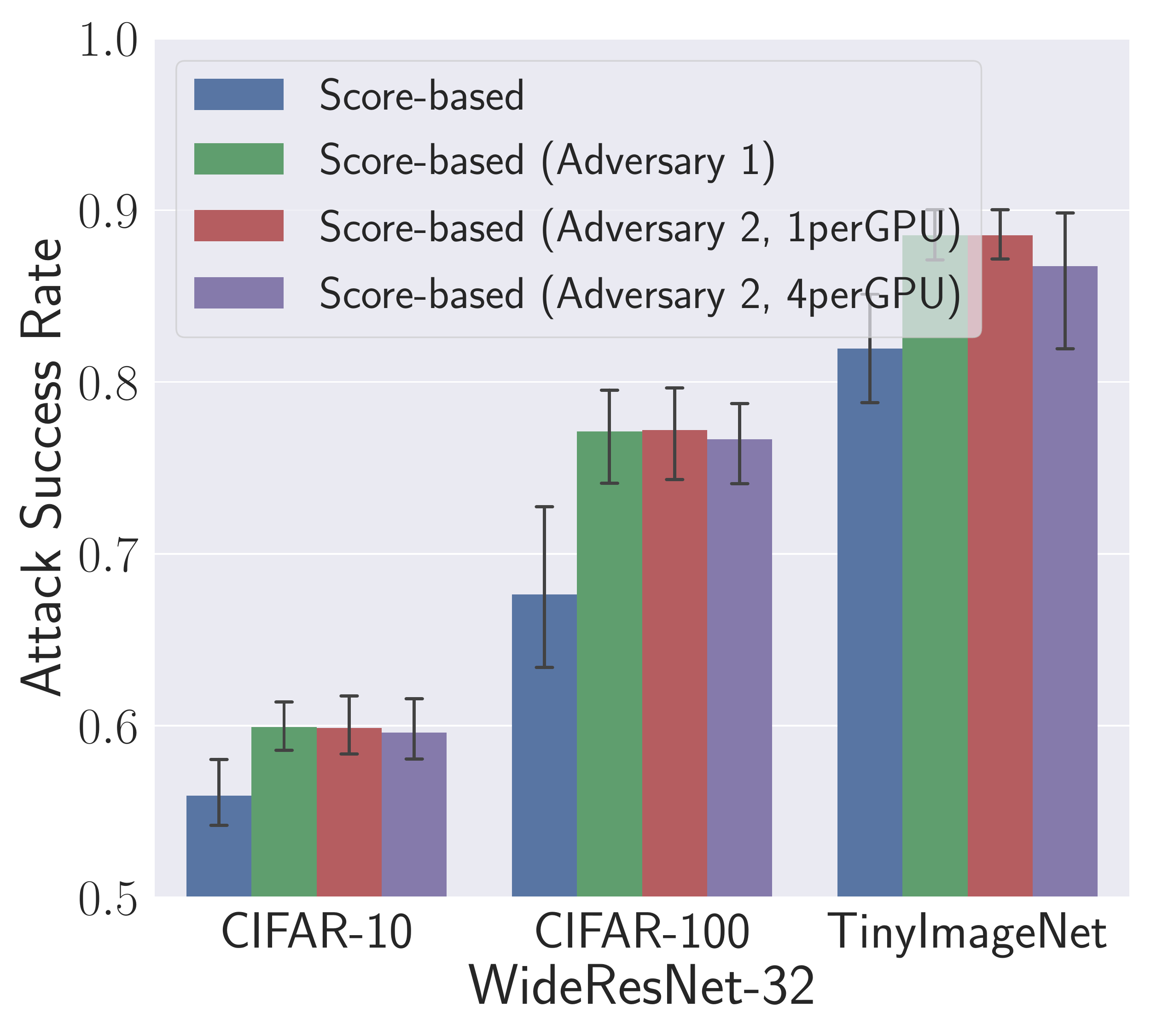}
\caption{Score-based}
\end{subfigure}
\begin{subfigure}{0.49\columnwidth}
\includegraphics[width=\columnwidth]{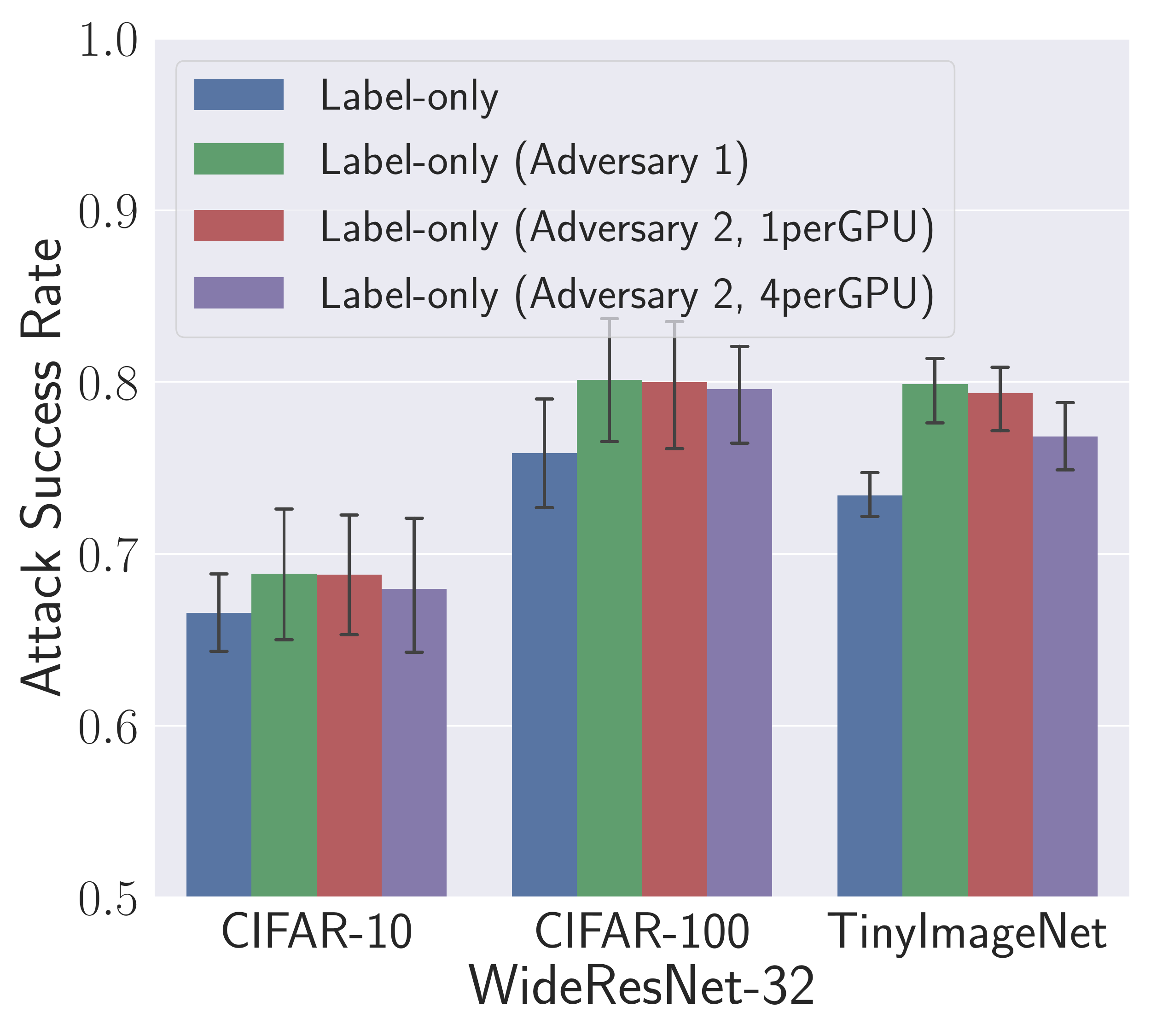}
\caption{Label-only}
\end{subfigure}
\caption{The attack performance of different membership inference attacks against multi-exit models.
    The x-axis represents different datasets. 
    The y-axis represents the attack \texttt{ASR} score. 1perGPU means one model running on a single GPU while 4perGPU means 4 models running on a single GPU.}
\label{fig:more_process}
\end{figure}

\begin{figure}[!t]
\centering
\begin{subfigure}{0.49\columnwidth}
\includegraphics[width=\columnwidth]{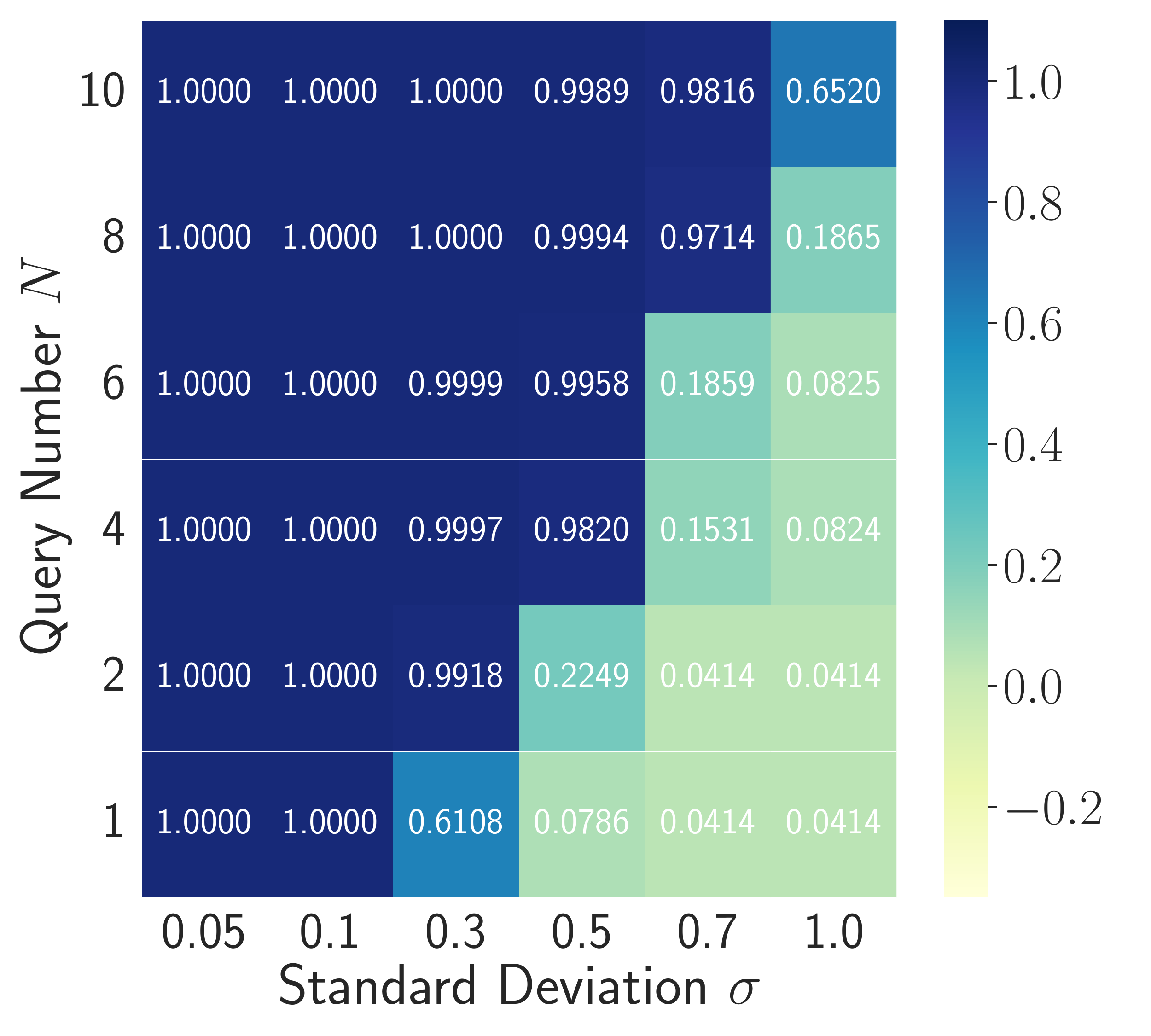}
\caption{Exit Prediction Accuracy}
\label{exit_prediction}
\end{subfigure}
\begin{subfigure}{0.49\columnwidth}
\includegraphics[width=\columnwidth]{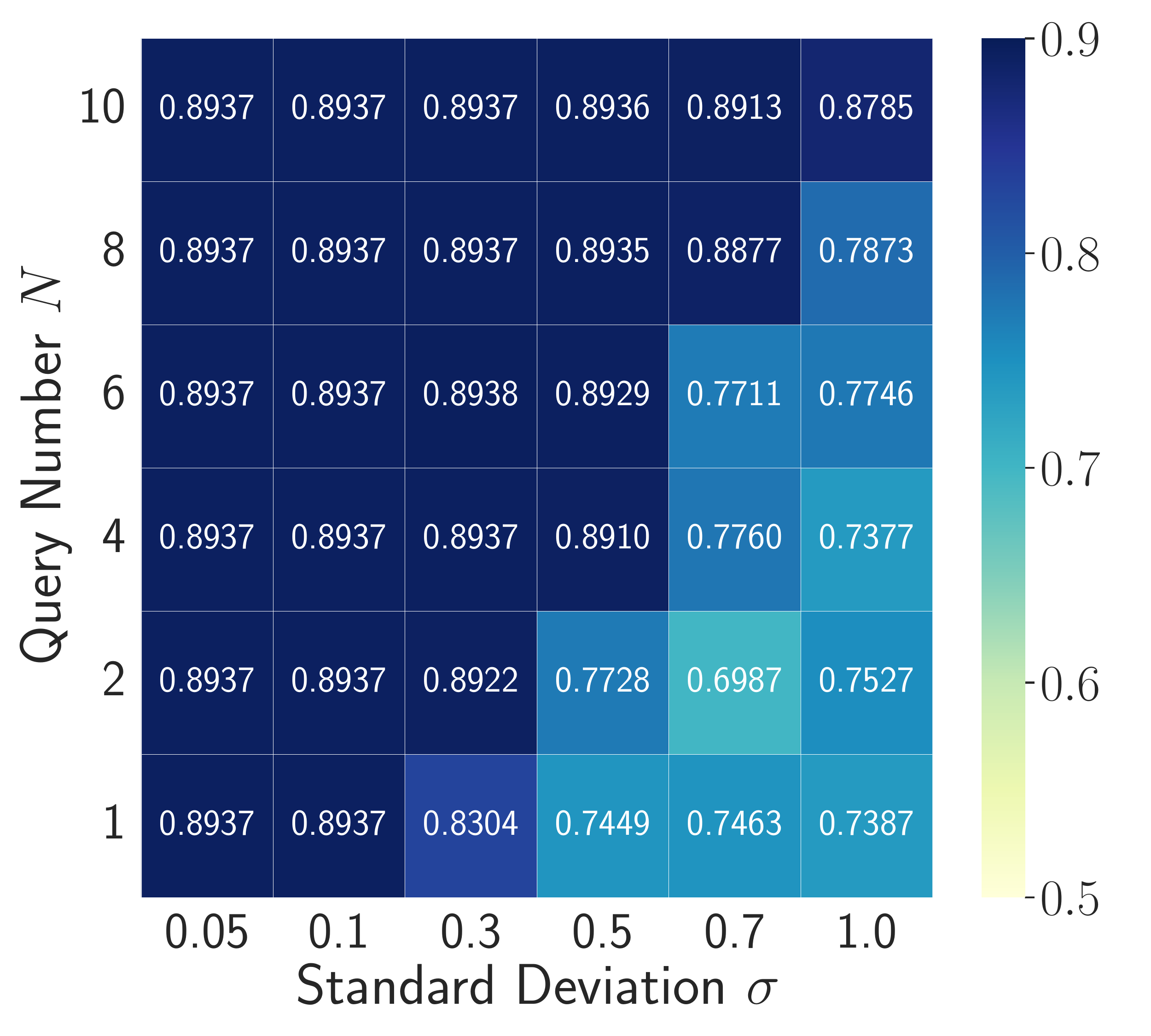}
\caption{\texttt{ASR} Score}
\label{asr_score}
\end{subfigure}
\caption{The exit prediction and attack performance under the effect of query numbers $N$ and standard deviation $\sigma$. 
    The model is WideResNet-32 trained on TinyImageNet.}
\label{fig:MLaaS}
\end{figure}

Next, we focus on the practicality of our hybrid attack against remotely deployed models, i.e., Machine Learning as a Service (MLaaS). 
This is a more challenging scenario where the communication channel can be very noisy.
To simulate the complex communication channel, we assume that the noise $z$ in the channel follows Gaussian distribution $\mathcal{N}(\mu, \sigma^{2})$.
More specifically, we first measure the clean inference time $t$, then sample the noise $z$ from $\mathcal{N}(\mu, \sigma^{2})$, and finally obtain the noisy inference time $t^{ \prime}=t+ z (z>0)$.
Here, the $z>0$ is to ensure that the noisy inference time $t^{ \prime}$ is larger than the clean inference time $t$.
To obtain a stable inference time, we propose a simple method that computes the inference time by averaging the noisy inference time 10 or more times, i.e., querying the remote model multiple times for each sample.
\autoref{fig:MLaaS} shows the prediction and attack performance under the effect of variance $\sigma$ of noise and query numbers for each sample. 
See more results in Appendix \autoref{fig:MLaaS_appendix}.
We can see that the highest prediction accuracy and \texttt{ASR} scores can be achieved if the number of queries is large enough, i.e., multiple queries can indeed eliminate the effect of noise.
Furthermore, as shown in \autoref{exit_prediction}, we can find that even if the prediction accuracy of the exit depth drops by more than 30\%, it still leads to high attack \texttt{ASR} scores.


\begin{figure}[t]
    \centering
    \includegraphics[width=0.7\linewidth]{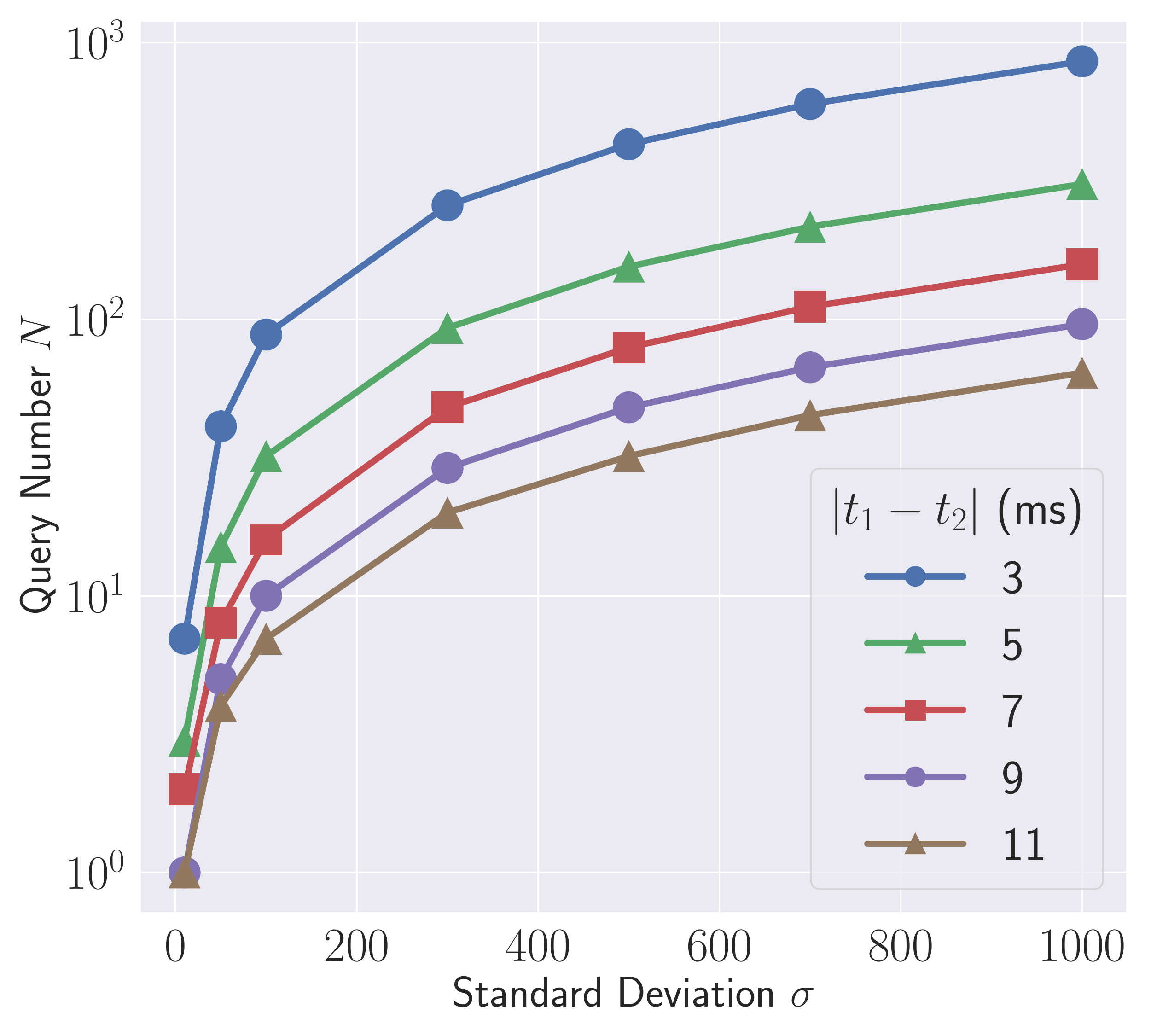}
    \caption{The relationship between query numbers $N$ and standard deviation $\sigma$. 
    The y-axis $N$ denotes the lower bound of query numbers that can guarantee to divide two adjacent exits $t_1$ and $t_2$ with more than 95\% confidence.}
    \label{fig:p_value}
\end{figure}


\begin{figure*}[!t]
\centering
\begin{subfigure}{0.5\columnwidth}
\includegraphics[width=\columnwidth]{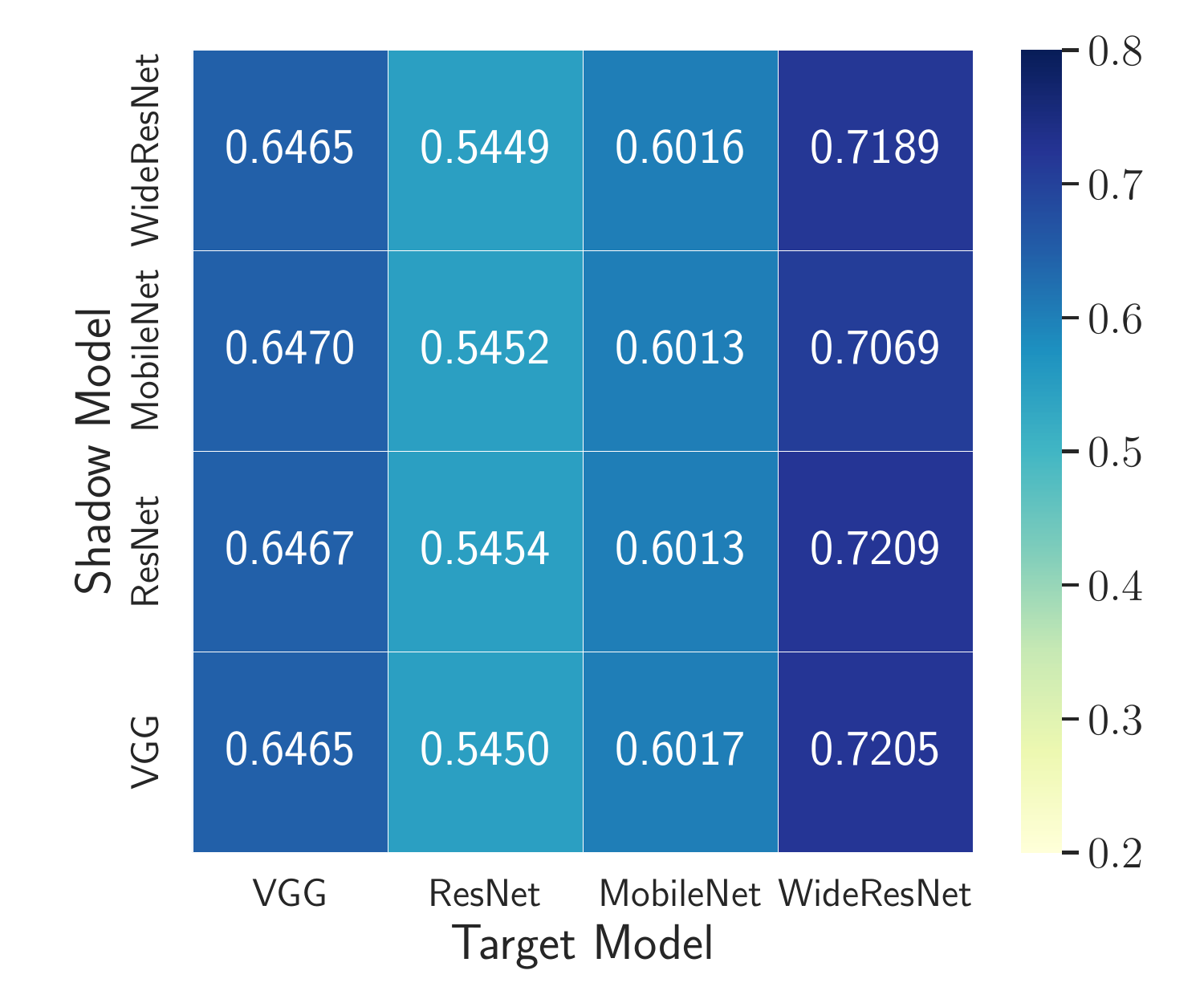}
\caption{Score-based}
\end{subfigure}
\begin{subfigure}{0.5\columnwidth}
\includegraphics[width=\columnwidth]{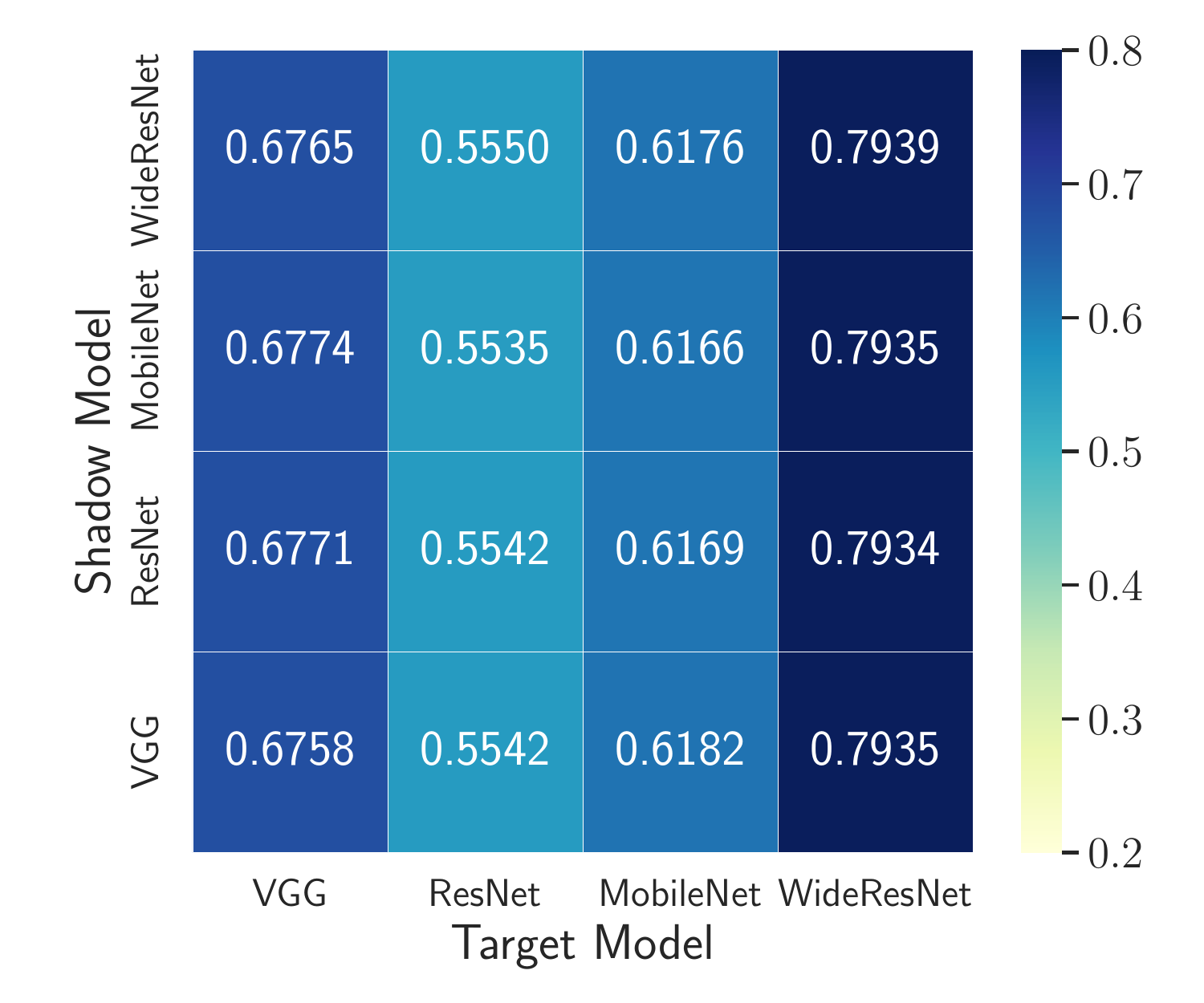}
\caption{Score-based (hybrid attack)}
\end{subfigure}
\begin{subfigure}{0.5\columnwidth}
\includegraphics[width=\columnwidth]{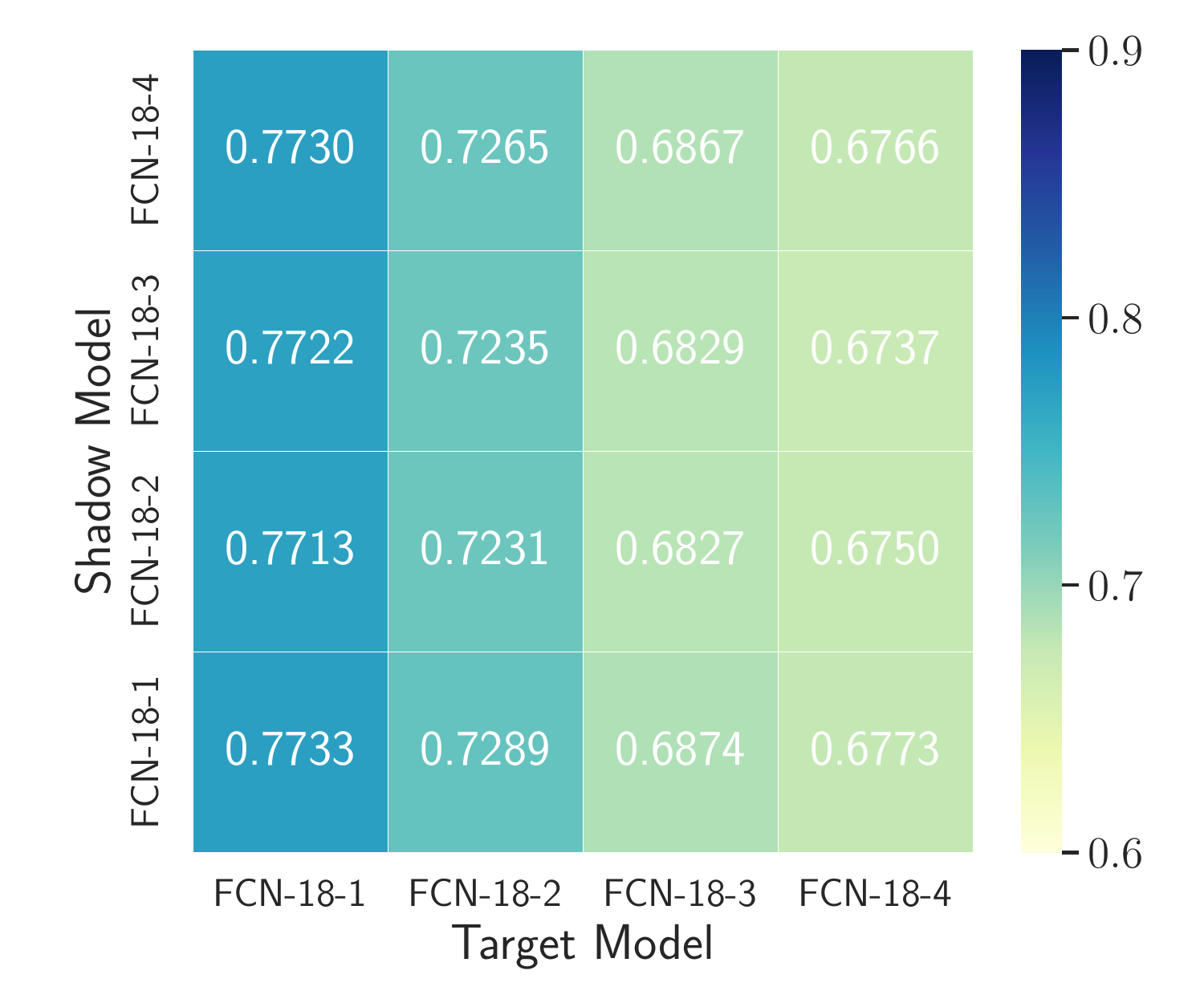}
\caption{Score-based}
\end{subfigure}
\begin{subfigure}{0.5\columnwidth}
\includegraphics[width=\columnwidth]{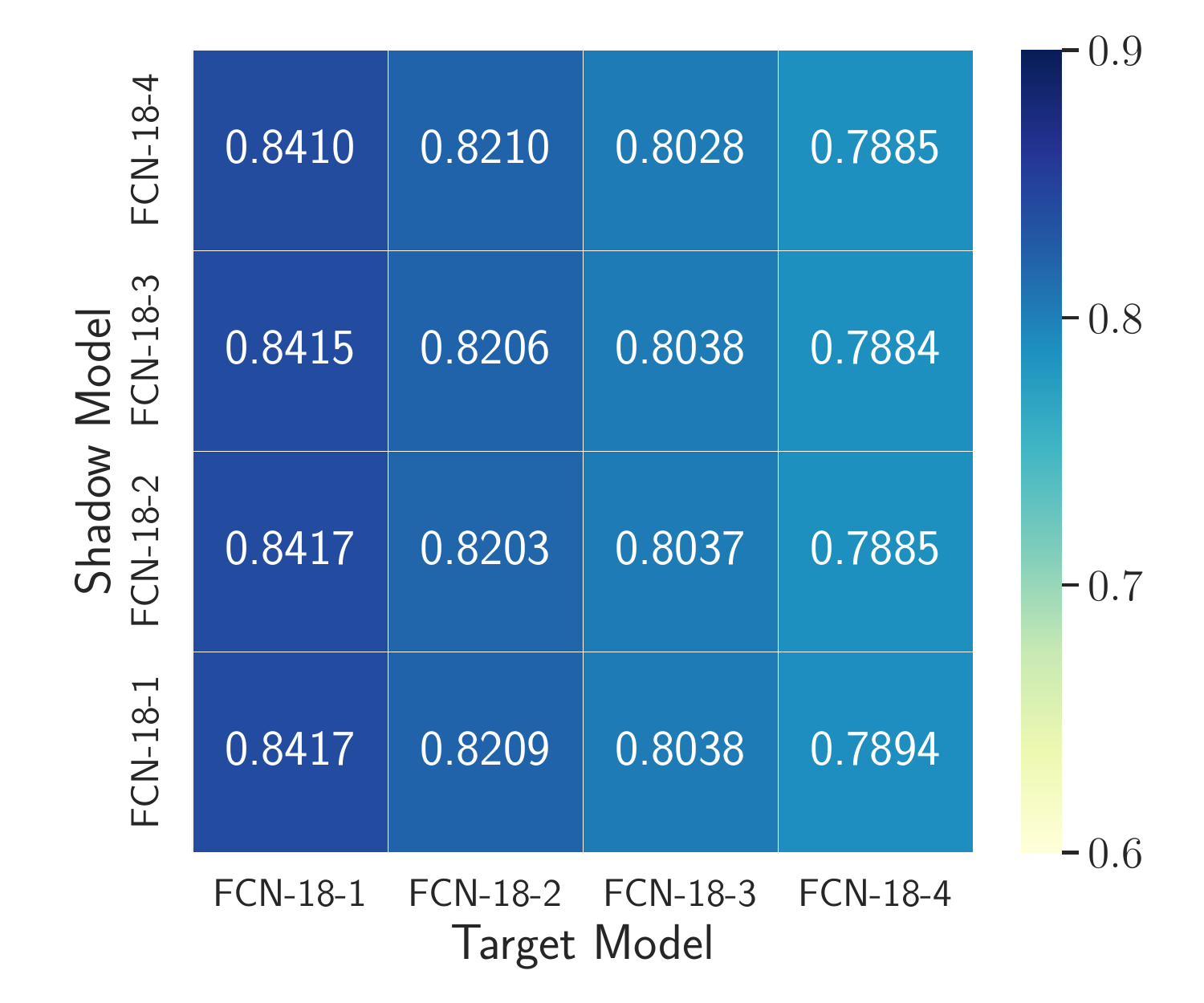}
\caption{Score-based (hybrid attack)}
\end{subfigure}
\caption{The attack performance when the shadow model has different architecture compared to the target model. These computer vision models (a and b) are trained on CIFAR-100, and these non-computer vision models (c and d) are trained on Purchases.}
\label{fig:model_transfer}
\end{figure*}


\begin{figure*}[!t]
\centering
\begin{subfigure}{0.5\columnwidth}
\includegraphics[width=\columnwidth]{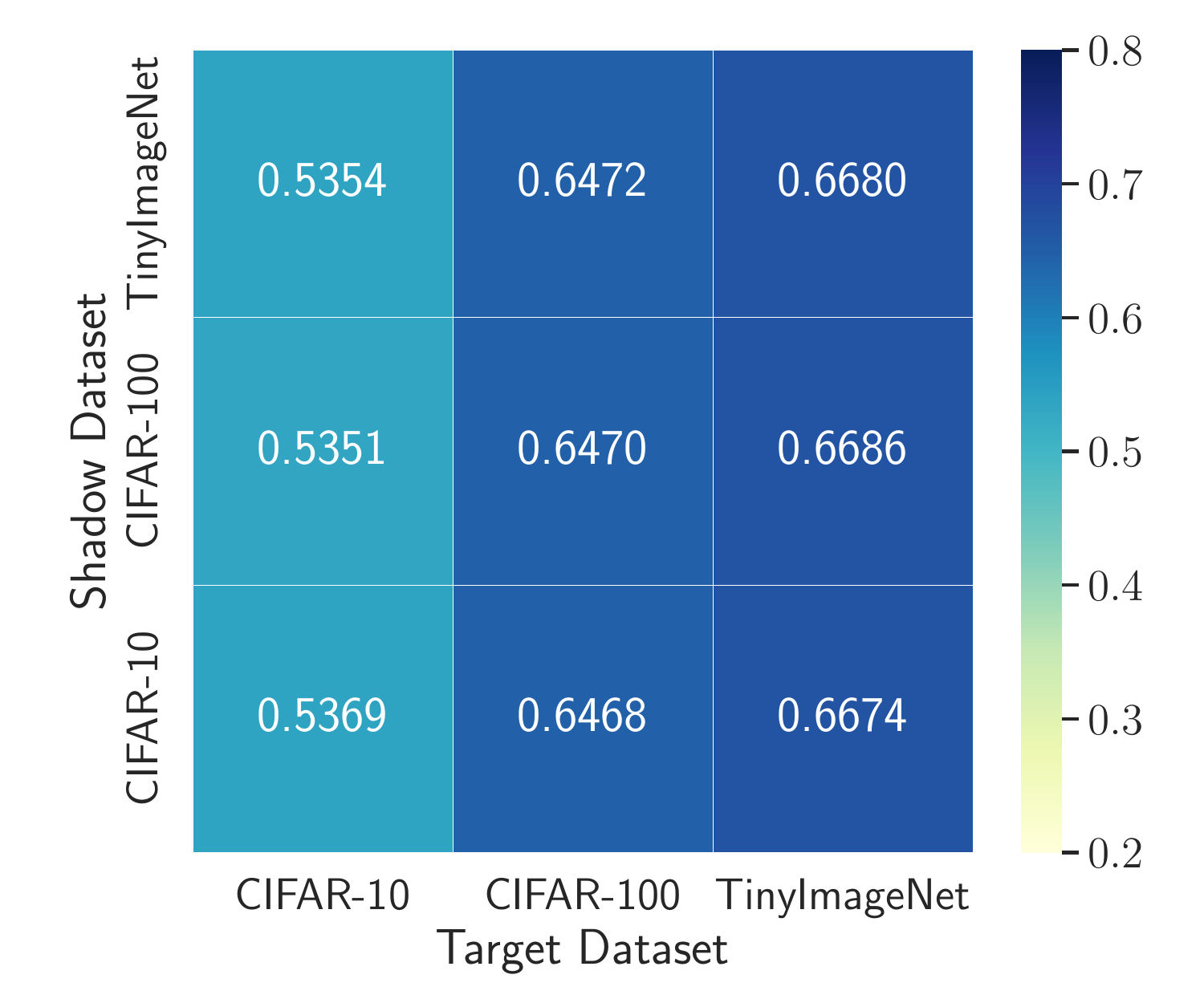}
\caption{Score-based}
\end{subfigure}
\begin{subfigure}{0.5\columnwidth}
\includegraphics[width=\columnwidth]{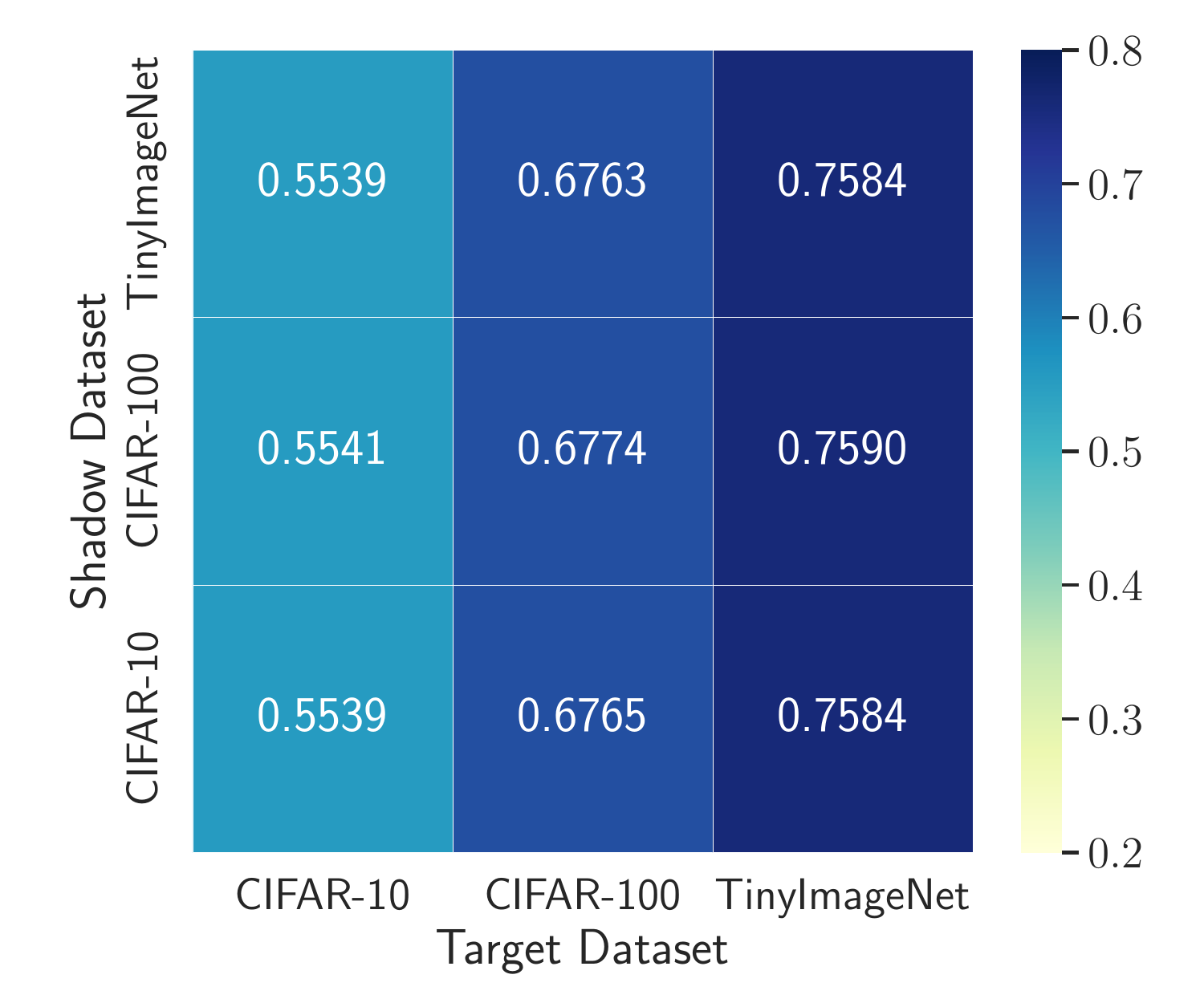}
\caption{Score-based (hybrid attack)}
\end{subfigure}
\begin{subfigure}{0.5\columnwidth}
\includegraphics[width=\columnwidth]{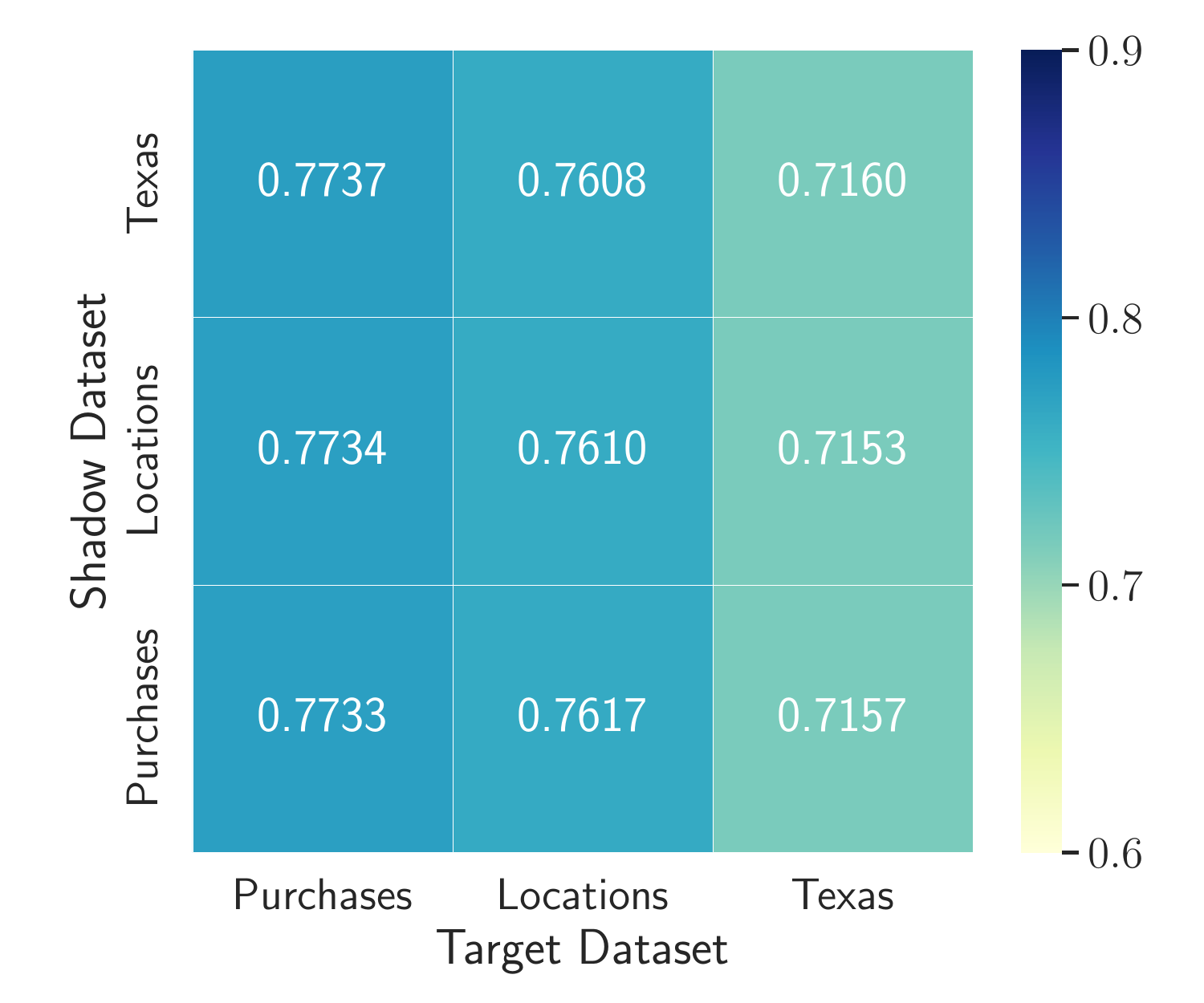}
\caption{Score-based}
\end{subfigure}
\begin{subfigure}{0.5\columnwidth}
\includegraphics[width=\columnwidth]{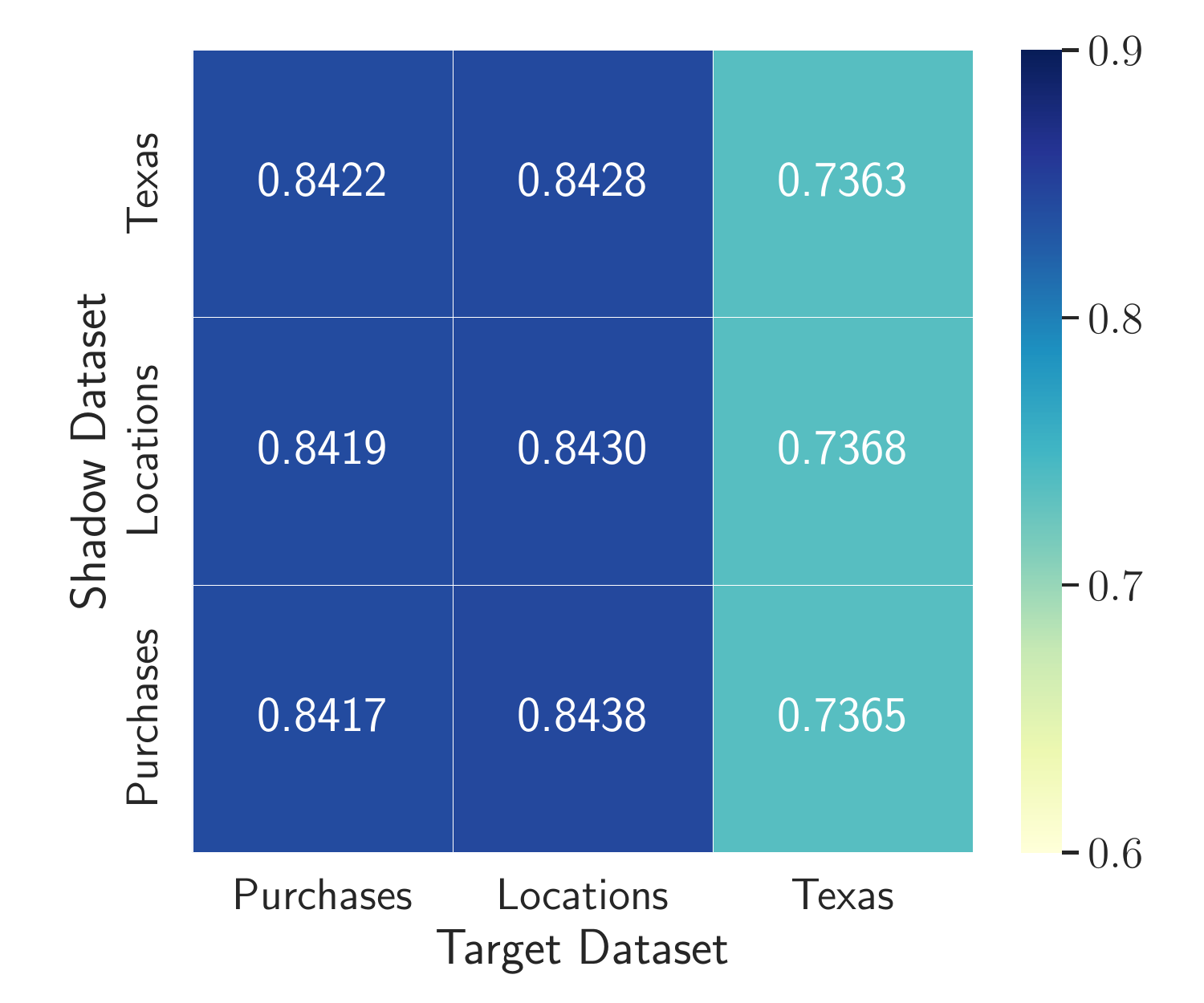}
\caption{Score-based (hybrid attack)}
\end{subfigure}
\caption{The attack performance when the shadow dataset comes from different distributions of the target dataset. The computer vision model (a and b) we used is MobileNet, and the non-computer vision model  (c and d) we used is FCN-18-1.}
\label{fig:dataset_transfer}
\end{figure*}

Furthermore, we delve more deeply into the lower bound of query numbers that can guarantee high attack performance. 
Consider the noise follows $\mathcal{N}(\mu, \sigma^{2})$, and two clean inference time $t_1$ and $t_2$ from two adjacent exits \textit{exit \#1} and \textit{exit \#2}, respectively.
Thus, the noisy inference time $t^{\prime}$ actually follows $\mathcal{N}(t+\mu, \sigma^{2})$.
The research question now is how many query numbers can guarantee the averaged noisy inference time $\bar{t^{\prime}_{1}}$ and $\bar{t^{\prime}_{2}}$ can be distinguished with high confidence. 
Here, we leverage Z-Test~\cite{Z_Test}, a statistical technique, to determine whether two population means $\bar{t^{\prime}_{1}}$ and $\bar{t^{\prime}_{2}}$ are significantly different.
To this end, we first query the target model with one certain sample many times (typically more than 100 times) to estimate the standard deviation $\sigma$.
Then we calculate the Z-Score by the following formula:
\begin{equation}
Z=\frac{\bar{t^{\prime}_{1}}-\bar{t^{\prime}_{2}}}{\sqrt{\frac{\sigma^2}{n_1}+\frac{\sigma^2}{n_2}}} = \frac{(t_1+\mu)-(t_2+\mu)}{\sqrt{\frac{\sigma^2}{N}+\frac{\sigma^2}{N}}} = \frac{(t_1-t_2)}{\sigma\sqrt{\frac{2}{N}}}
\end{equation}
where $n_1$ and $n_2$ represent the query numbers for $\bar{t^{\prime}_{1}}$ and $\bar{t^{\prime}_{2}}$, and we consider the same query numbers $N$ for all samples, i.e., $n_1=n_2=N$. 
Besides, as \autoref{fig:inference_time} shows, we consider the minimal time difference (ms) between two adjacent exit $|t_1-t_2|\in \{3,5,7,9,11\}$.
To satisfy $p \leqslant 0.05$, i.e., the average noise inference time $\bar{t^{\prime}_{1}}$ and $\bar{t^{\prime}_{2}}$ can be distinguished with more than 95\% confidence, we should ensure that $|Z| \geqslant 1.96$.\footnote{\url{https://pro.arcgis.com/en/pro-app/2.8/tool-reference/spatial-statistics/what-is-a-z-score-what-is-a-p-value.htm}}
Thus we can derive the relationship between $N$ and $\sigma$ as shown in \autoref{fig:p_value}.
Given $|t_1-t_2|$ and $\sigma$, the corresponding $N$ denotes the lower bound of query numbers that can guarantee to divide two adjacent exits with more than 95\% confidence.
Recall that as shown in \autoref{exit_prediction}, even if the prediction accuracy of the exit depth drops by more than 30\%, it still leads to a high attack \texttt{ASR} score, so the lower bound on the number of queries can also lead to high attack \texttt{ASR} score.

\subsection{Adversary 3}

Previous work~\cite{SZHBFB19,SSSS17,LZ21,SM21,LLR21,HZ21} has focused on the setup where the adversary trains a shadow model with the same architecture as the target model.
Here, we have to ask \textit{Does the same exit placement and the same architectural model lead to attack performance gains?}.
Therefore, here we investigate whether the exit information still leaks more membership information when we relax this assumption.
In addition, we also investigate the effect of the shadow dataset when we relax the assumption that the shadow dataset and target dataset are identically distributed.
In the following, we start with the threat model description. 
Then, we list the attack methodology.
In the end, we present the evaluation results.

\mypara{Threat Model.}
To challenge our hybrid attack, we remove the assumption that the adversary can build a shadow model with the same architecture and exit placement as the target model, which largely reduces the attack capabilities of the adversary. 
In addition, we perform the evaluation of the gain of the exit to attack performance by relaxing the assumption that the shadow and target datasets are identically distributed.

\mypara{Methodology.}
The strategy of the third adversary is very similar to the second adversary.
The only difference is that the third adversary uses a shadow model with a different architecture from the target model, which further inevitably leads to a different exit placement between shadow and target models.
For example, given a target model ResNet-56 with 6 exits, the adversary can only train a different model, like VGG-16 with 6 exits, to perform membership inference.
In this case, the placement of these 6 exits attached to the backbone model is different between ResNet-56 and VGG-16.

To relax the assumption on the same distribution between the shadow and target datasets, we use different datasets, e.g., CIFAR-10 as the target dataset and TinyImageNet as the shadow dataset, to launch our hybrid attack.

\mypara{Experimental Setup.}
We use the same settings as described in \autoref{Adversary_2}.

\mypara{Results.}
\autoref{fig:model_transfer} shows the attack performance when the shadow models are constructed by different architectures as the target models.
See Appendix \autoref{fig:model_transfer_cifar10_tinyimagenet} for more results.
First, we observe that the attack performance remains almost the same in both the original attack and hybrid attack, respectively.
More encouragingly, we can also find that the attack performance of our hybrid attack is clearly higher than that of the original attack.
For instance, when the target model is WideResNet-32 and the shadow model is VGG-16, the \texttt{ASR} score of our hybrid attack is 0.7935, while that of the original attack is only 0.7205.
Such observation indicates that we can relax the assumption that the shadow model has the same model architecture and exit placement as the target model.
See Appendix \autoref{reason_distribution} for the reason why we can relax this assumption.

Furthermore, we also investigate whether we can relax another assumption of the same distribution between the shadow dataset and target dataset.
\autoref{fig:dataset_transfer} shows the attack performance when the shadow dataset is distributed differently from the target dataset. 
More results can be found in Appendix \autoref{fig:model_transfer_cifar10_tinyimagenet}.
We observe that the performance of our hybrid attack is still better than the original attack even when the target and shadow datasets are different.
Such observation hints that we can also relax the assumption of a same-distribution shadow dataset.
See Appendix \autoref{reason_distribution} for the reason why we can relax this assumption.

In conclusion, we show that adversary 3 can free the attacker from knowing the target model (especially exit placements) and target dataset, which further enlarges the scope of the hybrid attack.
These results convincingly show that the corresponding risks are much more severe under the threats caused by our hybrid attack. 
Furthermore, the fact that privacy risks are much more severe shown by our hybrid attacks would hinder the process of green AI that aims at fast inference and energy-efficient computing.

\section{Possible Defenses}
\label{sec:defense}

In this section, we explore the possible defense and empirically conduct the evaluation.
Recall that the adversary determines the exit depths by observing the different magnitude of inference time, thus the intuition of our defense is to hide the difference in inference time for different exit points. We name our defense \textit{TimeGuard}.

\mypara{\textit{TimeGuard}.} 
The key point is that the multi-exit networks delays giving predictions, rather than giving them immediately. 
One simple but naive defense mechanism is delaying giving predictions to the maximum inference time, i.e., a sample passes forward through all the layers of the model, acting like a vanilla model without any exit inserted.
This behavior will make it impossible for the adversary to determine the exit information by observing inference time. 
The empirical evaluation can be found in Appendix \autoref{MaxTimeGuard}. 
However, this mechanism preserves privacy perfectly but is less efficient because it destroys one of the core ideas of multi-exit networks, which is to reduce the inference time for certain samples.

\begin{figure}[t]
    \centering
    \includegraphics[width=1\linewidth]{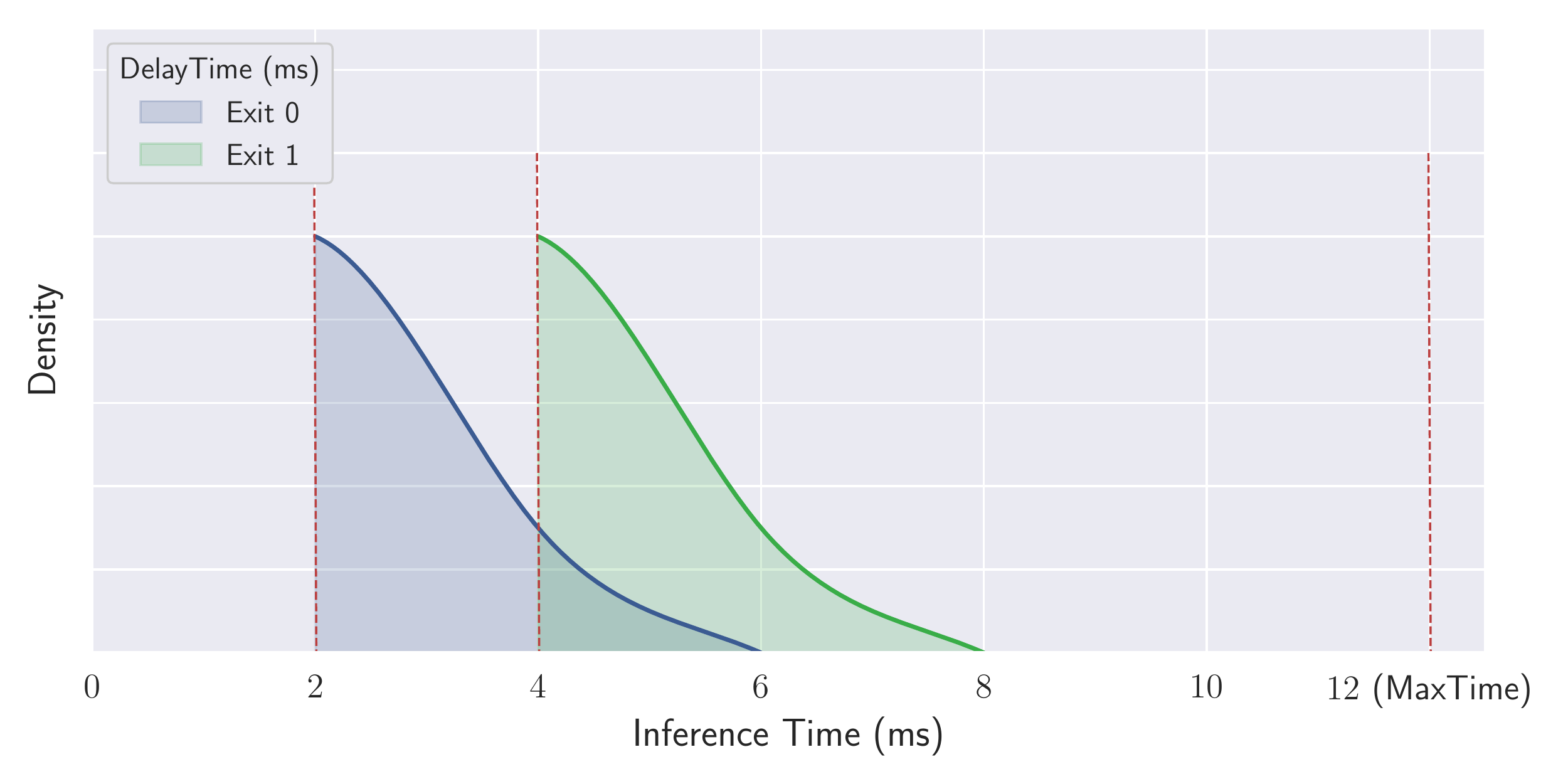}
    \caption{An illustration of how \textit{TimeGuard} works on a multi-exit network with 3 exits. 
    The y-axis represents the density of samples leaving at a certain delaytime among all samples at the same exit. 
    These delaytimes follow the right part of Gaussian distribution $\mathcal{N}(t, \sigma^{2})$.}
    \label{fig:randomdefense}
\end{figure}

To achieve a better trade-off between privacy and efficiency, here we propose a novel mechanism for \textit{TimeGuard} with high efficiency.
See \autoref{fig:randomdefense} for an illustration of \textit{TimeGuard} working on a 3-exit network. 
More concretely, consider the clean inference time $t$ of one certain exit, thus the delay inference times of all the samples leave at this exit follow the right part of Gaussian distribution $\mathcal{N}(t, \sigma^{2})$.
See algorithm of \textit{TimeGuard} in \autoref{alg:defense}.
\begin{algorithm}
\caption{\textit{TimeGuard} with high efficiency.}
\label{alg:defense}  
\KwIn{a data sample $x$, standard deviation $\sigma$, multi-exit model $\mathcal{M}$, a secret global seed $S$.}
\KwOut{delay time $t^{\prime}$ for $x$.}  
calculate hash $h$ by Hash($x$)\tcc*{Hash($x$) is Sha512 or ImageHash  for non-images or images.}
set random seed by random.seed(HKDF($h$, $S$))\tcc*{the seed of Gaussian noise is secret}
sample Gaussian noise $I$ by random.normal($t$, ${\sigma}^2$, size=1)\tcc*{$I$ is unique and repetitive for $x$} 
observe the exit depth where $x$ leaves by feeding $x$ to $\mathcal{M}$\;
obtain clean inference time $t$ of the exit\; 
calculate delay time $t^{\prime}=t+|t-I|$ \; 
return delay time $t^{\prime}$;  
\end{algorithm}  

\begin{figure}[!t]
\centering
\begin{subfigure}{0.49\columnwidth}
\includegraphics[width=\columnwidth]{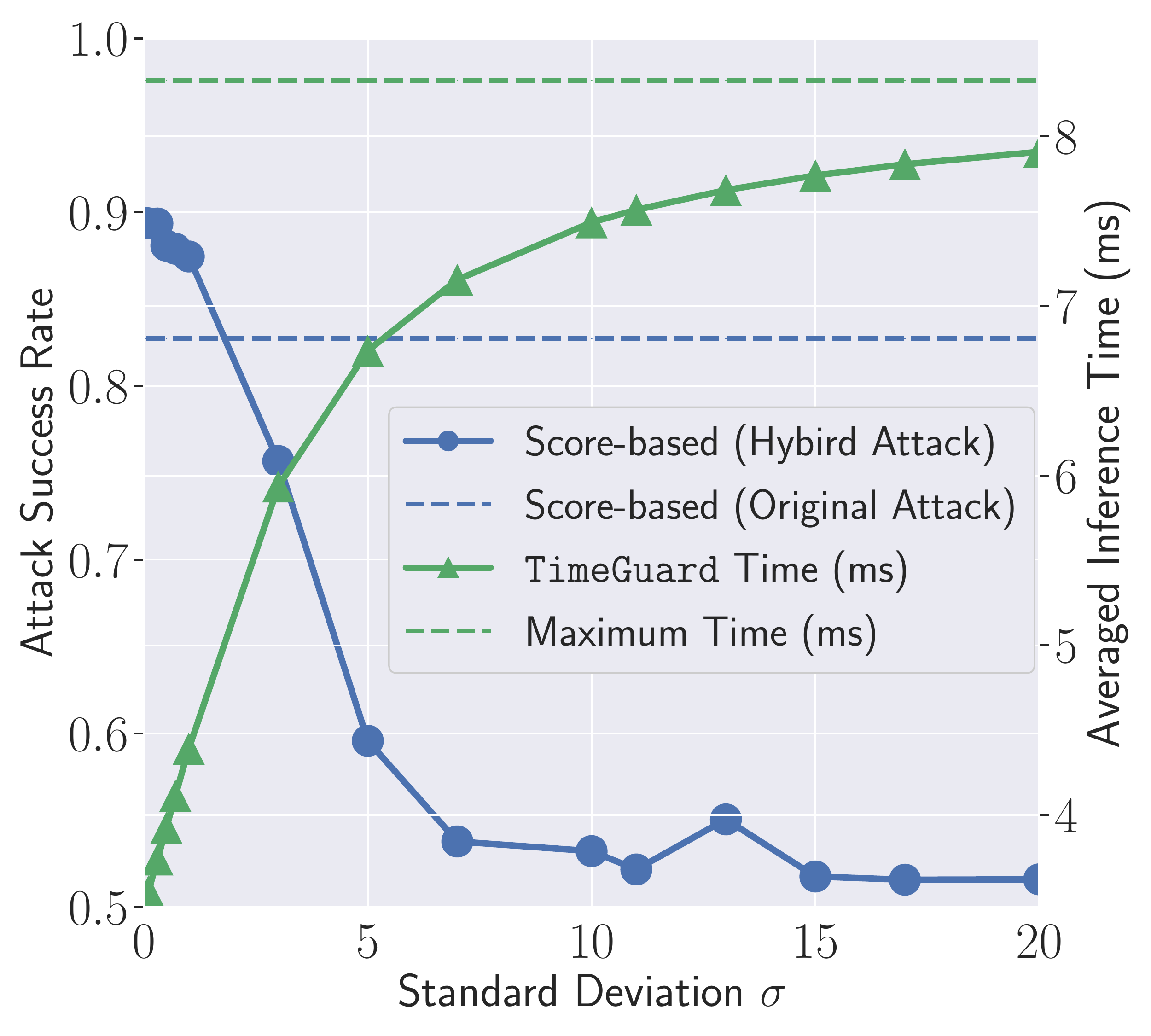}
\caption{TinyImageNet, WRNet-32}
\label{fig:ny_clu_stable}
\end{subfigure}
\begin{subfigure}{0.49\columnwidth}
\includegraphics[width=\columnwidth]{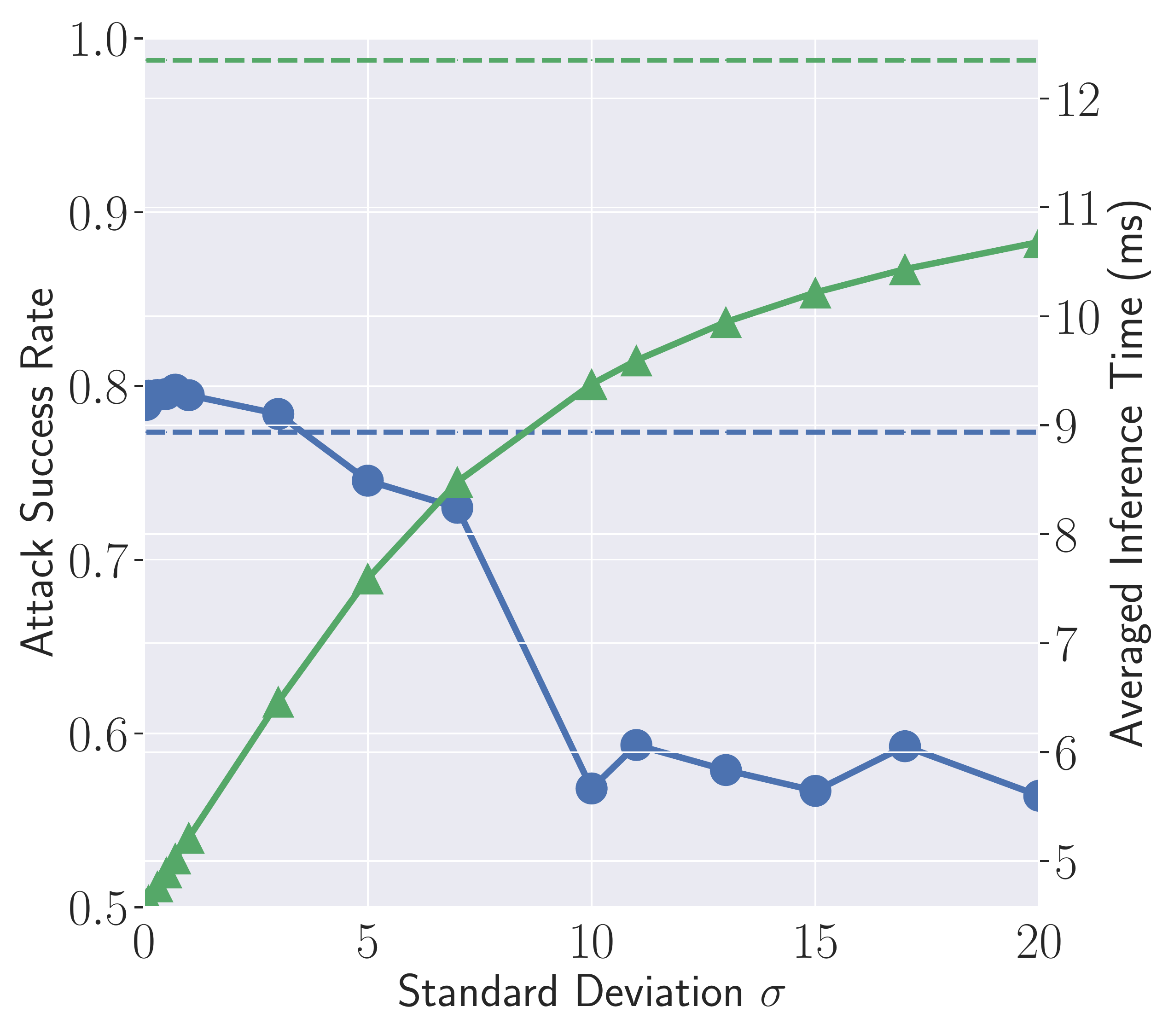}
\caption{Purchases, FCN-18-1}
\label{fig:ny_clu_increase}
\end{subfigure}
\caption{The attack performance and \textit{TimeGuard}'s efficiency under the effect of standard deviation used in \textit{TimeGuard}. Here, WRNet means WideResNet.}
\label{fig:randomdefense_trade_off}
\end{figure}

Here, we leverage ImageHash~\cite{ImageHash} or Sha512~\cite{N02} to calculate the unique hash $h$, and leverage HKDF~\cite{KE10} to generate the secret seed for Gaussian noise. 
In other words, we can obtain a fixed delay inference time $t^{\prime}$ for $x$ since the hash $h$ is unique (line 1) regardless of the number of queries.
Thus, multiple queries on a data sample always give us the same delayed inference time, which is different from the scenario of noise variance reduction.

To further investigate the trade-off between privacy and efficiency under the influence of the standard deviation, we report attack performance and averaged inference time of each sample by varying the standard deviation in \autoref{fig:randomdefense_trade_off}.
We can observe that as the standard deviation increases, the \texttt{ASR} score decreases, while the averaged inference time increases.
Since the \texttt{ASR} score of the original attack is the lower bound of the attack performance in both the original and hybrid attacks, the intersection of the two blue lines shown in \autoref{fig:randomdefense_trade_off} is the best defense scenario for the model.
In other words, the corresponding standard deviation is the optimal setting for \textit{TimeGuard}, which not only reduces the \texttt{ASR} score to the lower bound, but also maintains fast inference.

\section{Related Works}
\label{sec:related}

\subsection{Multi-exit Networks}

Multi-exit architecture is first proposed by Teerapittayanon et al.~\cite{TMK16}, which adds side branches to the backbone model and aims to reduce the inference run-time and energy by allowing certain samples to exit early. 
However, various modifications to the architecture are exploited to make the design more custom or more general. 
\cite{HCLWMW18} presents MSDNet, enabling anytime prediction via multi-scale feature maps and dense connectivity; ~\cite{KHD19} introduces a generic technique Shallow-Deep network (SDN), whilst identifying that standard DNNs are susceptible to overthinking.
In order to improve the classification accuracy, more efforts are focused on the training methods of multi-exit models.
Kaya et al.~\cite{KHD19} conduct both joint training and IC-only training; Phuong et al.~\cite{PL19} utilize knowledge distillation~\cite{HVD15,RBKCGB15} in the training objective, which makes the early exit learn from the last exit; Wang et al.~\cite{WL21} go a step further to allow each exit learn from all later exits and meanwhile it supports learning weight for each distillation loss adaptively.
Recently, more works begin to investigate the robustness of multi-exit models.
Hu et al.~\cite{HCWW20} propose RDI-Nets to achieve "Triple wins" between high accuracy and low inference cost and robust to adversarial attacks~\cite{BCMNSLGR13,SZSBEGF14,GSS15}; Hong et al.~\cite{HKMD21} perform an attack called DeepSloth to eliminate the computational saving which the multi-exit models provide; Dong et al.~\cite{DQZLLL21} give a DNN fingerprinting method via inference time to protect the IP of multi-exit models.
Besides, multi-exit networks are gaining significant attention and rapid development in industry~\cite{Multi_Intel,ZXGMXW20,HHSJCL20} that leverage multi-exit networks to accelerate forward inference.
There is also a growing body of applications deployed with multi-exit architectures, such as IoT scenarios and real-time systems~\cite{KHGRMMT17,LZC18,ZWTD19,HBWL19,JSZYZSH19}.

\subsection{Membership Inference Attacks}

Currently, membership inference is one of the major methods to evaluate privacy risks of machine learning models~\cite{SSSS17,YGFJ18,HMDC19,SZHBFB19,NSH19,SSM19,LF20,HWWBSZ21}.
Shokri et al.~\cite{SSSS17} propose the first membership inference attack against ML models. 
They train multiple attack models using a dataset constructed from multiple shadow models.
These attack models take the posterior of the target sample as input and predict its membership status, i.e., member or non-member.
Then Salem et al. ~\cite{SZHBFB19} propose a model and data-independent membership inference attack by gradually relaxing the assumption made by Shokri et al.~\cite{SSSS17}. 
Later, Nasr et al.~\cite{NSH19} focus on the privacy risk in centralized and federated learning scenarios, and conduct extensive experiments under both black-box and white-box settings.
Song et al.~\cite{SSM19} study the relationship between adversarial examples and the privacy risk caused by membership inference attacks.
Li and Zhang~\cite{LZ21} and Choquette-Choo et al.~\cite{CTCP21} propose the label-only membership inference attack by changing the predicted labels of the target model, then measuring the magnitude of the perturbation.
If the magnitude of the perturbation is larger than a predefined threshold, the adversary considers the data sample as a member and vice versa. 

Besides, there exists a wide range of other attacks, defenses and applications in machine learning domain~\cite{HLXCZ22,FJR15,HP21,KQ20,SHYBZ22,TZJRR16,GDG17,WYSLVZZ19,GWYGB18,ZCSZ22,PMSW18,PMGJCS17,PMJFCS16,TKPGBM17,CJW20,LXZYL20,GMXSX18,ACGMMTZ16,LHGH19,JOBLNL18,ACW18,CW17,LHZG19,CFL10,ABCPK18,RCK18,LV15,SRS17,VL14,YVCZZ17,ZE19,ZHRLPB18}.



\section{Conclusion}
\label{sec:con}

In this paper, we take the first step to audit the privacy risk of multi-exit networks through the lens of membership inference. 
We conduct extensive experiments and find that multi-exit networks are less susceptible to membership leakage and that exits (number and depth) are highly correlated with attack performance. 
We further propose a hybrid attack to improve the performance of existing membership inference attacks by using exit information as new adversary knowledge. 
We investigate three different adversarial settings for different adversary knowledge and end up with a model-free and data-free adversary, which shows that our hybrid attack is broadly applicable and thus the corresponding risk is much more severe than that shown by existing attacks. 
Finally, we present a simple but effective defense mechanism called \textit{TimeGuard} and empirically evaluate its effectiveness.

\section*{Acknowledgements}

This work is partially funded by the Helmholtz Association within the project ``Trustworthy Federated Data Analytics'' (TFDA) (funding number ZT-I-OO1 4).

\begin{small}
\bibliographystyle{plain}
\bibliography{normal_generated_py3}

\begin{thebibliography}{10}

\bibitem{GreenAICloud}
\url{https://greenai.cloud/}.

\bibitem{Multi_Intel}
\url{https://twitter.com/IntelAI/status/1089644305445208064}.

\bibitem{CIFAR}
\url{https://www.cs.toronto.edu/~kriz/cifar.html}.

\bibitem{TinyImageNet}
\url{https://www.kaggle.com/c/tiny-imagenet}.

\bibitem{Purchases}
\url{https://www.kaggle.com/c/acquire-valued-shoppers-challenge/data}.

\bibitem{Locations}
\url{https://sites.google.com/site/yangdingqi/home/foursquare-dataset}.

\bibitem{TexasHealthCare}
\url{https://www.dshs.texas.gov/thcic/hospitals/Inpatientpudf.shtm}.

\bibitem{Z_Test}
\url{https://en.wikipedia.org/wiki/Z-test}.

\bibitem{ImageHash}
\url{https://github.com/JohannesBuchner/imagehash}.

\bibitem{ACGMMTZ16}
Martin Abadi, Andy Chu, Ian Goodfellow, Brendan McMahan, Ilya Mironov, Kunal
  Talwar, and Li~Zhang.
\newblock {Deep Learning with Differential Privacy}.
\newblock In {\em {ACM SIGSAC Conference on Computer and Communications
  Security (CCS)}}, pages 308--318. ACM, 2016.

\bibitem{ABCPK18}
Yossi Adi, Carsten Baum, Moustapha Cisse, Benny Pinkas, and Joseph Keshet.
\newblock {Turning Your Weakness Into a Strength: Watermarking Deep Neural
  Networks by Backdooring}.
\newblock In {\em {USENIX Security Symposium (USENIX Security)}}, pages
  1615--1631. USENIX, 2018.

\bibitem{ACW18}
Anish Athalye, Nicholas Carlini, and David~A. Wagner.
\newblock {Obfuscated Gradients Give a False Sense of Security: Circumventing
  Defenses to Adversarial Examples}.
\newblock In {\em {International Conference on Machine Learning (ICML)}}, pages
  274--283. PMLR, 2018.

\bibitem{BCMNSLGR13}
Battista Biggio, Igino Corona, Davide Maiorca, Blaine Nelson, Nedim Srndic,
  Pavel Laskov, Giorgio Giacinto, and Fabio Roli.
\newblock {Evasion Attacks against Machine Learning at Test Time}.
\newblock In {\em {European Conference on Machine Learning and Principles and
  Practice of Knowledge Discovery in Databases (ECML/PKDD)}}, pages 387--402.
  Springer, 2013.

\bibitem{CW17}
Nicholas Carlini and David Wagner.
\newblock {Towards Evaluating the Robustness of Neural Networks}.
\newblock In {\em {IEEE Symposium on Security and Privacy (S\&P)}}, pages
  39--57. IEEE, 2017.

\bibitem{CJW20}
Jianbo Chen, Michael~I. Jordan, and Martin~J. Wainwright.
\newblock {HopSkipJumpAttack: {A} Query-Efficient Decision-Based Attack}.
\newblock In {\em {IEEE Symposium on Security and Privacy (S\&P)}}, pages
  1277--1294. IEEE, 2020.

\bibitem{CFL10}
James Cheng, Ada~Wai chee Fu, and Jia Liu.
\newblock {K-Isomorphism: Privacy Preserving Network Publication against
  Structural Attacks}.
\newblock In {\em {ACM SIGMOD International Conference on Management of Data
  (SIGMOD)}}, pages 459--470. ACM, 2010.

\bibitem{CTCP21}
Christopher A.~Choquette Choo, Florian Tram{\`e}r, Nicholas Carlini, and
  Nicolas Papernot.
\newblock {Label-Only Membership Inference Attacks}.
\newblock In {\em {International Conference on Machine Learning (ICML)}}, pages
  1964--1974. PMLR, 2021.

\bibitem{DQZLLL21}
Tian Dong, Han Qiu, Tianwei Zhang, Jiwei Li, Hewu Li, and Jialiang Lu.
\newblock {Fingerprinting Multi-exit Deep Neural Network Models via Inference
  Time}.
\newblock {\em {CoRR abs/2110.03175}}, 2021.

\bibitem{EKSX96}
Martin Ester, Hans{-}Peter Kriegel, J{\"{o}}rg Sander, and Xiaowei Xu.
\newblock {A Density-Based Algorithm for Discovering Clusters in Large Spatial
  Databases with Noise}.
\newblock In {\em {International Conference on Knowledge Discovery and Data
  Mining (KDD)}}, pages 226--231. AAAI, 1996.

\bibitem{FJR15}
Matt Fredrikson, Somesh Jha, and Thomas Ristenpart.
\newblock {Model Inversion Attacks that Exploit Confidence Information and
  Basic Countermeasures}.
\newblock In {\em {ACM SIGSAC Conference on Computer and Communications
  Security (CCS)}}, pages 1322--1333. ACM, 2015.

\bibitem{GWYGB18}
Karan Ganju, Qi~Wang, Wei Yang, Carl~A. Gunter, and Nikita Borisov.
\newblock {Property Inference Attacks on Fully Connected Neural Networks using
  Permutation Invariant Representations}.
\newblock In {\em {ACM SIGSAC Conference on Computer and Communications
  Security (CCS)}}, pages 619--633. ACM, 2018.

\bibitem{GPMXWOCB14}
Ian Goodfellow, Jean Pouget-Abadie, Mehdi Mirza, Bing Xu, David Warde-Farley,
  Sherjil Ozair, Aaron Courville, and Yoshua Bengio.
\newblock {Generative Adversarial Nets}.
\newblock In {\em {Annual Conference on Neural Information Processing Systems
  (NIPS)}}, pages 2672--2680. NIPS, 2014.

\bibitem{GSS15}
Ian Goodfellow, Jonathon Shlens, and Christian Szegedy.
\newblock {Explaining and Harnessing Adversarial Examples}.
\newblock In {\em {International Conference on Learning Representations
  (ICLR)}}, 2015.

\bibitem{GDG17}
Tianyu Gu, Brendan Dolan-Gavitt, and Siddharth Grag.
\newblock {Badnets: Identifying Vulnerabilities in the Machine Learning Model
  Supply Chain}.
\newblock {\em {CoRR abs/1708.06733}}, 2017.

\bibitem{GMXSX18}
Wenbo Guo, Dongliang Mu, Jun Xu, Purui Su, and Gang~Wang abd Xinyu~Xing.
\newblock {LEMNA: Explaining Deep Learning based Security Applications}.
\newblock In {\em {ACM SIGSAC Conference on Computer and Communications
  Security (CCS)}}, pages 364--379. ACM, 2018.

\bibitem{HMDC19}
Jamie Hayes, Luca Melis, George Danezis, and Emiliano~De Cristofaro.
\newblock {LOGAN: Evaluating Privacy Leakage of Generative Models Using
  Generative Adversarial Networks}.
\newblock {\em {Privacy Enhancing Technologies Symposium}}, 2019.

\bibitem{HZRS16}
Kaiming He, Xiangyu Zhang, Shaoqing Ren, and Jian Sun.
\newblock {Deep Residual Learning for Image Recognition}.
\newblock In {\em {IEEE Conference on Computer Vision and Pattern Recognition
  (CVPR)}}, pages 770--778. IEEE, 2016.

\bibitem{HLXCZ22}
Xinlei He, Zheng Li, Weilin Xu, Cory Cornelius, and Yang Zhang.
\newblock {Membership-Doctor: Comprehensive Assessment of Membership Inference
  Against Machine Learning Models}.
\newblock {\em {CoRR abs/2208.10445}}, 2022.

\bibitem{HWWBSZ21}
Xinlei He, Rui Wen, Yixin Wu, Michael Backes, Yun Shen, and Yang Zhang.
\newblock {Node-Level Membership Inference Attacks Against Graph Neural
  Networks}.
\newblock {\em {CoRR abs/2102.05429}}, 2021.

\bibitem{HZ21}
Xinlei He and Yang Zhang.
\newblock {Quantifying and Mitigating Privacy Risks of Contrastive Learning}.
\newblock In {\em {ACM SIGSAC Conference on Computer and Communications
  Security (CCS)}}, pages 845--863. ACM, 2021.

\bibitem{HVD15}
Geoffrey~E. Hinton, Oriol Vinyals, and Jeffrey Dean.
\newblock {Distilling the Knowledge in a Neural Network}.
\newblock {\em {CoRR abs/1503.02531}}, 2015.

\bibitem{HKMD21}
Sanghyun Hong, Yigitcan Kaya, Ionut{-}Vlad Modoranu, and Tudor Dumitras.
\newblock {A Panda? No, It's a Sloth: Slowdown Attacks on Adaptive Multi-Exit
  Neural Network Inference}.
\newblock In {\em {International Conference on Learning Representations
  (ICLR)}}, 2021.

\bibitem{HHSJCL20}
Lu~Hou, Zhiqi Huang, Lifeng Shang, Xin Jiang, Xiao Chen, and Qun Liu.
\newblock {DynaBERT: Dynamic {BERT} with Adaptive Width and Depth}.
\newblock In {\em {Annual Conference on Neural Information Processing Systems
  (NeurIPS)}}. NeurIPS, 2020.

\bibitem{HBWL19}
Chuang Hu, Wei Bao, Dan Wang, and Fengming Liu.
\newblock {Dynamic Adaptive {DNN} Surgery for Inference Acceleration on the
  Edge}.
\newblock In {\em {IEEE Conference on Computer Communications (INFOCOM)}},
  pages 1423--1431. IEEE, 2019.

\bibitem{HP21}
Hailong Hu and Jun Pang.
\newblock {Stealing Machine Learning Models: Attacks and Countermeasures for
  Generative Adversarial Networks}.
\newblock In {\em {Annual Computer Security Applications Conference (ACSAC)}},
  pages 1--16. ACM, 2021.

\bibitem{HCWW20}
Ting{-}Kuei Hu, Tianlong Chen, Haotao Wang, and Zhangyang Wang.
\newblock {Triple Wins: Boosting Accuracy, Robustness and Efficiency Together
  by Enabling Input-Adaptive Inference}.
\newblock In {\em {International Conference on Learning Representations
  (ICLR)}}, 2020.

\bibitem{HCLWMW18}
Gao Huang, Danlu Chen, Tianhong Li, Felix Wu, Laurens van~der Maaten, and
  Kilian~Q. Weinberger.
\newblock {Multi-Scale Dense Networks for Resource Efficient Image
  Classification}.
\newblock In {\em {International Conference on Learning Representations
  (ICLR)}}, 2018.

\bibitem{HYYBGC21}
Bo~Hui, Yuchen Yang, Haolin Yuan, Philippe Burlina, Neil~Zhenqiang Gong, and
  Yinzhi Cao.
\newblock {Practical Blind Membership Inference Attack via Differential
  Comparisons}.
\newblock In {\em {Network and Distributed System Security Symposium (NDSS)}}.
  Internet Society, 2021.

\bibitem{JOBLNL18}
Matthew Jagielski, Alina Oprea, Battista Biggio, Chang Liu, Cristina
  Nita-Rotaru, and Bo~Li.
\newblock {Manipulating Machine Learning: Poisoning Attacks and Countermeasures
  for Regression Learning}.
\newblock In {\em {IEEE Symposium on Security and Privacy (S\&P)}}, pages
  19--35. IEEE, 2018.

\bibitem{JE19}
Bargav Jayaraman and David Evans.
\newblock {Evaluating Differentially Private Machine Learning in Practice}.
\newblock In {\em {USENIX Security Symposium (USENIX Security)}}, pages
  1895--1912. USENIX, 2019.

\bibitem{JSBZG19}
Jinyuan Jia, Ahmed Salem, Michael Backes, Yang Zhang, and Neil~Zhenqiang Gong.
\newblock {MemGuard: Defending against Black-Box Membership Inference Attacks
  via Adversarial Examples}.
\newblock In {\em {ACM SIGSAC Conference on Computer and Communications
  Security (CCS)}}, pages 259--274. ACM, 2019.

\bibitem{JSZYZSH19}
Weiwen Jiang, Edwin~H.{-}M. Sha, Xinyi Zhang, Lei Yang, Qingfeng Zhuge, Yiyu
  Shi, and Jingtong Hu.
\newblock {Achieving Super-Linear Speedup across Multi-FPGA for Real-Time {DNN}
  Inference}.
\newblock {\em {Transactions on Embedded Computing Systems}}, 2019.

\bibitem{KHGRMMT17}
Yiping Kang, Johann Hauswald, Cao Gao, Austin Rovinski, Trevor Mudge, Jason
  Mars, and Lingjia Tang.
\newblock {Neurosurgeon: Collaborative Intelligence Between the Cloud and
  Mobile Edge}.
\newblock In {\em {International Conference on Architectural Support for
  Programming Languages and Operating Systems (ASPLOS)}}, pages 615--629. ACM,
  2017.

\bibitem{KQ20}
Sanjay Kariyappa and Moinuddin~K. Qureshi.
\newblock {Defending Against Model Stealing Attacks With Adaptive
  Misinformation}.
\newblock In {\em {IEEE Conference on Computer Vision and Pattern Recognition
  (CVPR)}}, pages 767--775. IEEE, 2020.

\bibitem{KHD19}
Yigitcan Kaya, Sanghyun Hong, and Tudor Dumitras.
\newblock {Shallow-Deep Networks: Understanding and Mitigating Network
  Overthinking}.
\newblock In {\em {International Conference on Machine Learning (ICML)}}, pages
  3301--3310. PMLR, 2019.

\bibitem{KVLL21}
Alexandros Kouris, Stylianos~I. Venieris, Stefanos Laskaridis, and Nicholas~D.
  Lane.
\newblock {Multi-Exit Semantic Segmentation Networks}.
\newblock {\em {CoRR abs/2106.03527}}, 2021.

\bibitem{KE10}
Hugo Krawczyk and Pasi Eronen.
\newblock {HMAC-based Extract-and-Expand Key Derivation Function {(HKDF)}}.
\newblock {\em {Request for Comments}}, 2010.

\bibitem{LVALL20}
Stefanos Laskaridis, Stylianos~I. Venieris, M{\'{a}}rio Almeida, Ilias
  Leontiadis, and Nicholas~D. Lane.
\newblock {{SPINN:} Synergistic Progressive Inference of Neural Networks over
  Device and Cloud}.
\newblock In {\em {Annual International Conference on Mobile Computing and
  Networking (MobiCom)}}, pages 37:1--37:15. ACM, 2020.

\bibitem{LF20}
Klas Leino and Matt Fredrikson.
\newblock {Stolen Memories: Leveraging Model Memorization for Calibrated
  White-Box Membership Inference}.
\newblock In {\em {USENIX Security Symposium (USENIX Security)}}, pages
  1605--1622. USENIX, 2020.

\bibitem{LV15}
Bo~Li and Yevgeniy Vorobeychik.
\newblock {Scalable Optimization of Randomized Operational Decisions in
  Adversarial Classification Settings}.
\newblock In {\em {International Conference on Artificial Intelligence and
  Statistics (AISTATS)}}, pages 599--607. JMLR, 2015.

\bibitem{LZC18}
En~Li, Zhi Zhou, and Xu~Chen.
\newblock {Edge Intelligence: On-Demand Deep Learning Model Co-Inference with
  Device-Edge Synergy}.
\newblock In {\em {ACM Special Interest Group on Data Communication
  (SIGCOMM)}}, pages 31--36. ACM, 2018.

\bibitem{LXZYL20}
Huichen Li, Xiaojun Xu, Xiaolu Zhang, Shuang Yang, and Bo~Li.
\newblock {{QEBA:} Query-Efficient Boundary-Based Blackbox Attack}.
\newblock In {\em {IEEE Conference on Computer Vision and Pattern Recognition
  (CVPR)}}, pages 1218--1227. IEEE, 2020.

\bibitem{LLR21}
Jiacheng Li, Ninghui Li, and Bruno Ribeiro.
\newblock {Membership Inference Attacks and Defenses in Classification Models}.
\newblock In {\em {ACM Conference on Data and Application Security and Privacy
  (CODASPY)}}, pages 5--16. ACM, 2021.

\bibitem{LHGH19}
Zheng Li, Ge~Han, Shanqing Guo, and Chengyu Hu.
\newblock {DeepKeyStego: Protecting Communication by Key-Dependent
  Steganography with Deep Networks}.
\newblock In {\em {International Conference on High Performance Computing and
  Communications (HPCC)}}, pages 1937--1944. IEEE, 2019.

\bibitem{LHZG19}
Zheng Li, Chengyu Hu, Yang Zhang, and Shanqing Guo.
\newblock {How to Prove Your Model Belongs to You: A Blind-Watermark based
  Framework to Protect Intellectual Property of DNN}.
\newblock In {\em {Annual Computer Security Applications Conference (ACSAC)}},
  pages 126--137. ACM, 2019.

\bibitem{LZ21}
Zheng Li and Yang Zhang.
\newblock {Membership Leakage in Label-Only Exposures}.
\newblock In {\em {ACM SIGSAC Conference on Computer and Communications
  Security (CCS)}}, pages 880--895. ACM, 2021.

\bibitem{LZWZDJ20}
Weijie Liu, Peng Zhou, Zhiruo Wang, Zhe Zhao, Haotang Deng, and Qi~Ju.
\newblock {FastBERT: a Self-distilling {BERT} with Adaptive Inference Time}.
\newblock In {\em {Annual Meeting of the Association for Computational
  Linguistics (ACL)}}, pages 6035--6044. ACL, 2020.

\bibitem{L82}
Stuart~P. Lloyd.
\newblock {Least Squares Quantization in {PCM}}.
\newblock {\em {IEEE Transactions on Information Theory}}, 1982.

\bibitem{NDG03}
Subhas~C. Nandy, Sandip Das, and Partha~P. Goswami.
\newblock {An Efficient K Nearest Neighbors Searching Algorithm for A Query
  Line}.
\newblock {\em {Theoretical Computer Science}}, 2003.

\bibitem{NSH18}
Milad Nasr, Reza Shokri, and Amir Houmansadr.
\newblock {Machine Learning with Membership Privacy using Adversarial
  Regularization}.
\newblock In {\em {ACM SIGSAC Conference on Computer and Communications
  Security (CCS)}}, pages 634--646. ACM, 2018.

\bibitem{NSH19}
Milad Nasr, Reza Shokri, and Amir Houmansadr.
\newblock {Comprehensive Privacy Analysis of Deep Learning: Passive and Active
  White-box Inference Attacks against Centralized and Federated Learning}.
\newblock In {\em {IEEE Symposium on Security and Privacy (S\&P)}}, pages
  1021--1035. IEEE, 2019.

\bibitem{N02}
NIST.
\newblock {\em {FIPS 180-2: Secure Hash Standard}}.
\newblock FIPS, 2002.

\bibitem{PMSW18}
Nicolas Papernot, Patrick McDaniel, Arunesh Sinha, and Michael Wellman.
\newblock {SoK: Towards the Science of Security and Privacy in Machine
  Learning}.
\newblock In {\em {IEEE European Symposium on Security and Privacy (Euro
  S\&P)}}, pages 399--414. IEEE, 2018.

\bibitem{PMGJCS17}
Nicolas Papernot, Patrick~D. McDaniel, Ian Goodfellow, Somesh Jha, Z.~Berkay
  Celik, and Ananthram Swami.
\newblock {Practical Black-Box Attacks Against Machine Learning}.
\newblock In {\em {ACM Asia Conference on Computer and Communications Security
  (ASIACCS)}}, pages 506--519. ACM, 2017.

\bibitem{PMJFCS16}
Nicolas Papernot, Patrick~D. McDaniel, Somesh Jha, Matt Fredrikson, Z.~Berkay
  Celik, and Ananthram Swami.
\newblock {The Limitations of Deep Learning in Adversarial Settings}.
\newblock In {\em {IEEE European Symposium on Security and Privacy (Euro
  S\&P)}}, pages 372--387. IEEE, 2016.

\bibitem{PL19}
Mary Phuong and Christoph Lampert.
\newblock {Distillation-Based Training for Multi-Exit Architectures}.
\newblock In {\em {IEEE International Conference on Computer Vision (ICCV)}},
  pages 1355--1364. IEEE, 2019.

\bibitem{ROF20}
Shadi Rahimian, Tribhuvanesh Orekondy, and Mario Fritz.
\newblock {Differential Privacy Defenses and Sampling Attacks for Membership
  Inference}.
\newblock In {\em {PriML Workshop (PriML)}}. NeurIPS, 2020.

\bibitem{RBKCGB15}
Adriana Romero, Nicolas Ballas, Samira~Ebrahimi Kahou, Antoine Chassang, Carlo
  Gatta, and Yoshua Bengio.
\newblock {FitNets: Hints for Thin Deep Nets}.
\newblock In {\em {International Conference on Learning Representations
  (ICLR)}}, 2015.

\bibitem{R56}
Murray Rosenblatt.
\newblock {Remarks on Some Nonparametric Estimates of a Density Function}.
\newblock {\em {The Annals of Mathematical Statistics}}, 1956.

\bibitem{RCK18}
Bita~Darvish Rouhani, Huili Chen, and Farinaz Koushanfar.
\newblock {DeepSigns: A Generic Watermarking Framework for IP Protection of
  Deep Learning Models}.
\newblock {\em {CoRR abs/1804.00750}}, 2018.

\bibitem{SZHBFB19}
Ahmed Salem, Yang Zhang, Mathias Humbert, Pascal Berrang, Mario Fritz, and
  Michael Backes.
\newblock {ML-Leaks: Model and Data Independent Membership Inference Attacks
  and Defenses on Machine Learning Models}.
\newblock In {\em {Network and Distributed System Security Symposium (NDSS)}}.
  Internet Society, 2019.

\bibitem{SHZZC18}
Mark Sandler, Andrew~G. Howard, Menglong Zhu, Andrey Zhmoginov, and
  Liang{-}Chieh Chen.
\newblock {MobileNetV2: Inverted Residuals and Linear Bottlenecks}.
\newblock In {\em {IEEE Conference on Computer Vision and Pattern Recognition
  (CVPR)}}, pages 4510--4520. IEEE, 2018.

\bibitem{SDSE20}
Roy Schwartz, Jesse Dodge, Noah~A. Smith, and Oren Etzioni.
\newblock {Green {AI}}.
\newblock {\em {Commun. of the ACM}}, 2020.

\bibitem{SHYBZ22}
Zeyang Sha, Xinlei He, Ning Yu, Michael Backes, and Yang Zhang.
\newblock {Can't Steal? Cont-Steal! Contrastive Stealing Attacks Against Image
  Encoders}.
\newblock {\em {CoRR abs/2201.07513}}, 2022.

\bibitem{SH21}
Virat Shejwalkar and Amir Houmansadr.
\newblock {Membership Privacy for Machine Learning Models Through Knowledge
  Transfer}.
\newblock In {\em {AAAI Conference on Artificial Intelligence (AAAI)}}. AAAI,
  2021.

\bibitem{SSSS17}
Reza Shokri, Marco Stronati, Congzheng Song, and Vitaly Shmatikov.
\newblock {Membership Inference Attacks Against Machine Learning Models}.
\newblock In {\em {IEEE Symposium on Security and Privacy (S\&P)}}, pages
  3--18. IEEE, 2017.

\bibitem{SZ15}
Karen Simonyan and Andrew Zisserman.
\newblock {Very Deep Convolutional Networks for Large-Scale Image Recognition}.
\newblock In {\em {International Conference on Learning Representations
  (ICLR)}}, 2015.

\bibitem{SRS17}
Congzheng Song, Thomas Ristenpart, and Vitaly Shmatikov.
\newblock {Machine Learning Models that Remember Too Much}.
\newblock In {\em {ACM SIGSAC Conference on Computer and Communications
  Security (CCS)}}, pages 587--601. ACM, 2017.

\bibitem{SM21}
Liwei Song and Prateek Mittal.
\newblock {Systematic Evaluation of Privacy Risks of Machine Learning Models}.
\newblock In {\em {USENIX Security Symposium (USENIX Security)}}. USENIX, 2021.

\bibitem{SSM19}
Liwei Song, Reza Shokri, and Prateek Mittal.
\newblock {Privacy Risks of Securing Machine Learning Models against
  Adversarial Examples}.
\newblock In {\em {ACM SIGSAC Conference on Computer and Communications
  Security (CCS)}}, pages 241--257. ACM, 2019.

\bibitem{SZSBEGF14}
Christian Szegedy, Wojciech Zaremba, Ilya Sutskever, Joan Bruna, Dumitru Erhan,
  Ian Goodfellow, and Rob Fergus.
\newblock {Intriguing Properties of Neural Networks}.
\newblock In {\em {International Conference on Learning Representations
  (ICLR)}}, 2014.

\bibitem{TMK16}
Surat Teerapittayanon, Bradley McDanel, and H.~T. Kung.
\newblock {BranchyNet: Fast Inference via Early Exiting from Deep Neural
  Networks}.
\newblock In {\em {International Conference on Pattern Recognition (ICPR)}},
  pages 2464--2469, 2016.

\bibitem{TKPGBM17}
Florian Tram{\`e}r, Alexey Kurakin, Nicolas Papernot, Ian Goodfellow, Dan
  Boneh, and Patrick McDaniel.
\newblock {Ensemble Adversarial Training: Attacks and Defenses}.
\newblock In {\em {International Conference on Learning Representations
  (ICLR)}}, 2017.

\bibitem{TZJRR16}
Florian Tram{\`e}r, Fan Zhang, Ari Juels, Michael~K. Reiter, and Thomas
  Ristenpart.
\newblock {Stealing Machine Learning Models via Prediction APIs}.
\newblock In {\em {USENIX Security Symposium (USENIX Security)}}, pages
  601--618. USENIX, 2016.

\bibitem{VL14}
Yevgeniy Vorobeychik and Bo~Li.
\newblock {Optimal Randomized Classification in Adversarial Settings}.
\newblock In {\em {International Conference on Autonomous Agents and
  Multi-agent Systems (AAMAS)}}, pages 485--492. IFAAMAS/ACM, 2014.

\bibitem{WYSLVZZ19}
Bolun Wang, Yuanshun Yao, Shawn Shan, Huiying Li, Bimal Viswanath, Haitao
  Zheng, and Ben~Y. Zhao.
\newblock {Neural Cleanse: Identifying and Mitigating Backdoor Attacks in
  Neural Networks}.
\newblock In {\em {IEEE Symposium on Security and Privacy (S\&P)}}, pages
  707--723. IEEE, 2019.

\bibitem{WL21}
Xinglu Wang and Yingming Li.
\newblock {Harmonized Dense Knowledge Distillation Training for Multi-Exit
  Architectures}.
\newblock In {\em {AAAI Conference on Artificial Intelligence (AAAI)}}, pages
  10218--10226. AAAI, 2021.

\bibitem{WSHXNBWL20}
Yue Wang, Jianghao Shen, Ting{-}Kuei Hu, Pengfei Xu, Tan~M. Nguyen, Richard~G.
  Baraniuk, Zhangyang Wang, and Yingyan Lin.
\newblock {Dual Dynamic Inference: Enabling More Efficient, Adaptive, and
  Controllable Deep Inference}.
\newblock {\em {Journal of Selected Topics in Signal Processing}}, 2020.

\bibitem{XTLYL20}
Ji~Xin, Raphael Tang, Jaejun Lee, Yaoliang Yu, and Jimmy Lin.
\newblock {DeeBERT: Dynamic Early Exiting for Accelerating {BERT} Inference}.
\newblock In {\em {Annual Meeting of the Association for Computational
  Linguistics (ACL)}}, pages 2246--2251. ACL, 2020.

\bibitem{YVCZZ17}
Yuanshun Yao, Bimal Viswanath, Jenna Cryan, Haitao Zheng, and Ben~Y. Zhao.
\newblock {Automated Crowdturfing Attacks and Defenses in Online Review
  Systems}.
\newblock In {\em {ACM SIGSAC Conference on Computer and Communications
  Security (CCS)}}, pages 1143--1158. ACM, 2017.

\bibitem{YGFJ18}
Samuel Yeom, Irene Giacomelli, Matt Fredrikson, and Somesh Jha.
\newblock {Privacy Risk in Machine Learning: Analyzing the Connection to
  Overfitting}.
\newblock In {\em {IEEE Computer Security Foundations Symposium (CSF)}}, pages
  268--282. IEEE, 2018.

\bibitem{ZK16}
Sergey Zagoruyko and Nikos Komodakis.
\newblock {Wide Residual Networks}.
\newblock In {\em {Proceedings of the British Machine Vision Conference
  (BMVC)}}. {BMVA} Press, 2016.

\bibitem{ZE19}
Xiao Zhang and David Evans.
\newblock {Cost-Sensitive Robustness against Adversarial Examples}.
\newblock In {\em {International Conference on Learning Representations
  (ICLR)}}, 2019.

\bibitem{ZHRLPB18}
Yang Zhang, Mathias Humbert, Tahleen Rahman, Cheng-Te Li, Jun Pang, and Michael
  Backes.
\newblock {Tagvisor: A Privacy Advisor for Sharing Hashtags}.
\newblock In {\em {The Web Conference (WWW)}}, pages 287--296. ACM, 2018.

\bibitem{ZCSZ22}
Junhao Zhou, Yufei Chen, Chao Shen, and Yang Zhang.
\newblock {Property Inference Attacks Against GANs}.
\newblock In {\em {Network and Distributed System Security Symposium (NDSS)}}.
  Internet Society, 2022.

\bibitem{ZWTD19}
Li~Zhou, Hao Wen, Radu Teodorescu, and David H.~C. Du.
\newblock {Distributing Deep Neural Networks with Containerized Partitions at
  the Edge}.
\newblock In {\em {Hot Topics in Edge Computing (HotEdge)}}. USENIX, 2019.

\bibitem{ZXGMXW20}
Wangchunshu Zhou, Canwen Xu, Tao Ge, Julian~J. McAuley, Ke~Xu, and Furu Wei.
\newblock {{BERT} Loses Patience: Fast and Robust Inference with Early Exit}.
\newblock In {\em {Annual Conference on Neural Information Processing Systems
  (NeurIPS)}}. NeurIPS, 2020.

\end{thebibliography}
\end{small}

\newpage
\appendix

\section{Appendix}

\subsection{Datasets Description}
\label{appendix:datasets}

Here we briefly describe each dataset.

\mypara{CIFAR-10/CIFAR-100.}
CIFAR-10~\cite{CIFAR} and CIFAR-100~\cite{CIFAR} are benchmark datasets used to evaluate image recognition algorithms. 
CIFAR-10 is composed of 32$\times$32 color images in 10 classes, with 6000 images per class. 
In total, there are 50000 training images and 10000 test images. 
CIFAR-100 has the same format as CIFAR-10, but it has 100 classes containing 600 images each. 
There are 500 training images and 100 testing images per class. 

\mypara{TinyImageNet.} 
TinyImageNet is another benchmark dataset used to evaluate image recognition algorithms.
It contains 100000 images of 200 classes (500 for each class) downsized to 64×64 colored images. Each class has 500 training images, 50 validation images, and 50 test images.

\mypara{Purchases.}
This dataset is based on the “acquire valued
shopper” challenge from Kaggle, which consists of 197,000 customer records with 600 binary features representing the customer purchase history. 
The records are clustered into 100 classes, each representing a unique purchase style, such that the goal is to predict a customer’s purchase style.
This dataset is also widely used to evaluate membership inference attack in ~\cite{SSSS17,HYYBGC21,JE19,LLR21,NSH18,ROF20,CTCP21}.

\mypara{Locations.}
This dataset was pre-processed from the Foursquare dataset\footnote{\url{https://sites.google.com/site/yangdingqi/home/foursquare-dataset}} which has 5,010 data samples with 446 binary features, each of which represents whether a user visited a particular region or location type. 
This dataset represents a 30-class classification problem, and the goal is to predict the user’s geosocial type given his or her record.
This dataset is used to evaluate membership inference attack in ~\cite{JSBZG19, SSSS17, CTCP21}

\mypara{Texas.} 
This dataset is based on the Discharge Data public use files published by the Texas Department of State Health Services.\footnote{\url{https://www.dshs.texas.gov/THCIC/Hospitals/Download.shtm}}
The dataset has 67,330 data samples with 6,170 binary features. These features
represent the external causes of injury (e.g., suicide, drug misuse),
the diagnosis, the procedures the patient underwent, and some
generic information (e.g., gender, age, and race). Similar to [58], we
focus on the 100 most frequent procedures (i.e., 100 classes) and the classification task is to predict a procedure for a patient using the patient’s data.
This dataset is used to evaluate membership inference attack in ~\cite{HYYBGC21,LLR21,JSBZG19,CTCP21,SSSS17,SH21}.

\subsection{The Architecture of FCN-18}
\label{appendix:fcn-18}

First, we designed a block module consisting of three layers, namely Linear, BatchNorm1d, and ReLU.
For the backbone architecture, we build it by stacking 5 blocks one by one and finally connecting them to the three Linear layers.
For each exit module architecture, we build it by two Linear layers and place it after each block module.

For extensive evaluation, we build four FCN-18 with different linear layers of different numbers of hidden neural units (1024, 2048, 3072, 4096), and named FCN-18-1/2/3/4 throughout the paper.

\subsection{Possible Reason for Relaxing Same-distribution and Same-architecture}
\label{reason_distribution}

Here, we mainly discuss the reason for relaxing same distribution between shadow and target datasets, and same architecture between shadow and target models.

The attack pipeline is that we train an attack model based on the output of shadow model, then utilize the trained attack model to predict the output of target model.
The key to the success of the attack model is that there is a significant difference between the outputs of the members and non-members from the shadow model, and the attack model learns how to distinguish between them.
Also, there is a significant difference between the outputs of members and non-members from the target model, so a trained attack model can also distinguish between them.

In general, such ‘‘difference’’ on which the attack model relies always exits in the target and shadow model respectively, regardless of whether the target and shadow datasets are identically distributed, or the target and shadow models have the same architecture.
Therefore, even if we relax the assumption of same-distribution and same-architecture, the attack still works.

\begin{table}[!htbp]
\definecolor{mygray}{gray}{0.9}
\centering
\caption{The threshold $\tau$ set for computer vision tasks.}
\label{table:end_threshold_cv}
\setlength{\tabcolsep}{2.5pt}
\scalebox{0.8}{
\begin{tabular}
{c|c|c|c|c|c}
\toprule
\multirow{2}{*}{Dataset}&Exit&  \multicolumn{4}{c}{Model Architecture}\\
 &Number&VGG&ResNet&MobileNet&WideResNet\\
\midrule
\multirow{5}{*}{CIFAR-10}& 2 & 0.9& 0.7&  0.6& 0.85\\
                         & 3 & 0.9& 0.7&  0.6& 0.85\\
                         & 4 & 0.9& 0.7&  0.6& 0.85\\
                         & 5 & 0.9& 0.7&  0.6& 0.85\\
                         & 6 & 0.9& 0.7&  0.6& 0.85\\
\midrule
\multirow{5}{*}{CIFAR-100}& 2 &  0.2& 0.3&  0.4& 0.8\\
                          & 3 &  0.2& 0.3&  0.4& 0.8\\
                          & 4 &  0.2& 0.3&  0.4& 0.8\\
                          & 5 &  0.2& 0.3&  0.4& 0.8\\
                          & 6 &  0.2& 0.3&  0.4& 0.8\\
\midrule
\multirow{5}{*}{TinyImageNet}& 2 &  0.4& 0.25&  0.55& 0.85\\
                             & 3 &  0.4& 0.25&  0.55& 0.85\\
                             & 4 &  0.4& 0.25&  0.55& 0.85\\
                             & 5 &  0.4& 0.25&  0.55& 0.85\\
                             & 6 &  0.4& 0.25&  0.55& 0.85\\
\bottomrule
\end{tabular}
}
\end{table}

\begin{table}[!htbp]
\definecolor{mygray}{gray}{0.9}
\centering
\caption{The threshold $\tau$ set for non-computer vision tasks.}
\label{table:end_threshold_non_cv}
\setlength{\tabcolsep}{2pt}
\scalebox{0.8}{
\begin{tabular}{c|c|c|c|c|c}
\toprule
\multirow{2}{*}{Dataset}&Exit&  \multicolumn{4}{c}{Model Architecture}\\
 &Number&FCN-18-1&FCN-v-2&FCN-18-3&FCN-18-4\\
\midrule
Purchases& 2/3/4/5/6 & 0.7& 0.7&  0.7& 0.7\\
\midrule
Locations& 2/3/4/5/6 &  0.5& 0.5&  0.5& 0.5\\
\midrule
Texas       & 2/3/4/5/6 &  0.7& 0.7&  0.7& 0.7\\
\bottomrule
\end{tabular}
}
\end{table}

\begin{figure}[!htbp]
    \centering
    \subfloat[Classification Performance]{\includegraphics[width=0.5\linewidth]{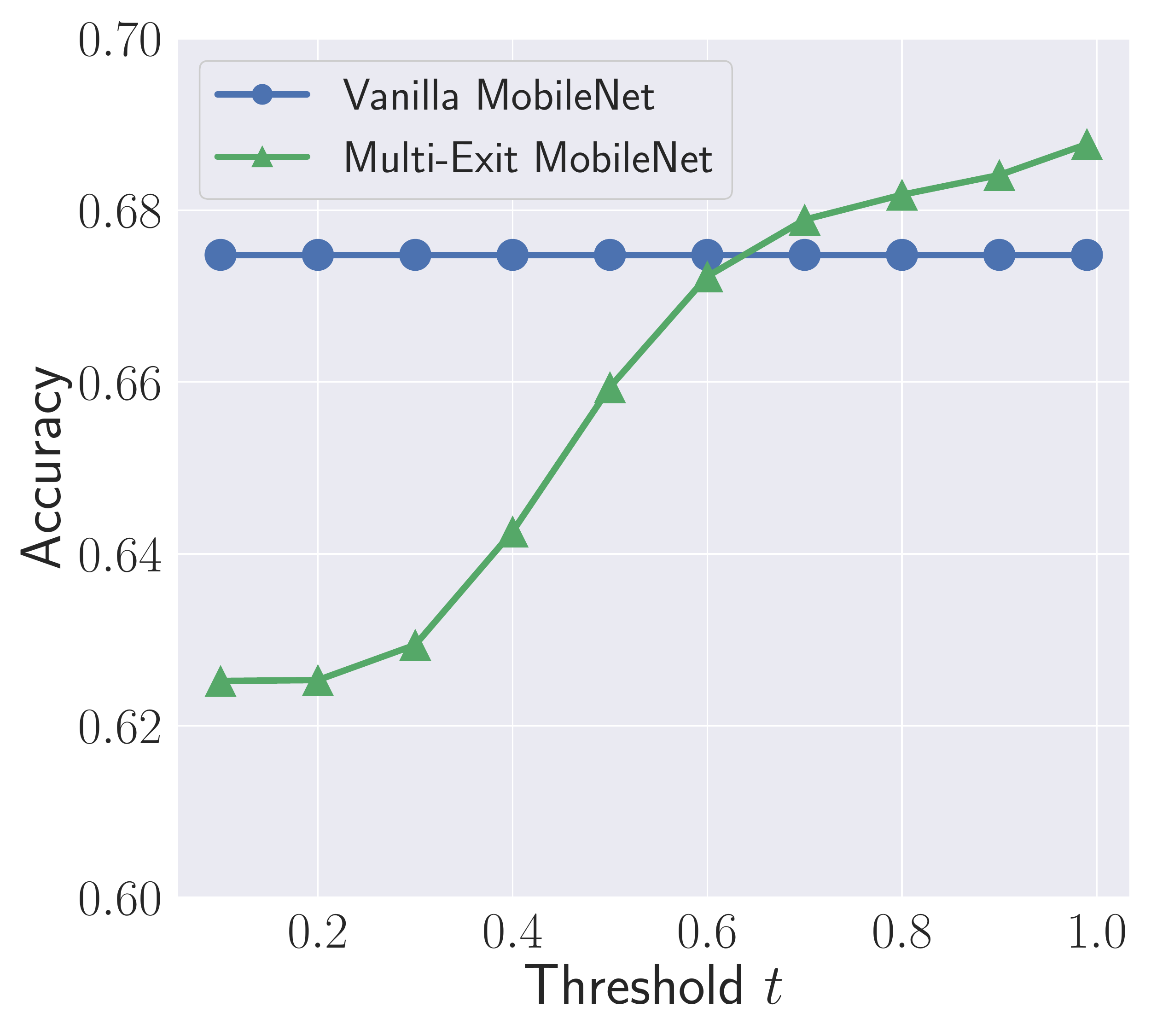}}
    \subfloat[Computational Cost]{\includegraphics[width=0.5\linewidth]{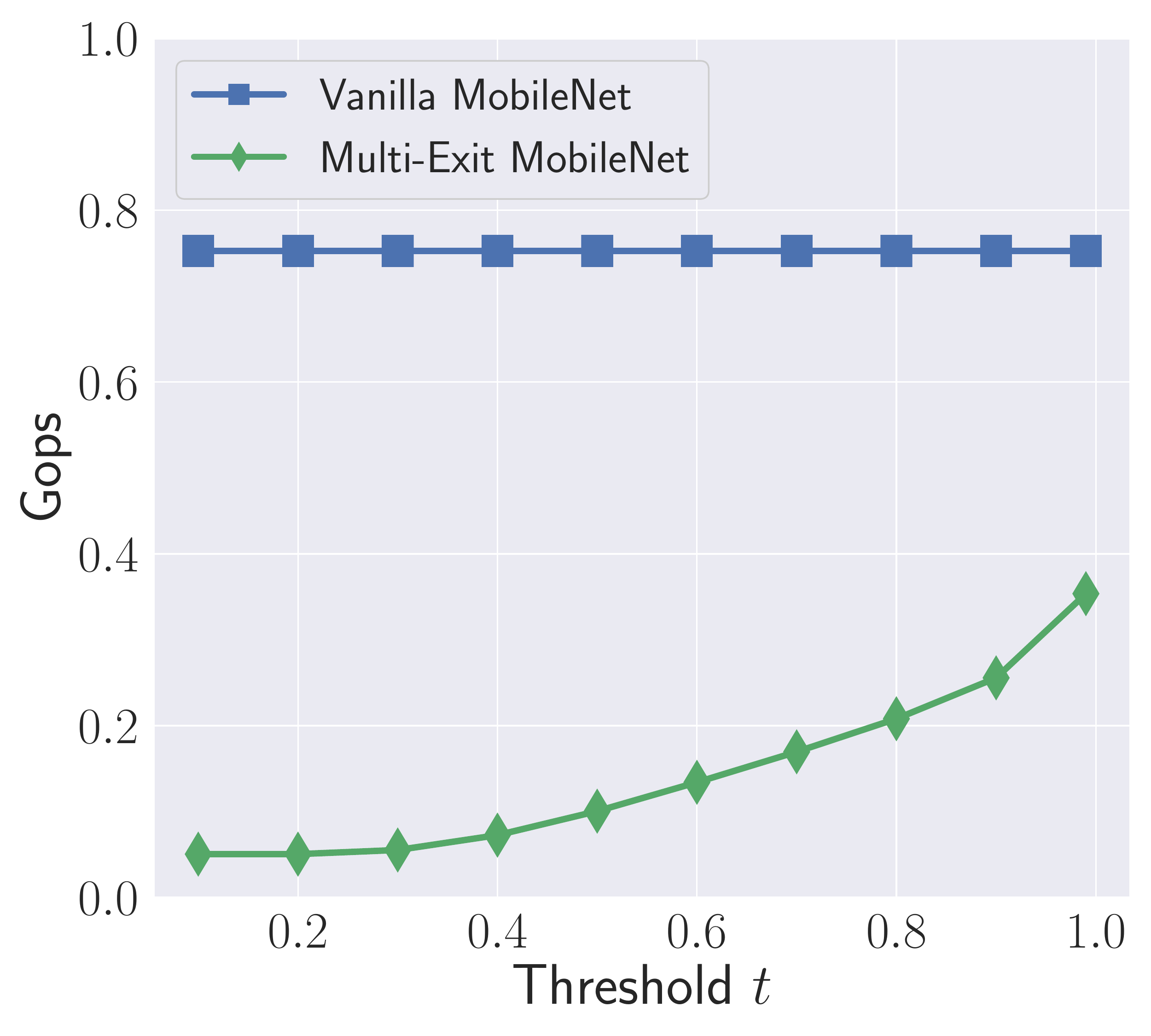}}
    \caption{Prediction accuracy and computational cost with respect to the threshold from 0 to 1. }
    \label{fig:choose_threshold}
\end{figure}

\begin{figure}[!htbp]
    \centering
    \subfloat[ResNet-56]{\includegraphics[width=0.5\linewidth]{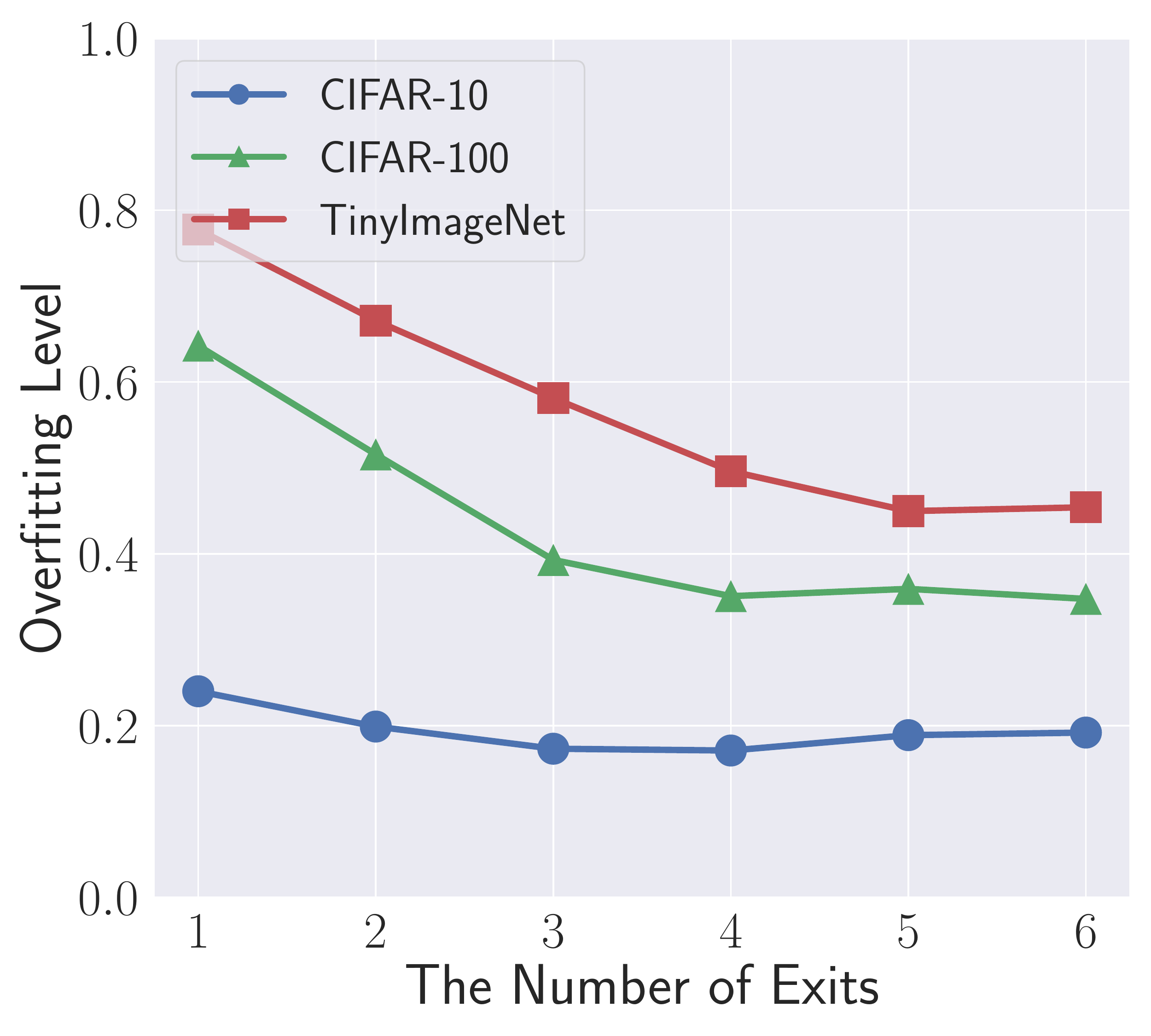}}
    \subfloat[MobileNet]{\includegraphics[width=0.5\linewidth]{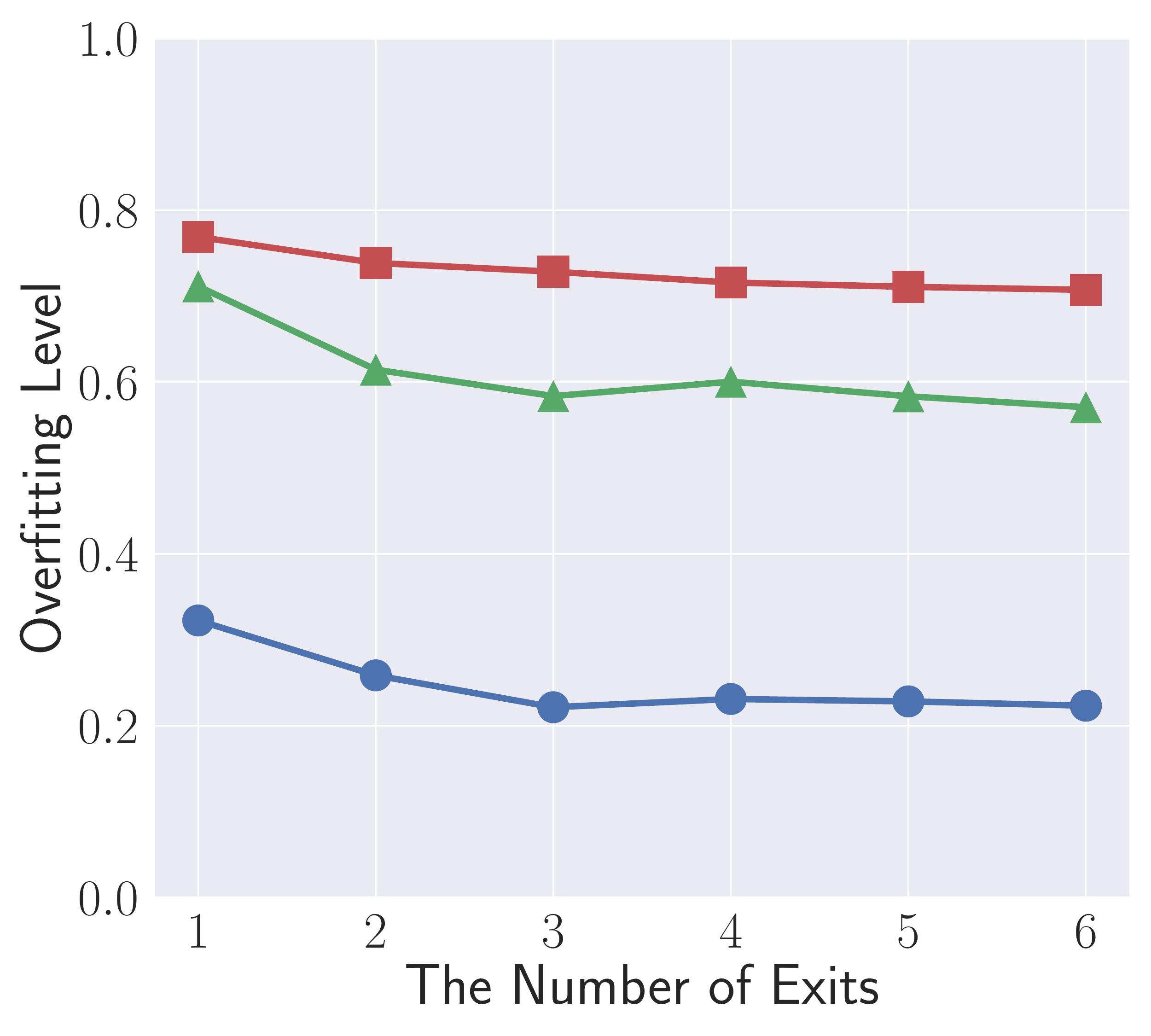}}
    \caption{Comparison of overfitting levels between vanilla and multi-exit model.}
    \label{fig:overfitting_rest}
\end{figure}

\begin{figure}[t]
    \centering
    \subfloat[CIFAR-10]{\includegraphics[width=0.5\linewidth]{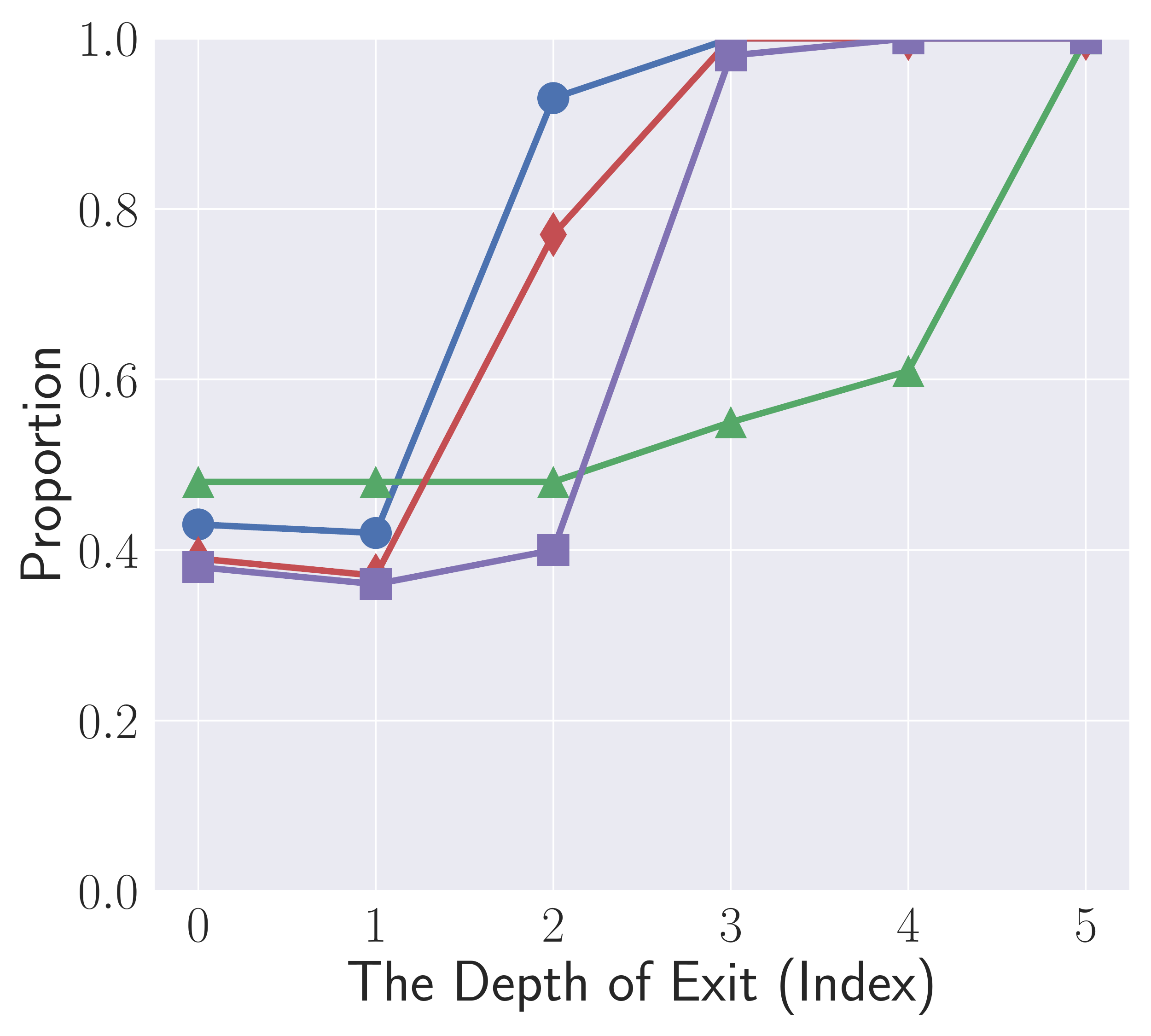}}
    \subfloat[Locations]{\includegraphics[width=0.5\linewidth]{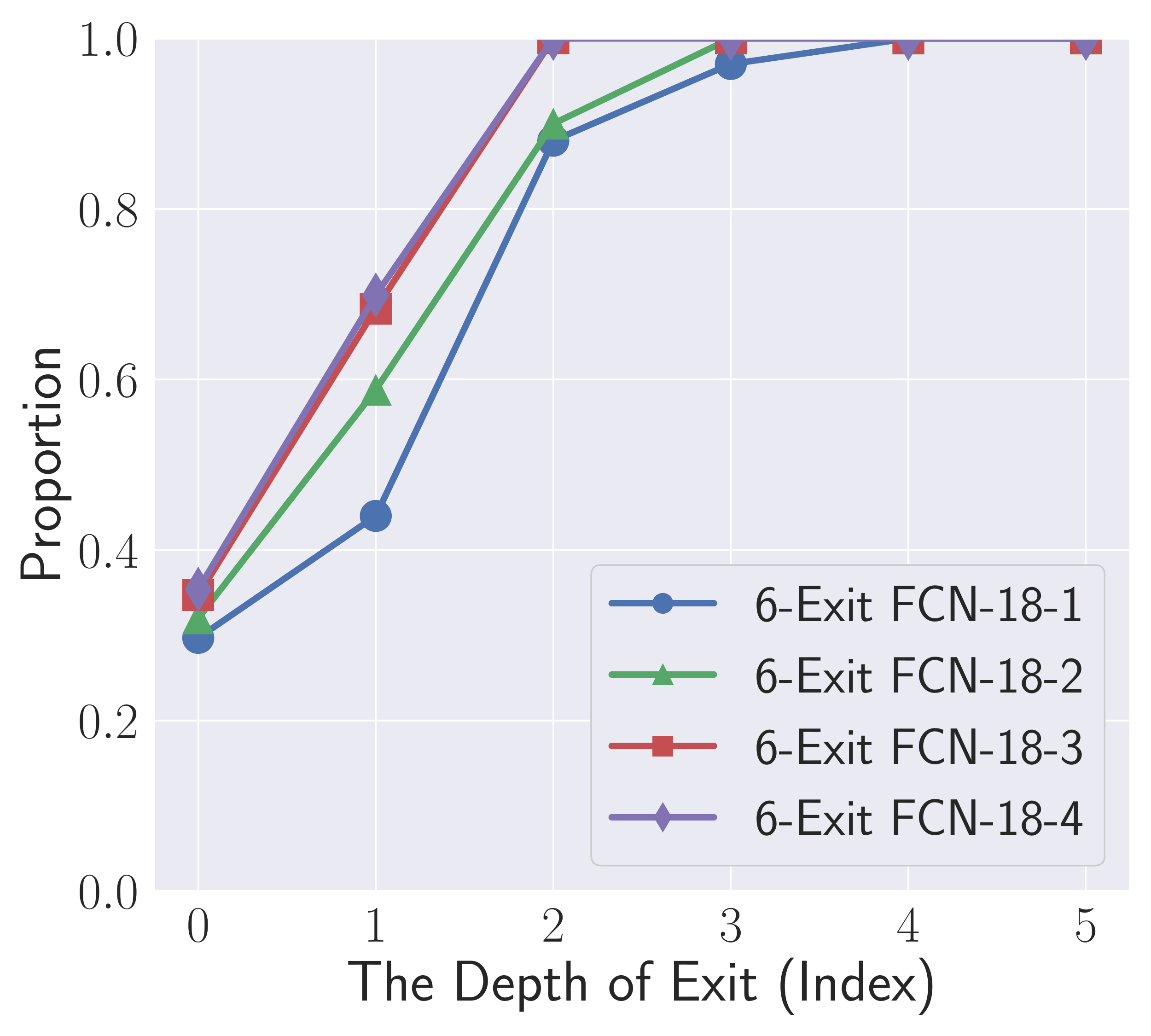}}
    \caption{Proportion of non-members in all samples leaving at each exit.}
\label{fig:appendix_ratio} 
\end{figure}

\begin{figure}[!htbp]
    \centering
    \subfloat[VGG-16]{\includegraphics[width=0.5\linewidth]{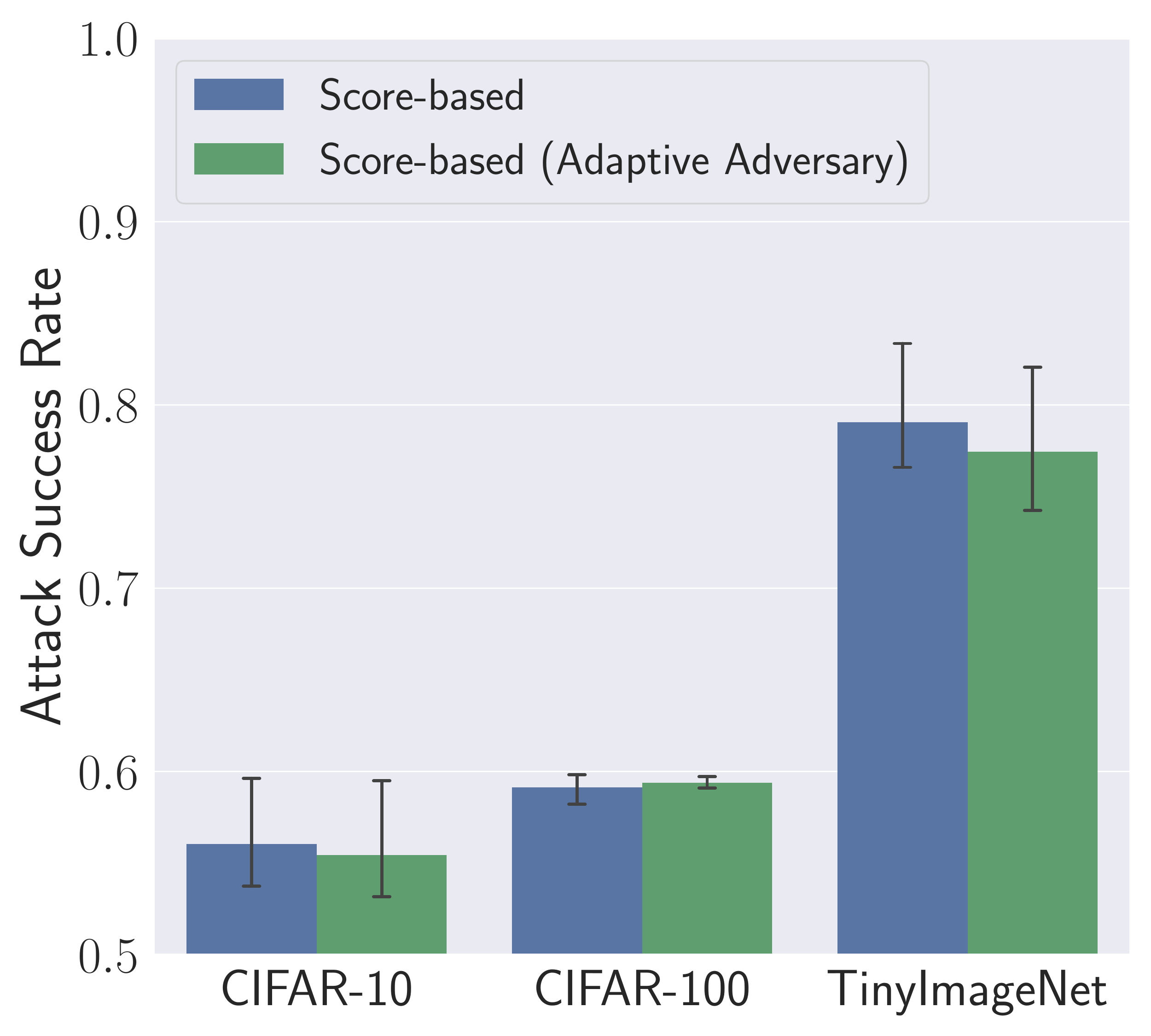}}
    \subfloat[MobileNet]{\includegraphics[width=0.5\linewidth]{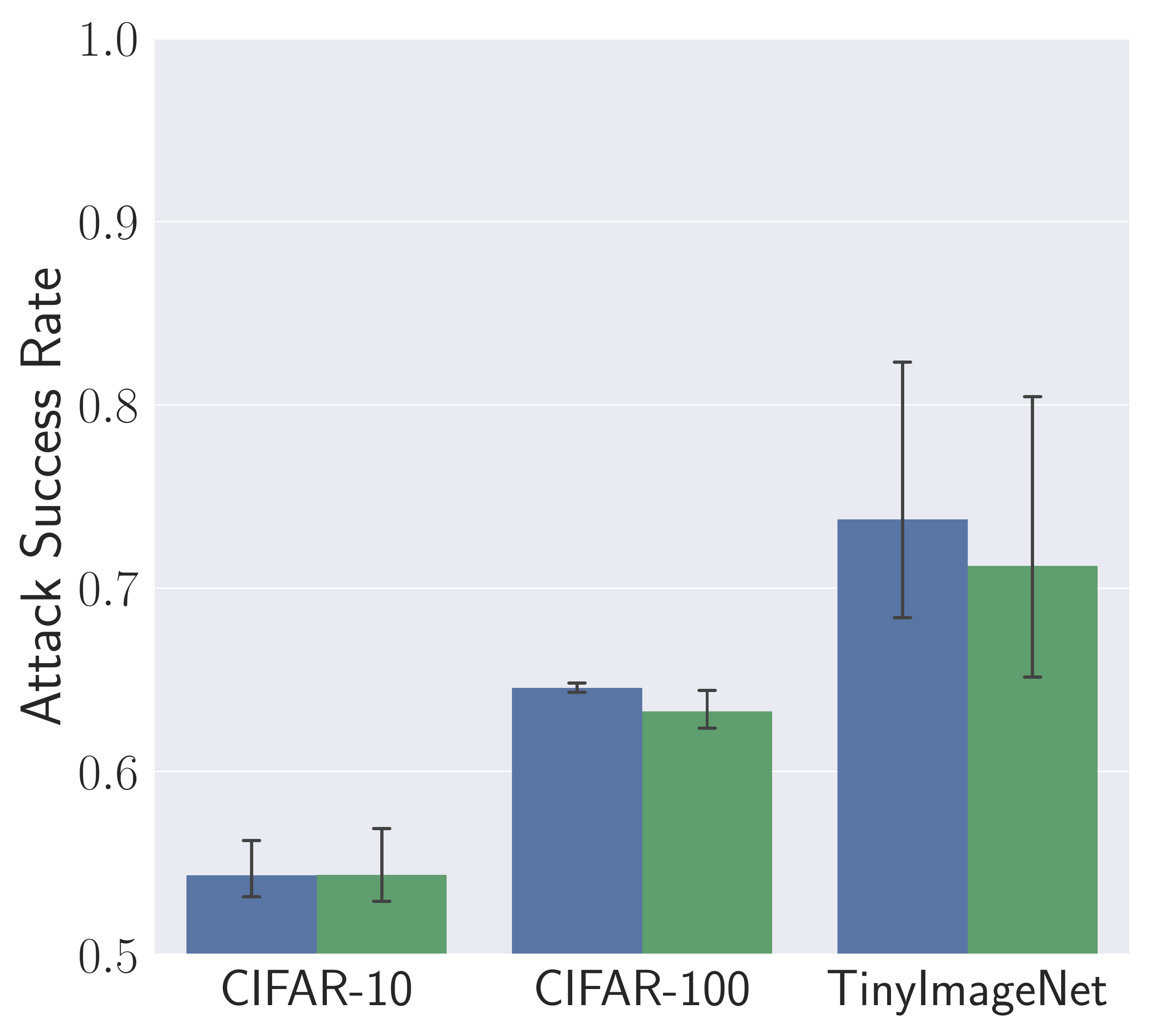}}
    \caption{The attack performance of original attack and adaptive attack against our defense \textit{TimeGuard}.}
    \label{fig:adaptive_adversary}
\end{figure}
\begin{figure}[!ht]
    \centering
    \subfloat[Exit Prediction Accuracy]{\includegraphics[width=0.5\linewidth]{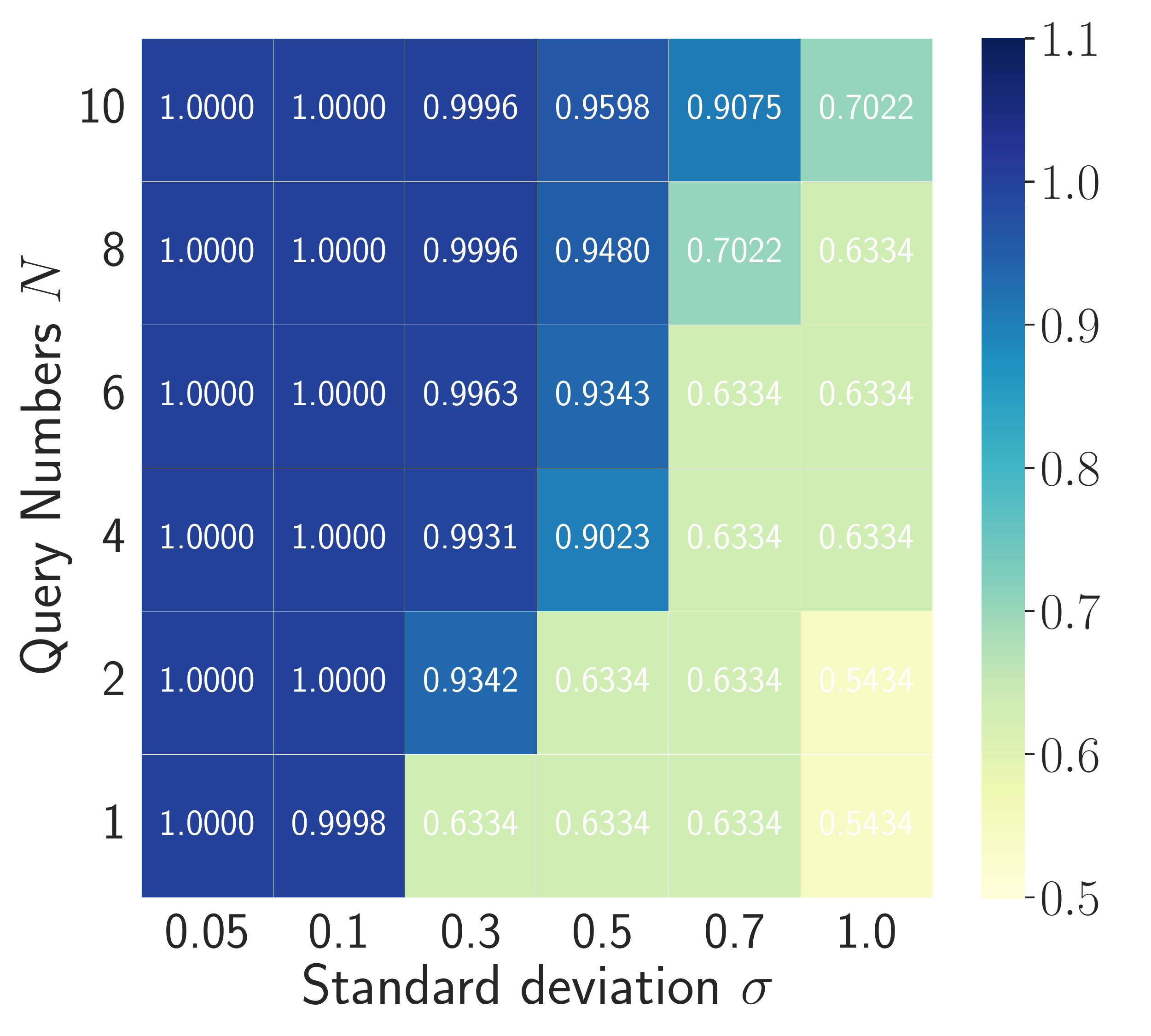}}
    \subfloat[\texttt{ASR} Score]{\includegraphics[width=0.5\linewidth]{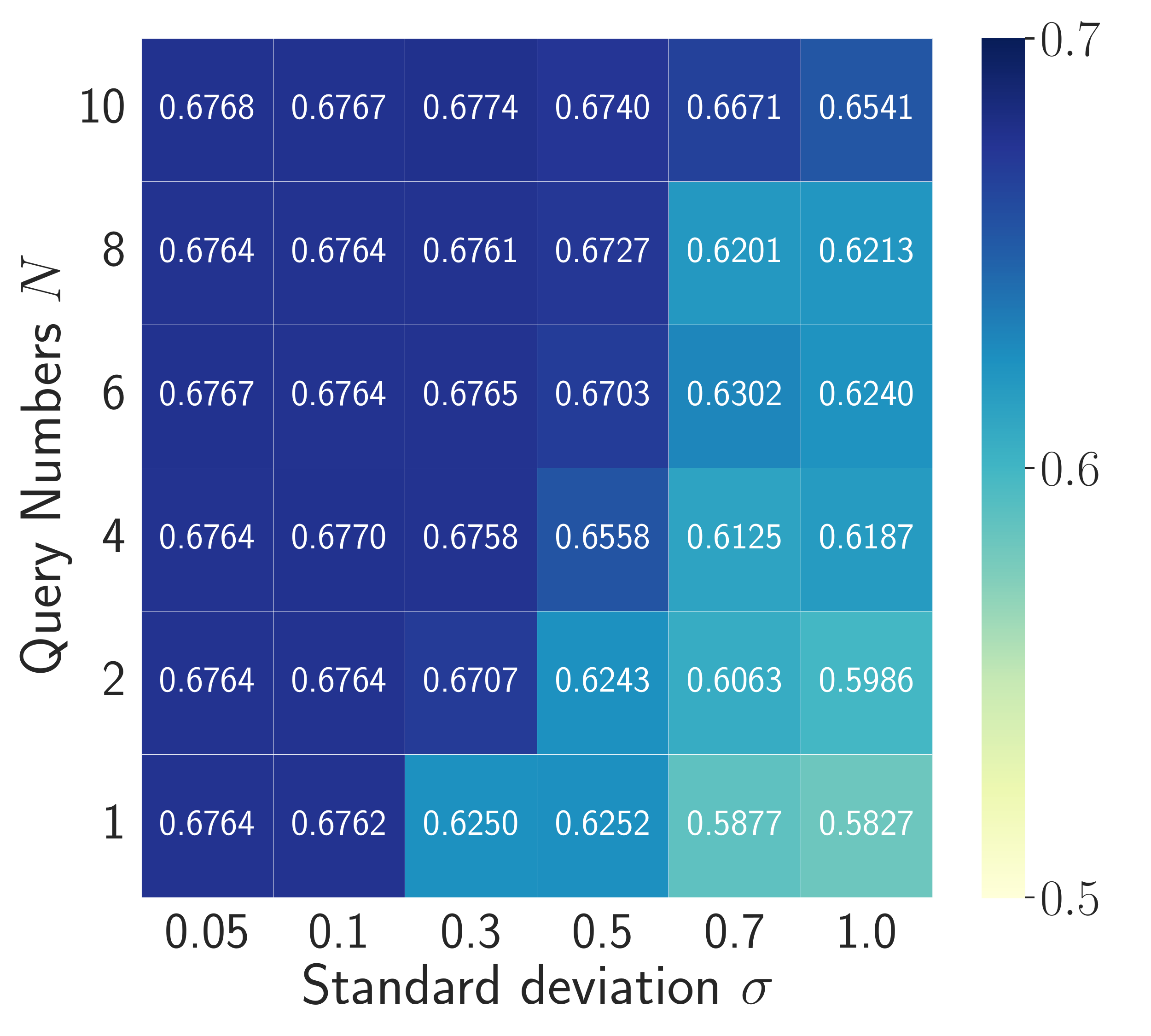}}
    \caption{The exit prediction and attack performance under the effect of query numbers $N$ and standard deviation $\sigma$. 
    The model is MobileNet trained on CIFAR-100.}
    \label{fig:MLaaS_appendix}
\end{figure}

\begin{figure*}[htb]
    \centering
    \subfloat[Vanilla ResNet-56]{\includegraphics[width=0.25\linewidth]{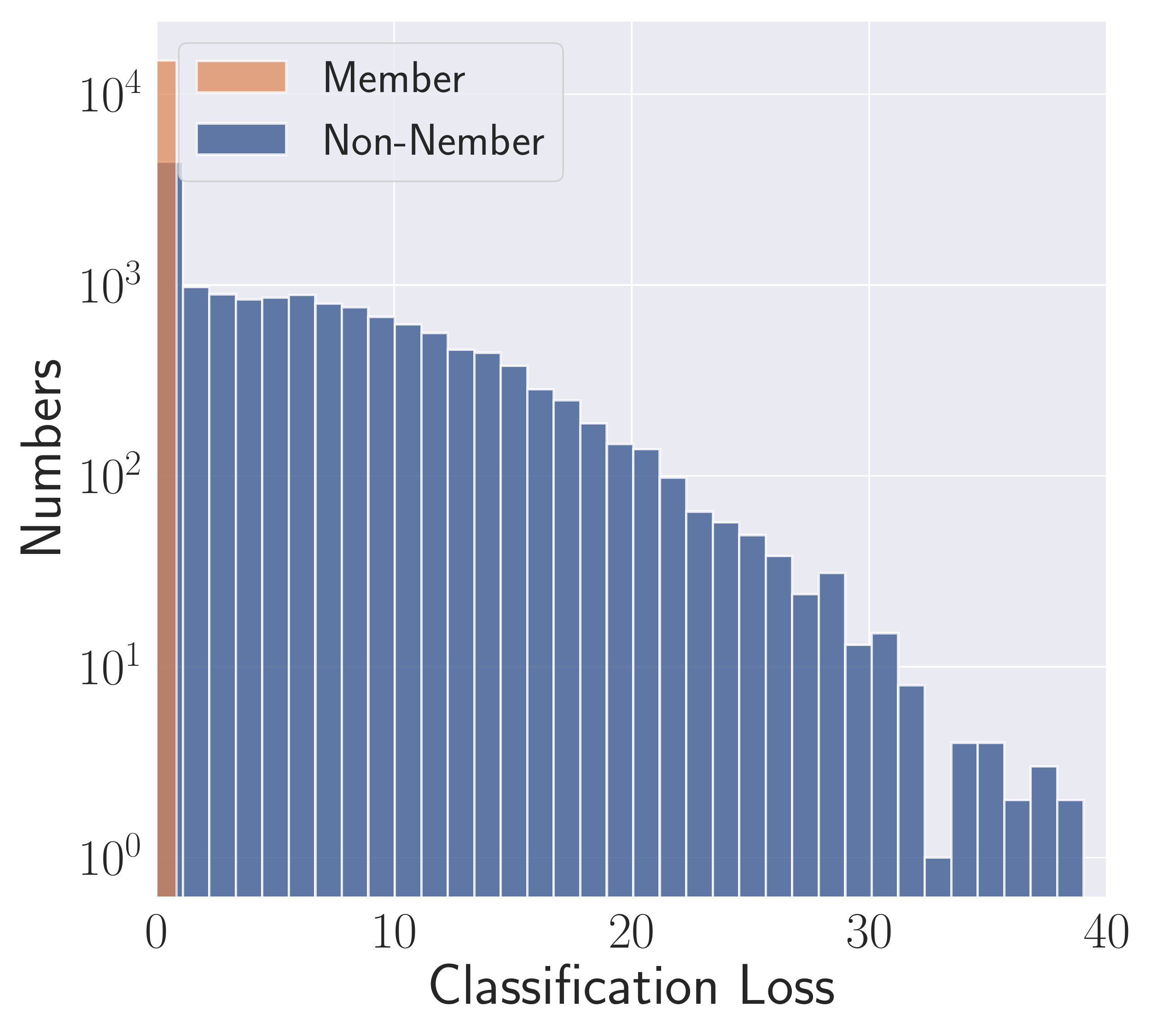}}
    \subfloat[2-Exit ResNet-56]{\includegraphics[width=0.25\linewidth]{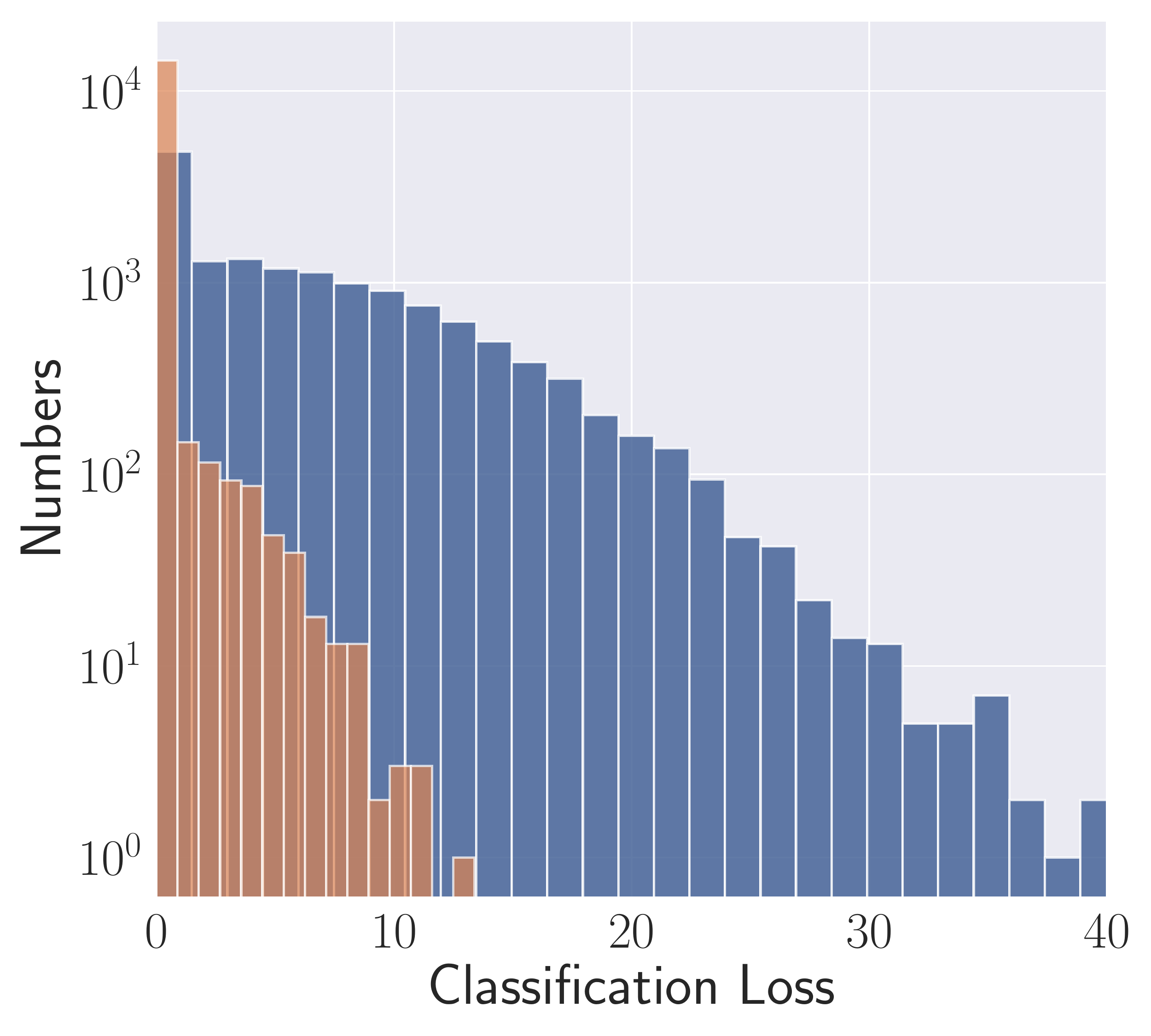}}
    \subfloat[4-Exit ResNet-56]{\includegraphics[width=0.25\linewidth]{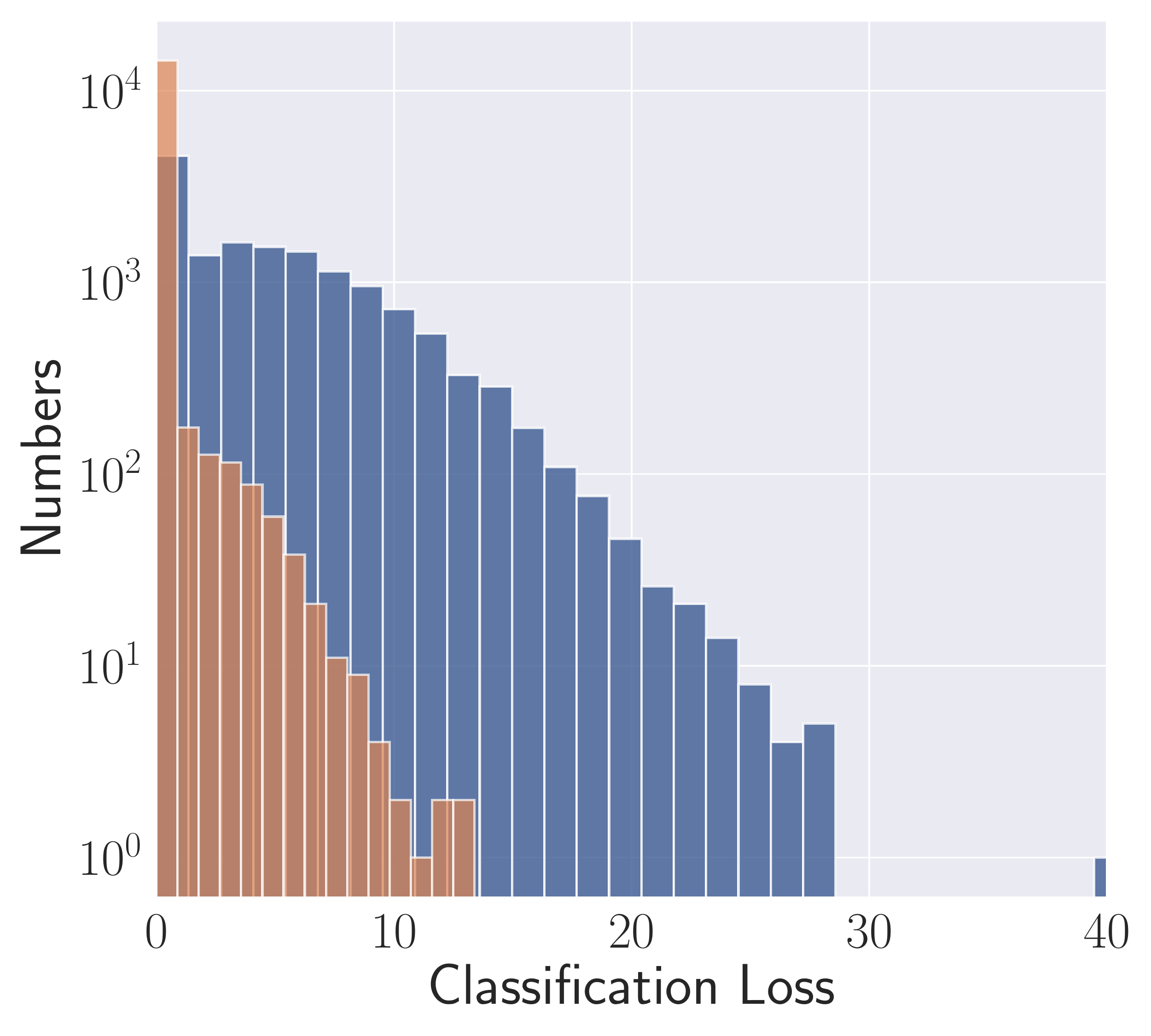}}
    \subfloat[6-Exit ResNet-56]{\includegraphics[width=0.25\linewidth]{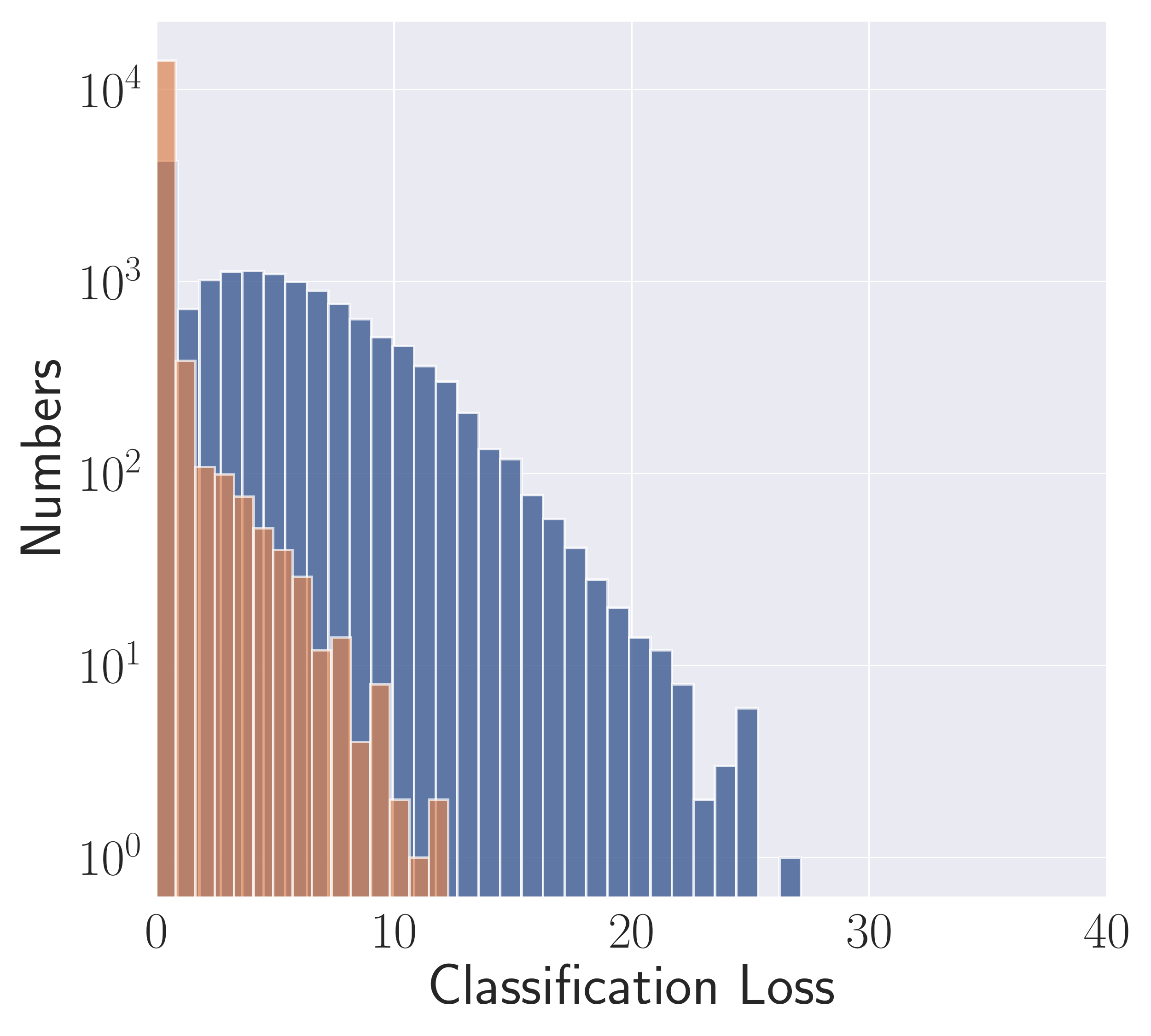}}
    \caption{The distribution of loss with respect to original classification tasks for member and non-member samples for both the vanilla model and the multi-exit model with ResNet-56 on TinyImageNet. 
    The x-axis represents each sample’s classification loss. The y-axis represents the number of member and nonmember samples.}
    \label{fig:vgg_TN_loss_distribution}
\end{figure*}


\begin{figure*}[t]
    \centering
    \subfloat[Classification Accuracy ]{\includegraphics[width=0.33\linewidth]{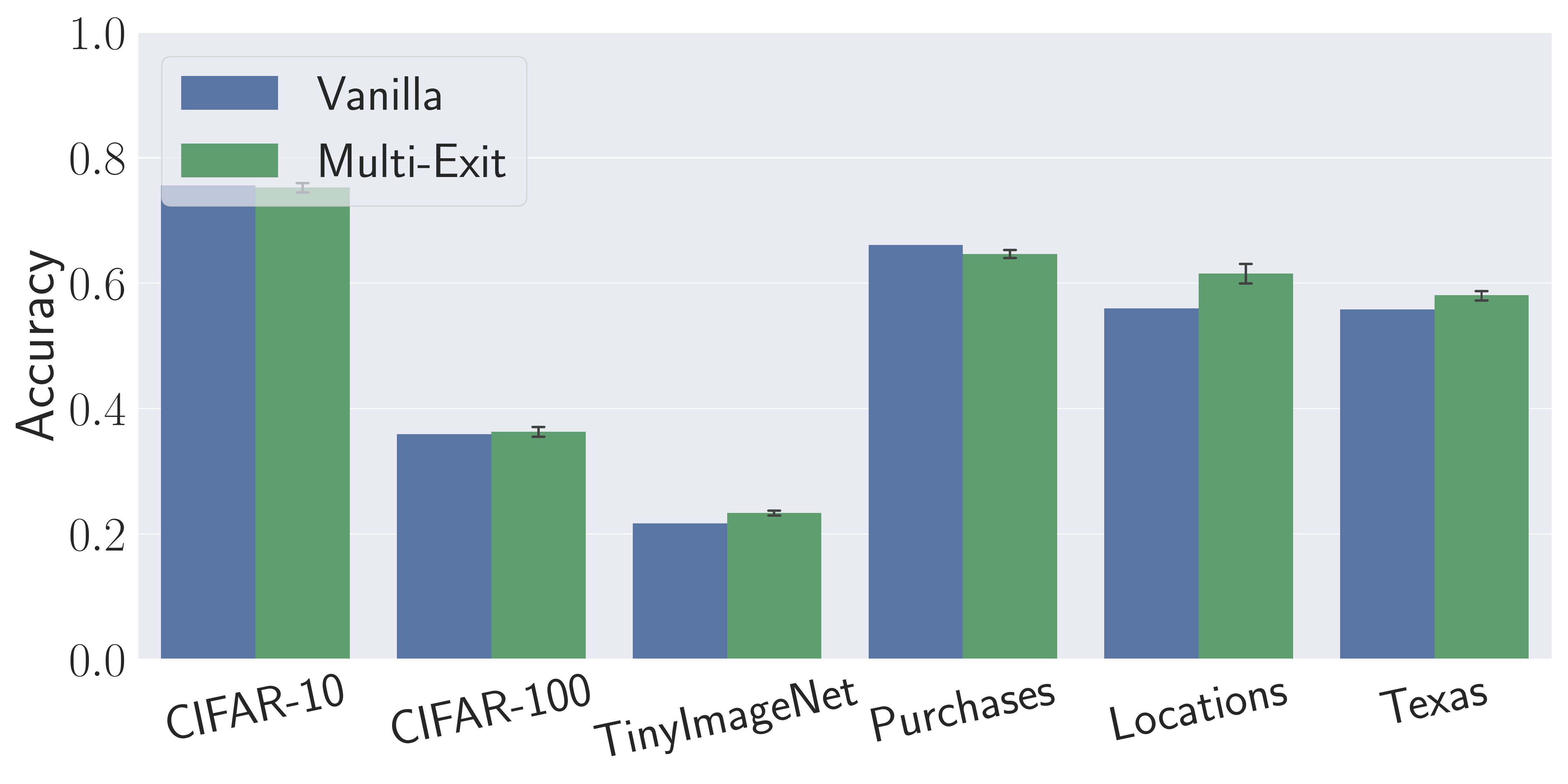}
    \includegraphics[width=0.33\linewidth]{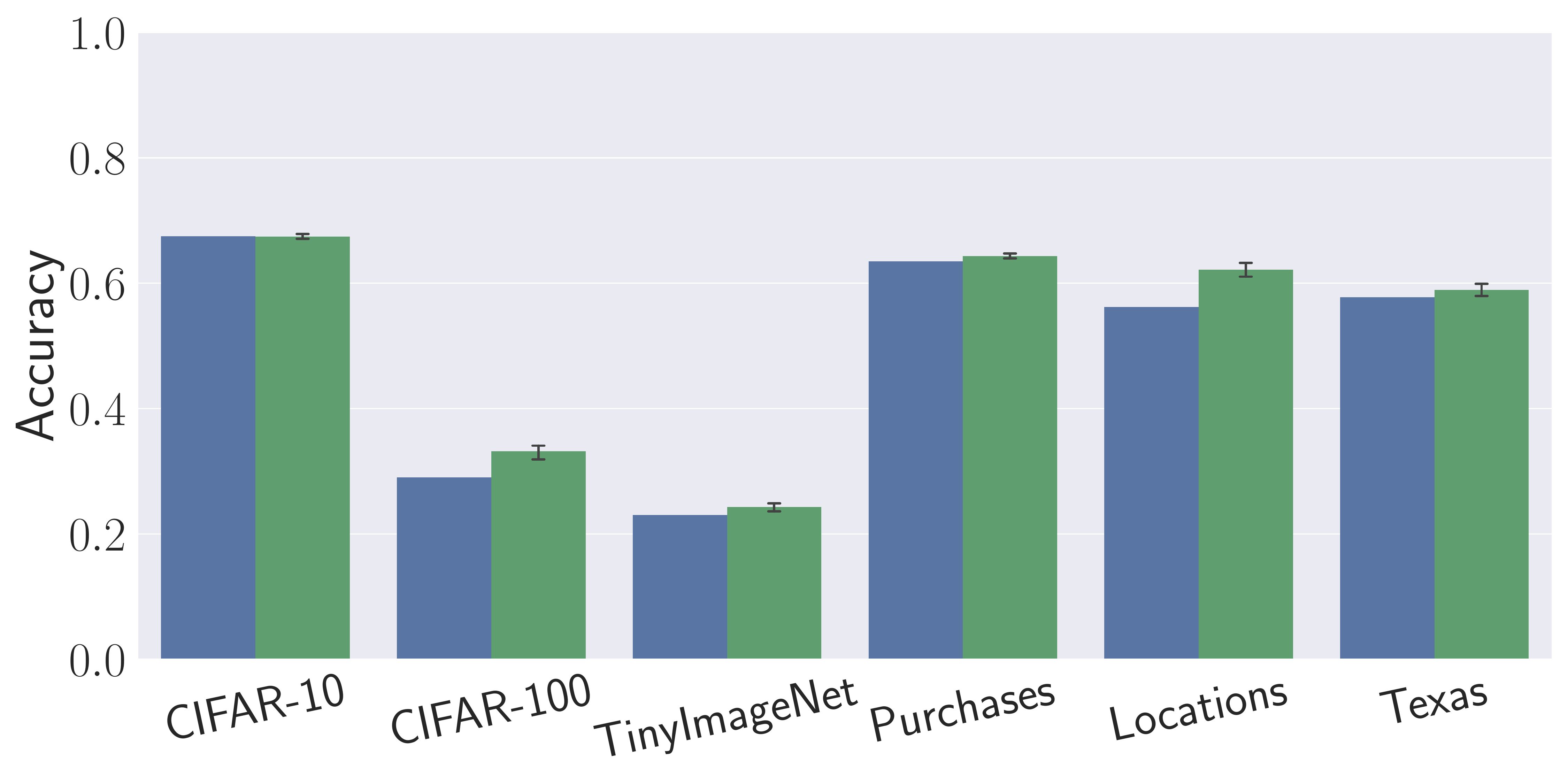}
    \includegraphics[width=0.33\linewidth]{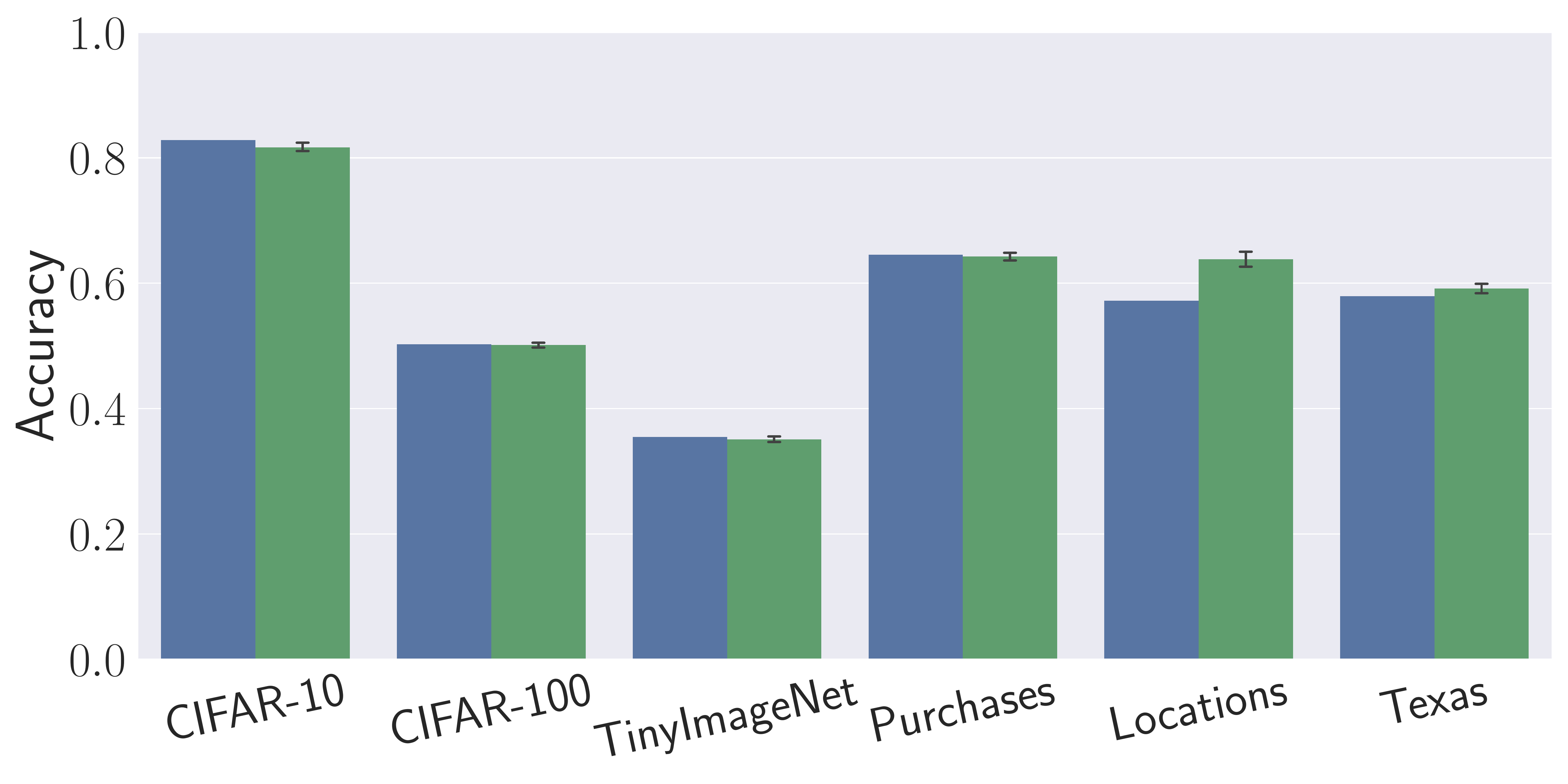}
    }
    
    \subfloat[Computational Cost]{
    \includegraphics[width=0.33\linewidth]{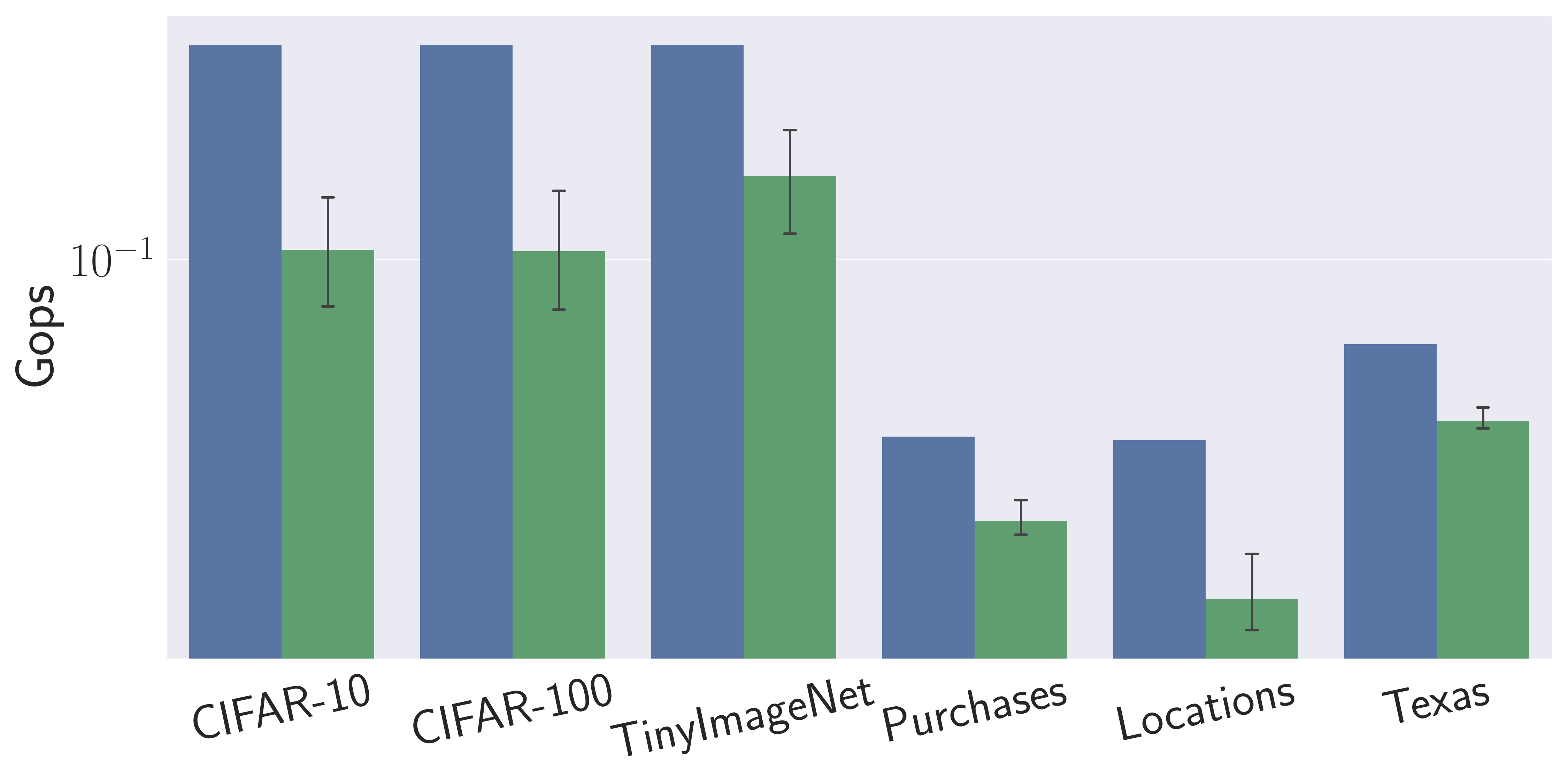}
    \includegraphics[width=0.33\linewidth]{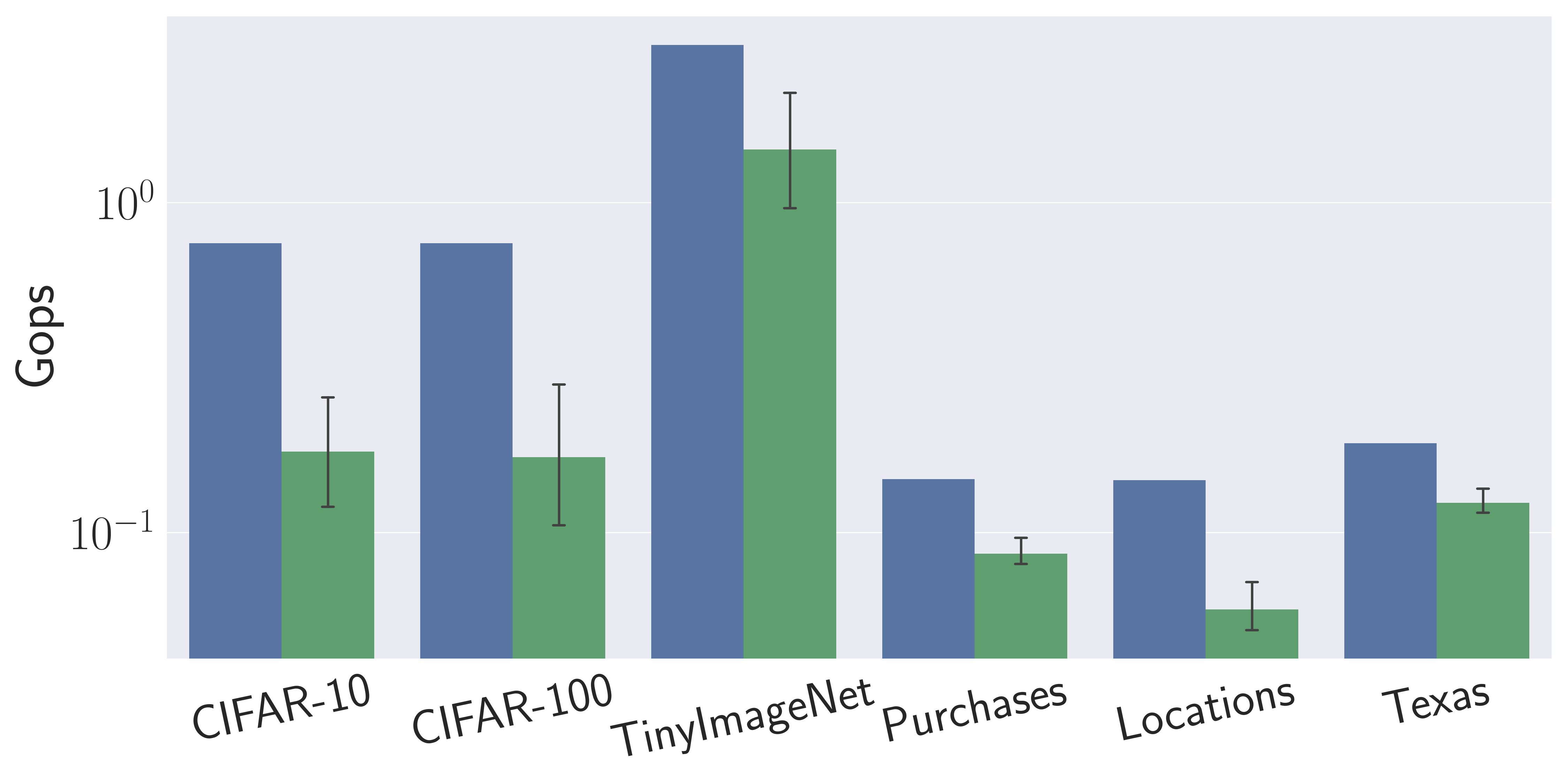}
    \includegraphics[width=0.33\linewidth]{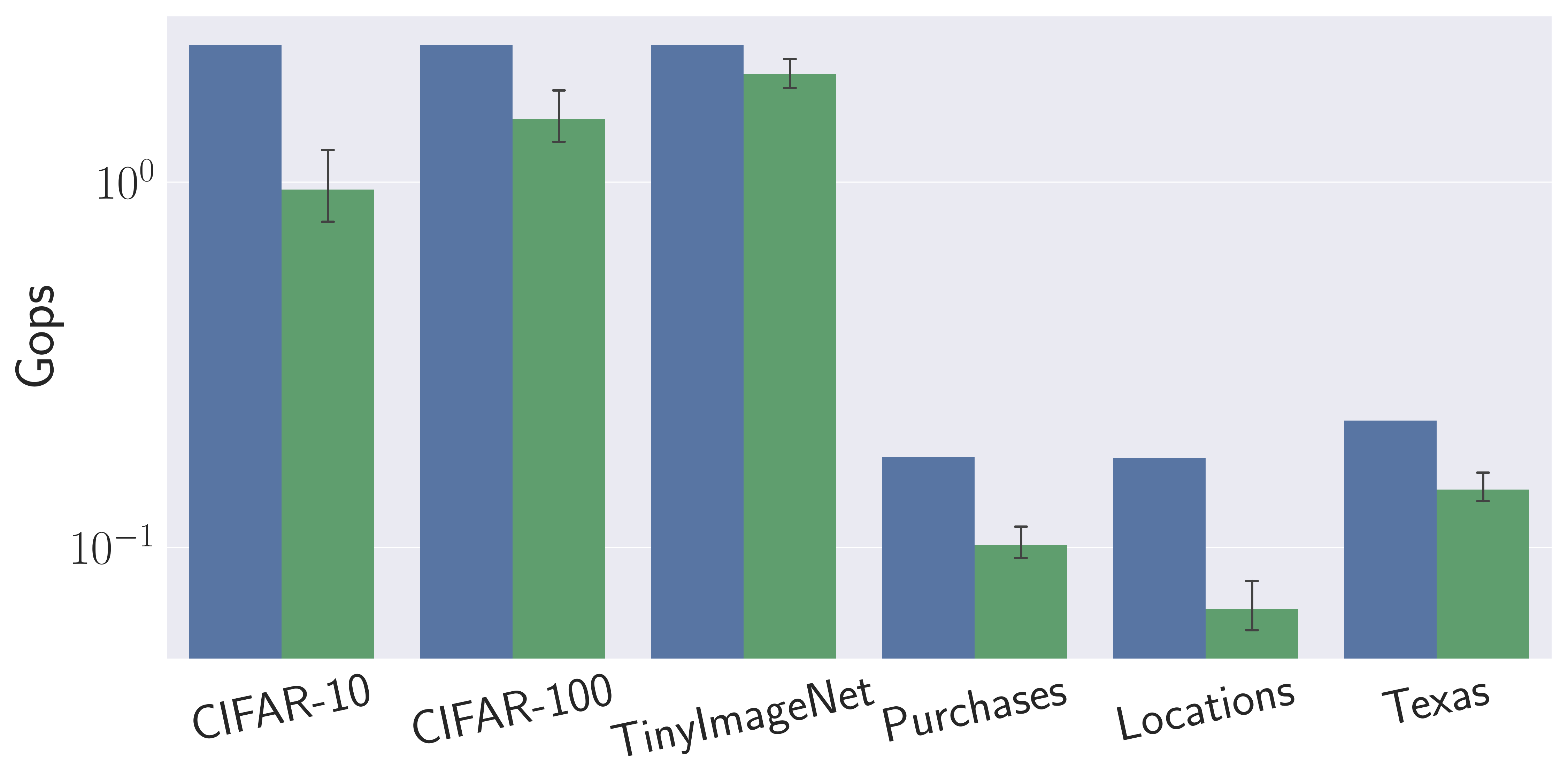}}
    \caption{The performance of original classification tasks and computational costs for both vanilla and multi-exit models. 
    From left to right (first/second/third columns), computer vision tasks on VGG-16/ResNet-56/WideResNet-32 and non-computer vision tasks on FCN-18-2/FCN-18-3/FCN-18-4.}
\label{fig:appendix_classification_performance}
\end{figure*}

\begin{figure*}[!htbp]
    \centering
    \subfloat[Score-based]{\includegraphics[width=0.25\linewidth]{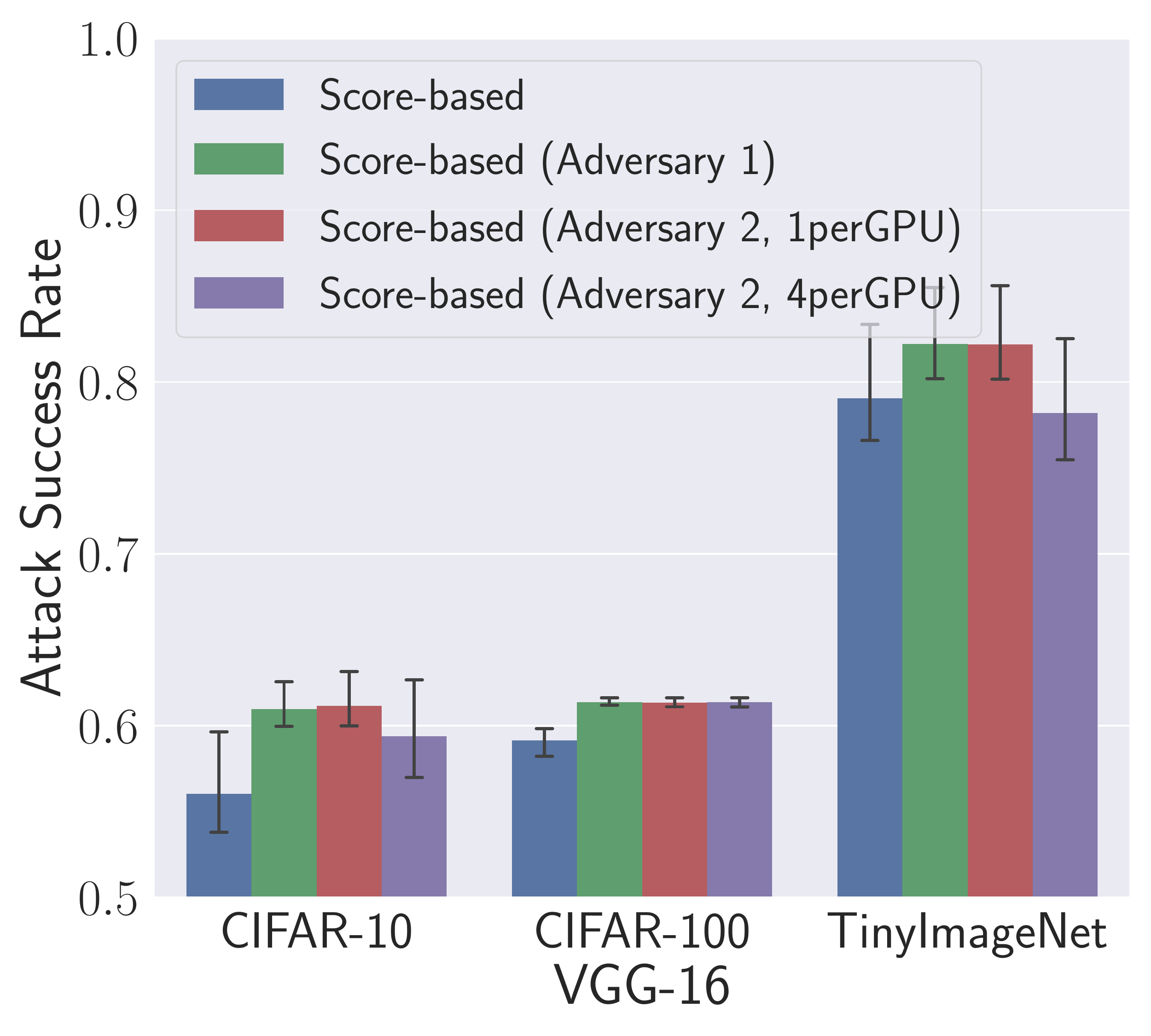}
    \includegraphics[width=0.25\linewidth]{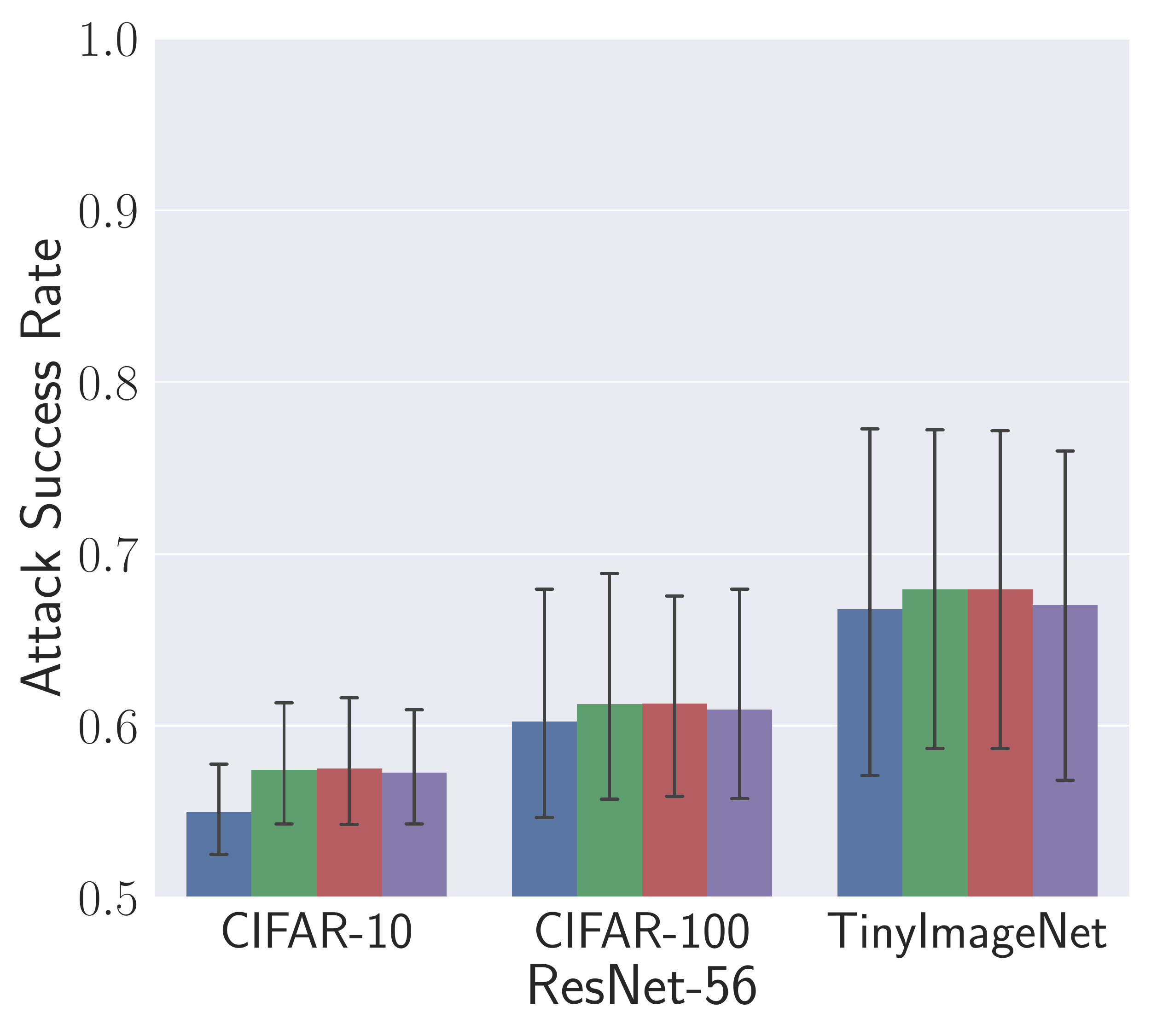}
    }
    \subfloat[Label-only]{\includegraphics[width=0.25\linewidth]{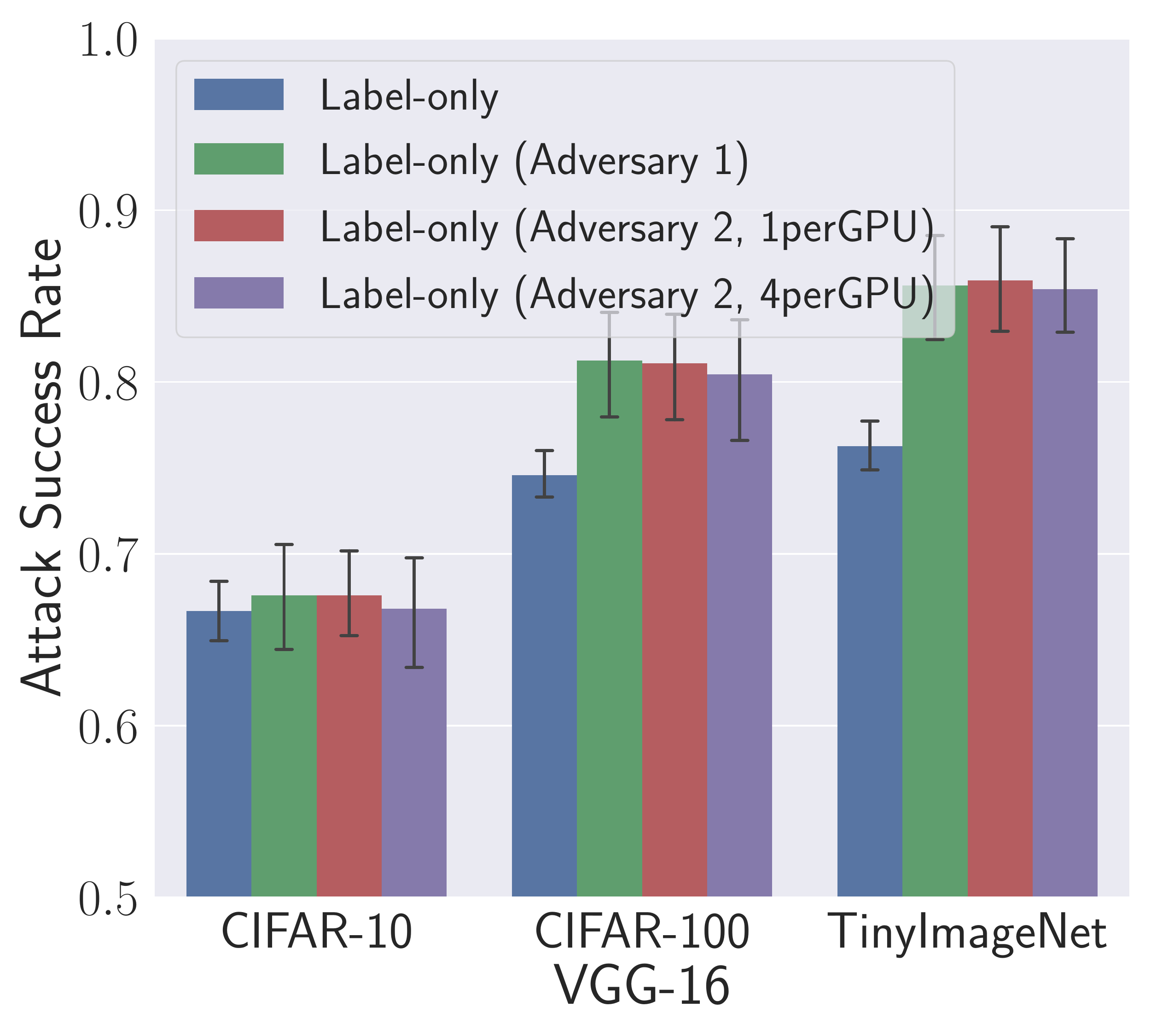}
    \includegraphics[width=0.25\linewidth]{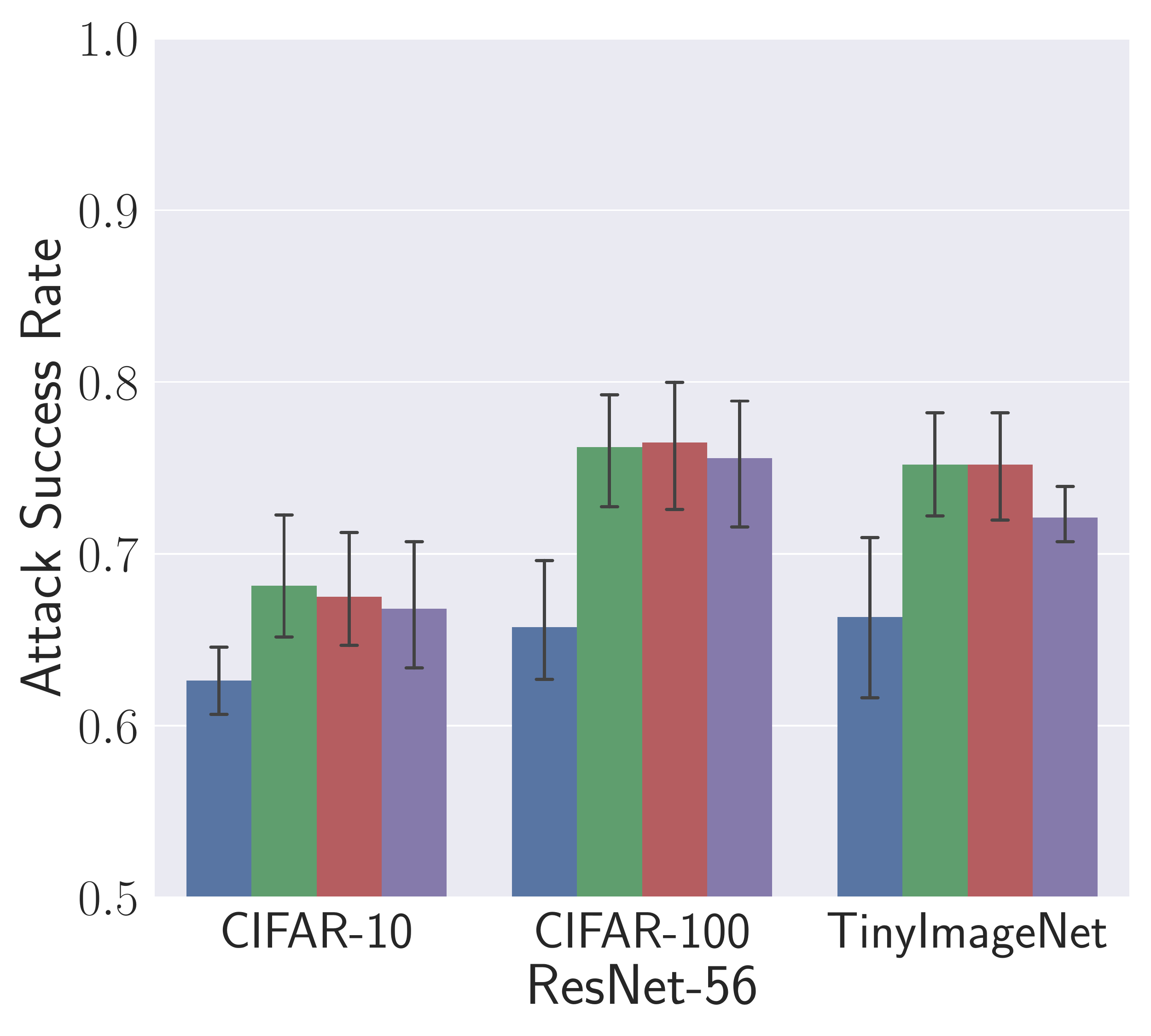}}
    \caption{The performance of different membership inference attacks against multi-exit models.}
    \label{fig:appendix_more_process}
\end{figure*}

\begin{figure*}[t]
    \centering
    \subfloat[Gradient-based]{\includegraphics[width=0.33\linewidth]{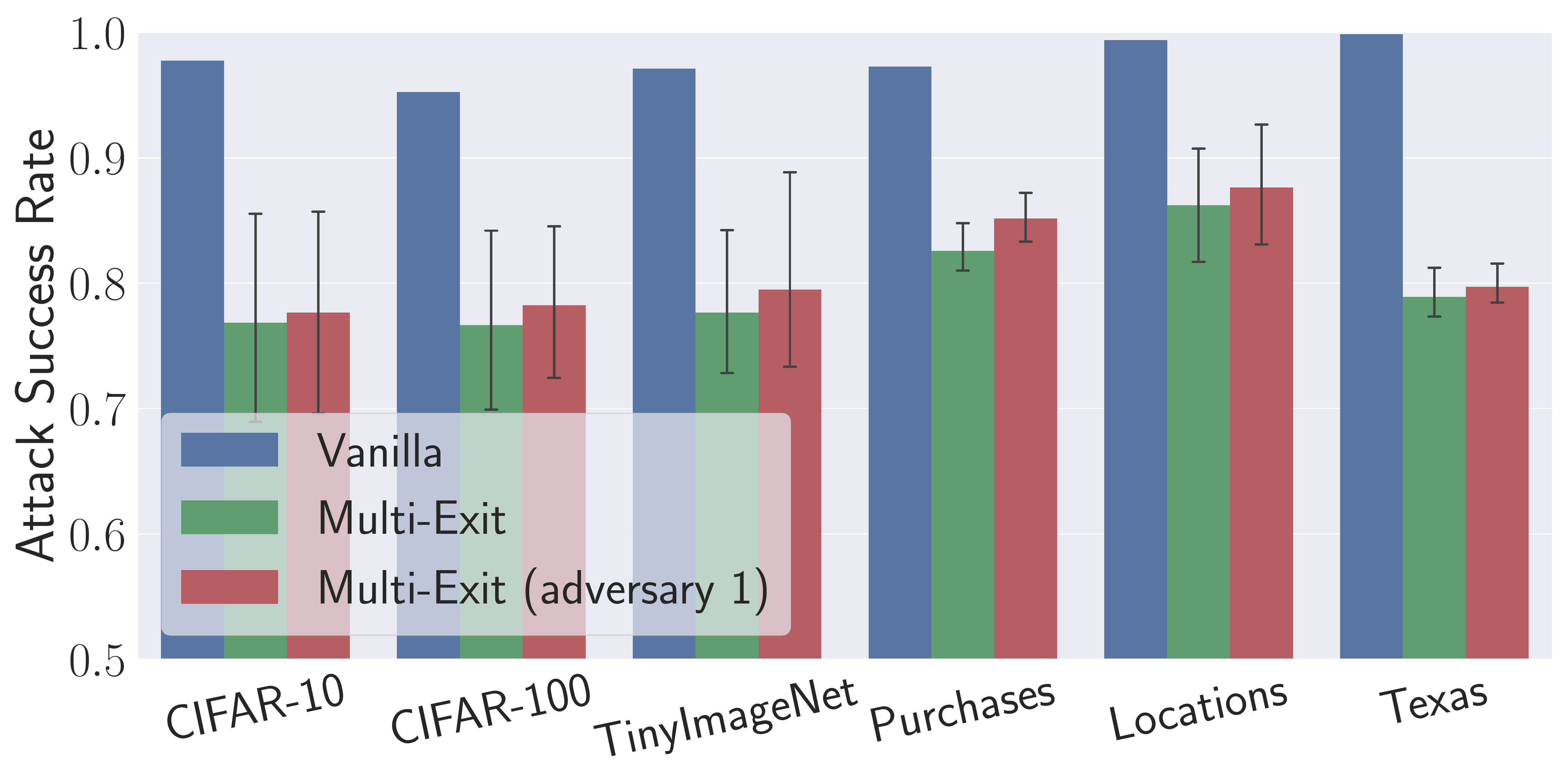}
    \includegraphics[width=0.33\linewidth]{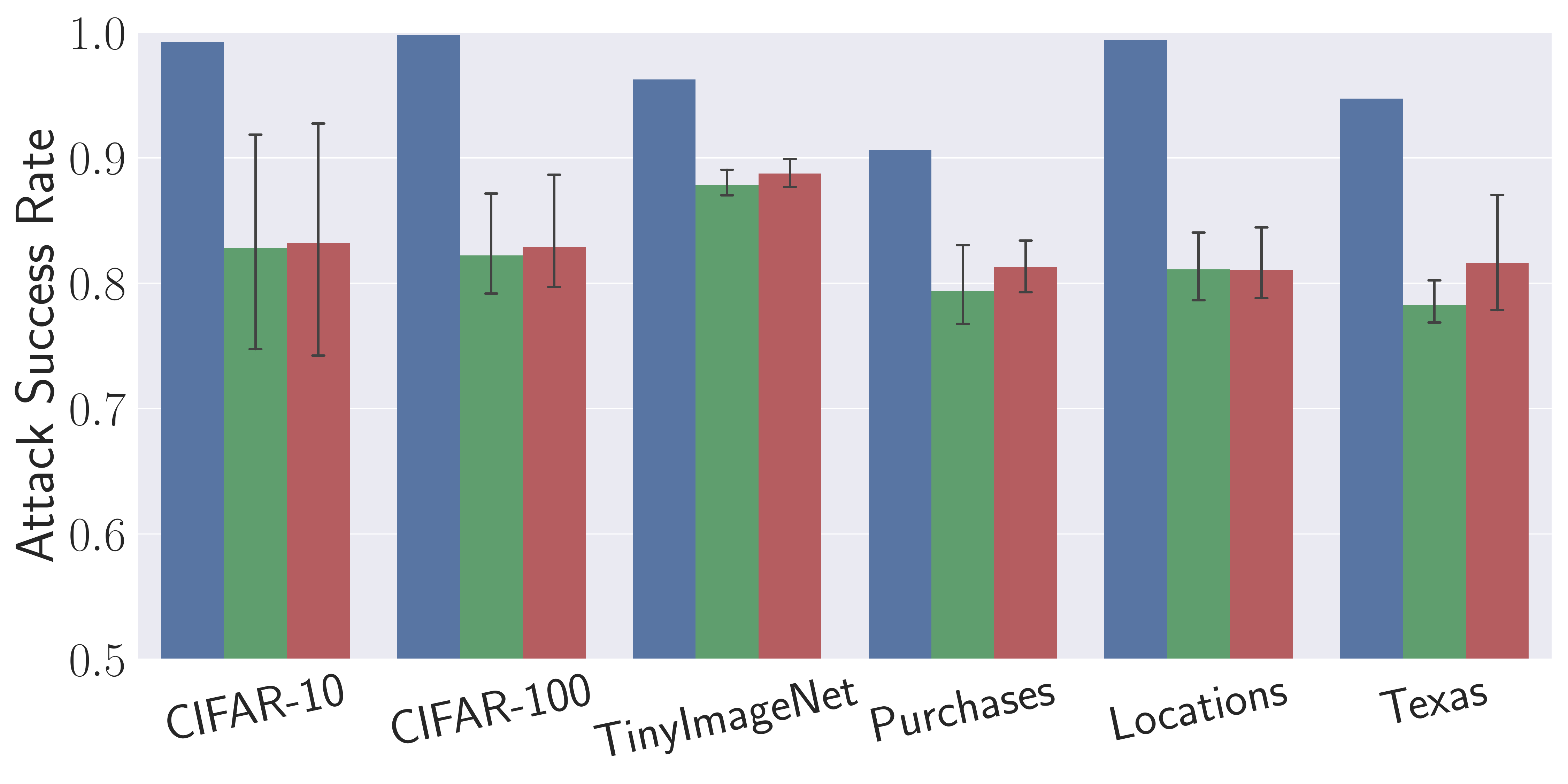}
    \includegraphics[width=0.33\linewidth]{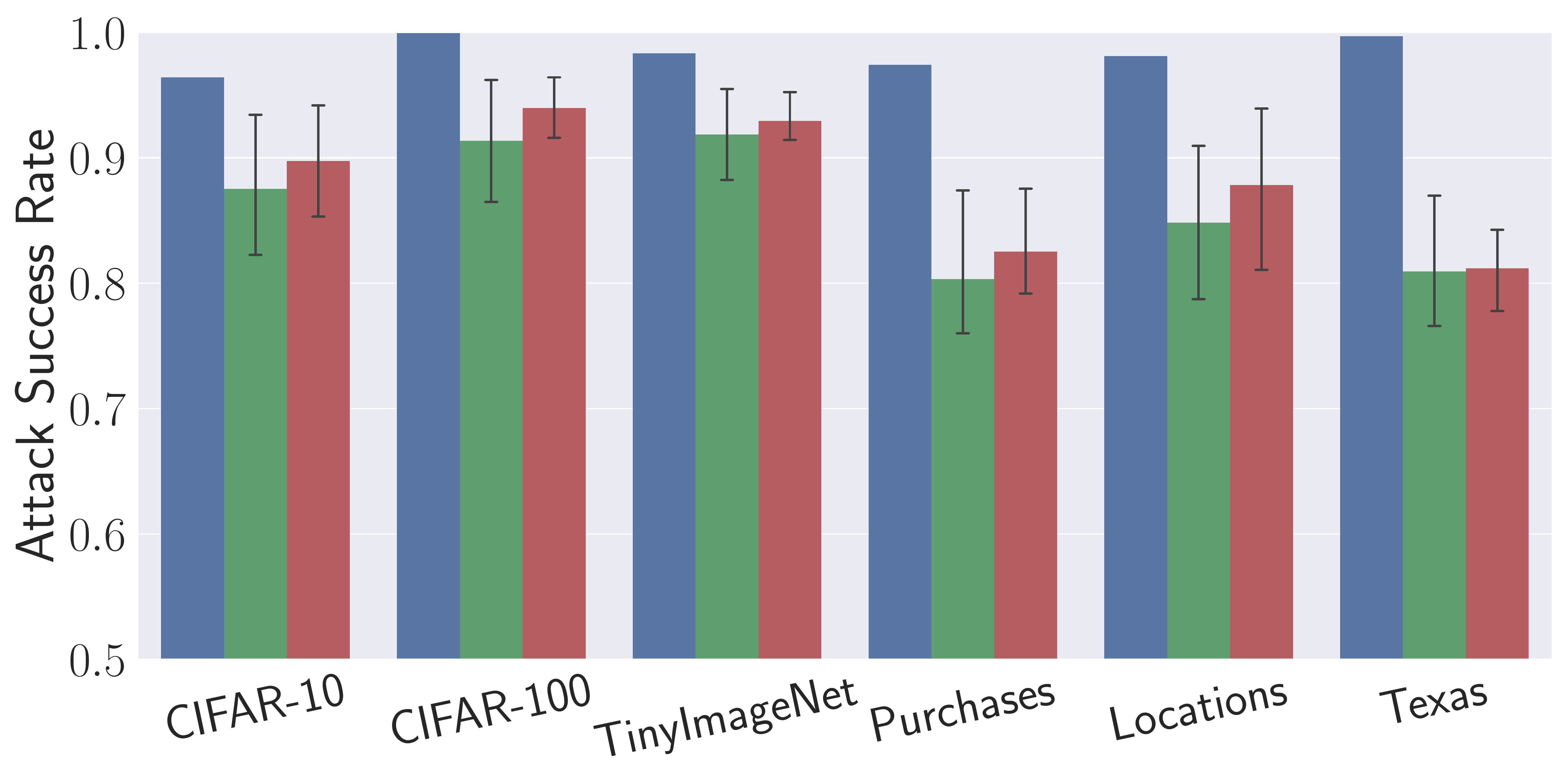}}

    \subfloat[Score-based]{
    \includegraphics[width=0.33\linewidth]{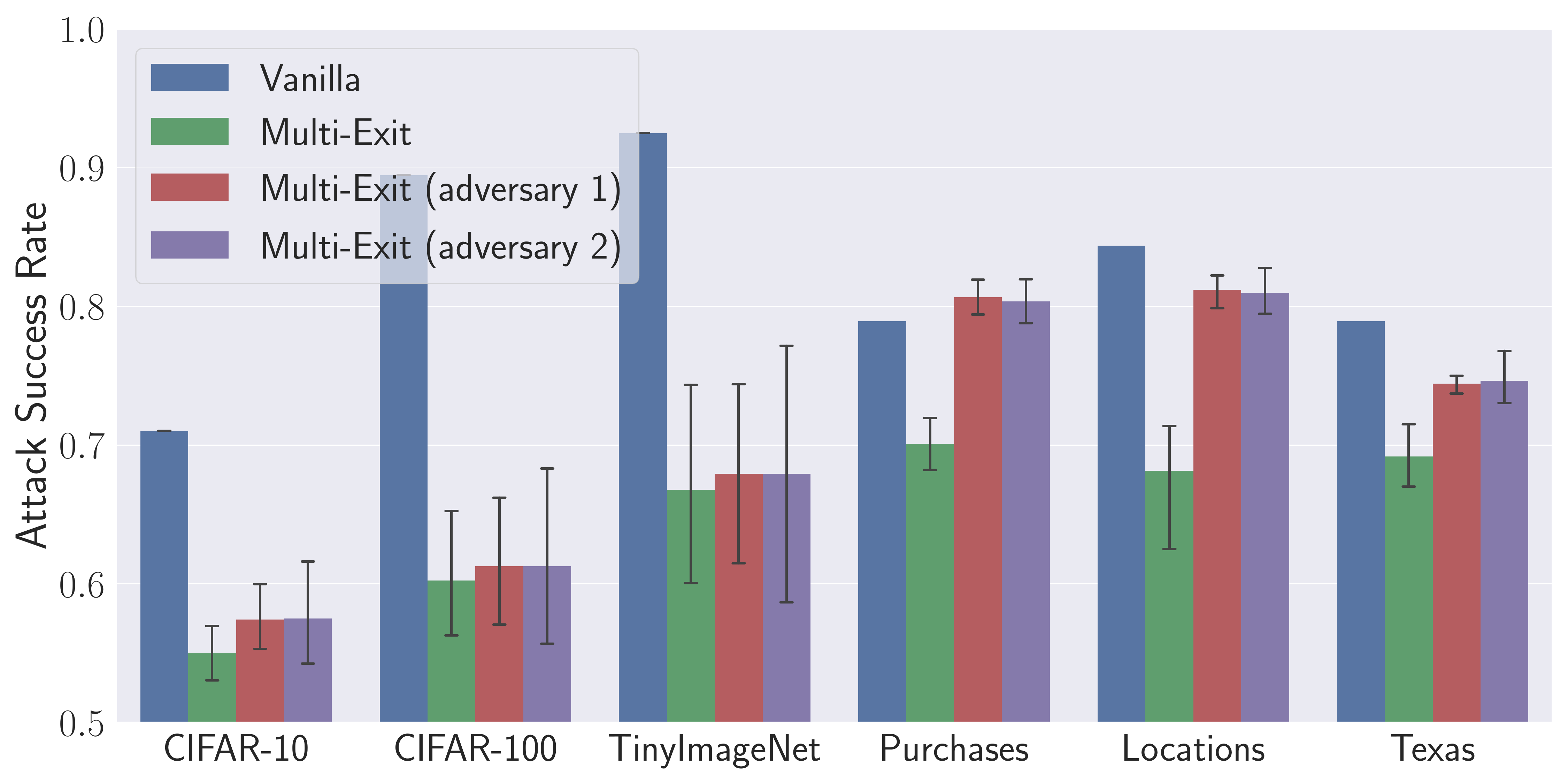}\includegraphics[width=0.33\linewidth]{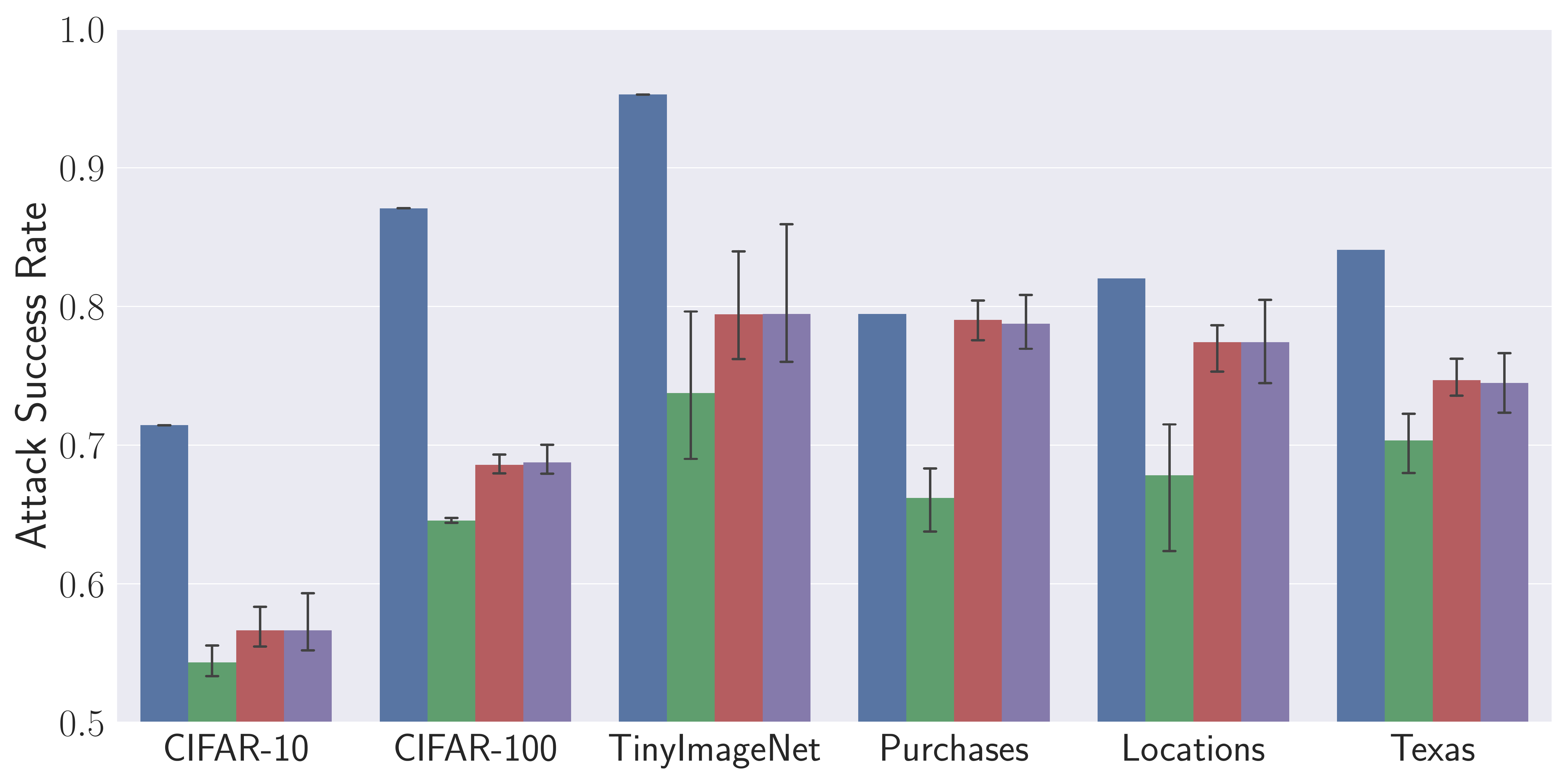}
    \includegraphics[width=0.33\linewidth]{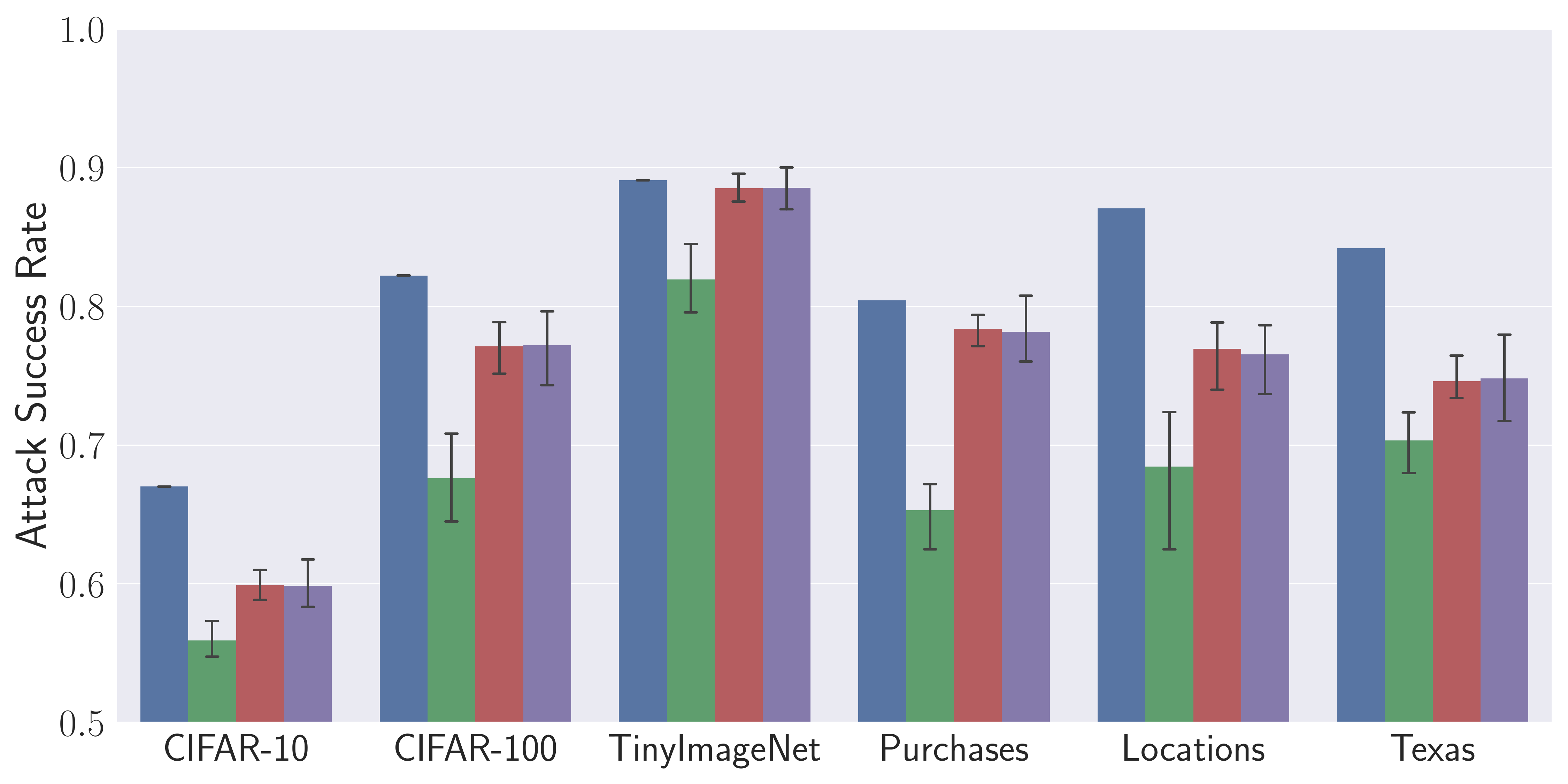}}

    \subfloat[Label-only]{
    \includegraphics[width=0.33\linewidth]{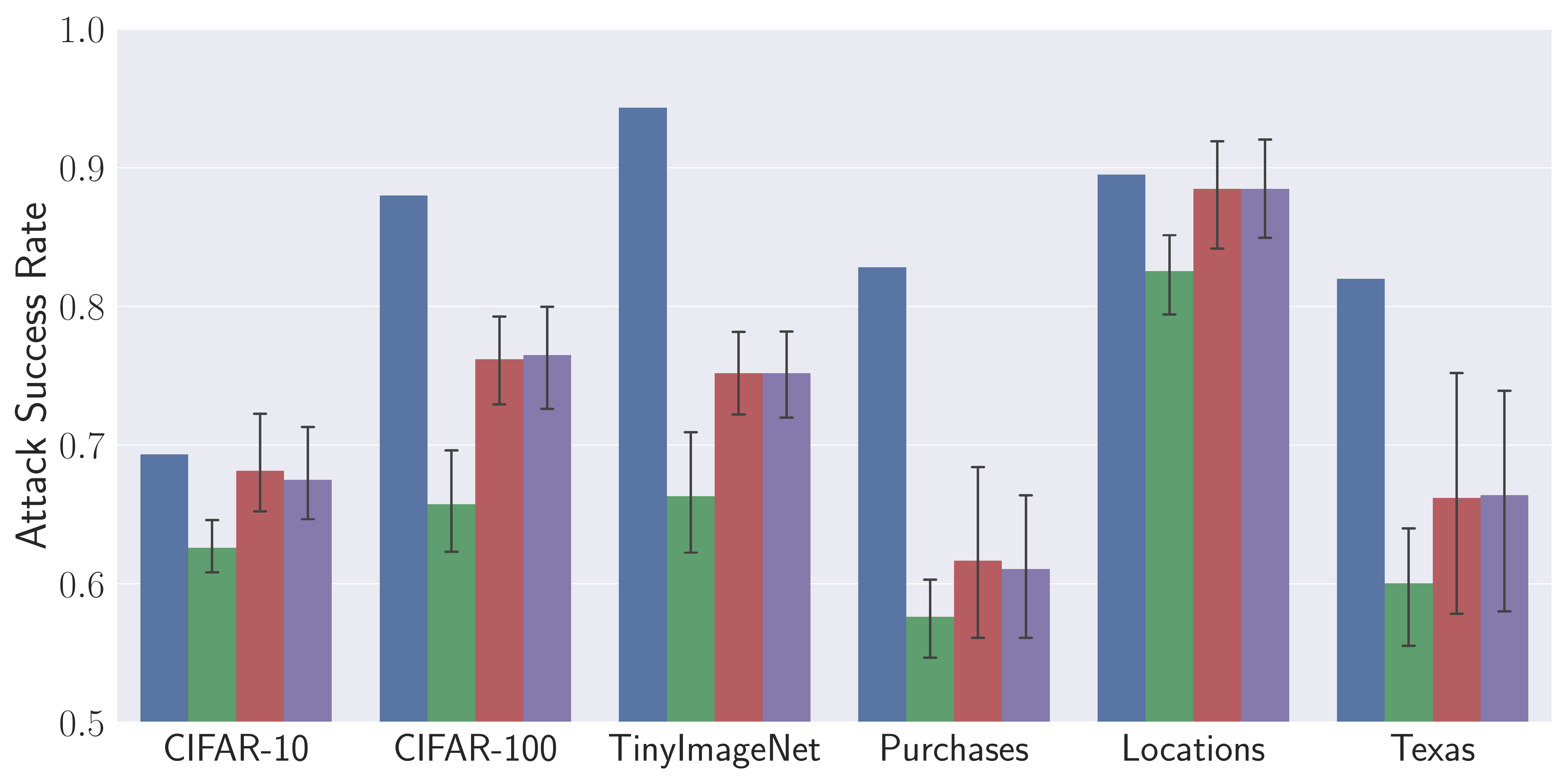}
    \includegraphics[width=0.33\linewidth]{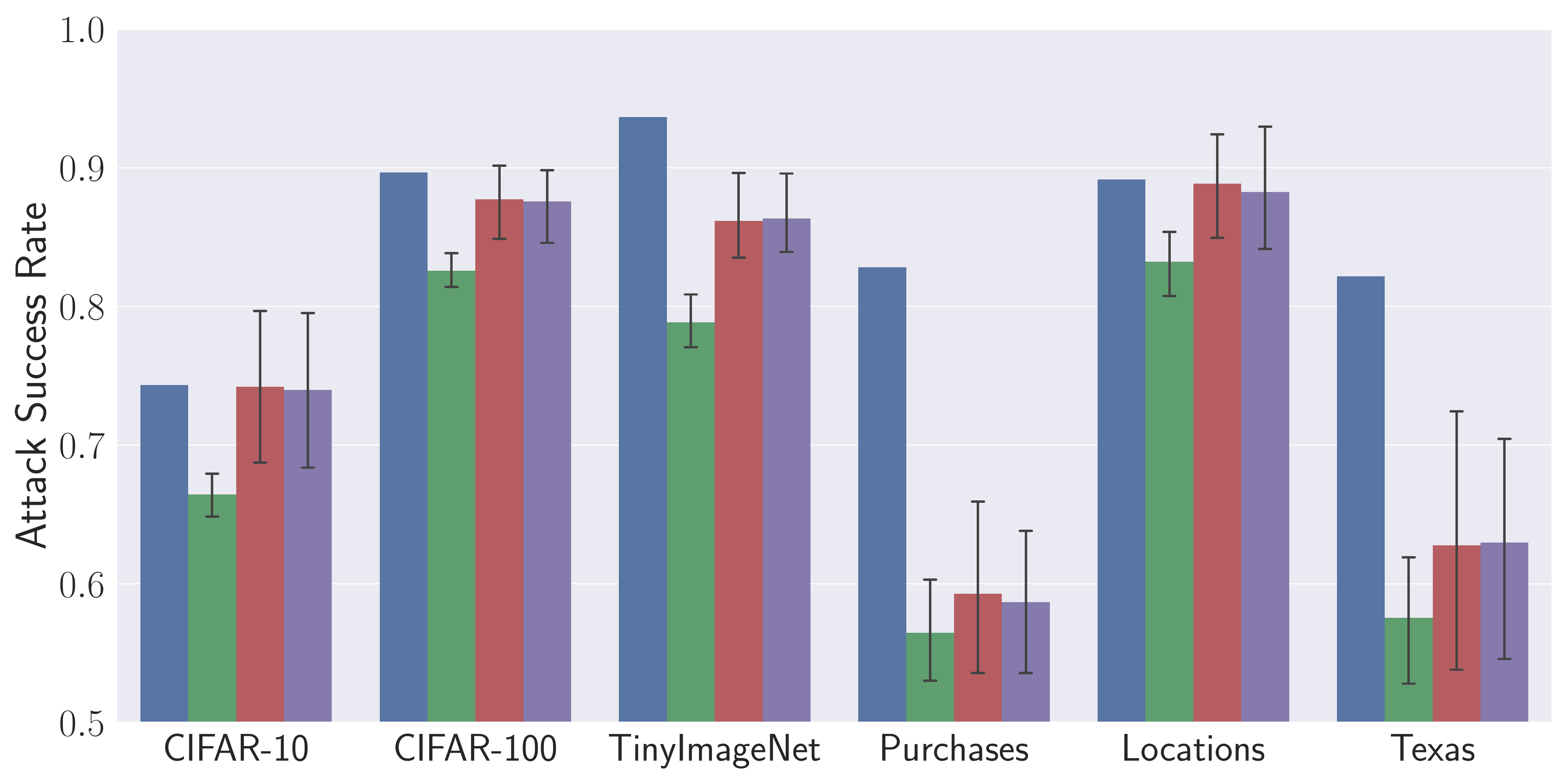}
    \includegraphics[width=0.33\linewidth]{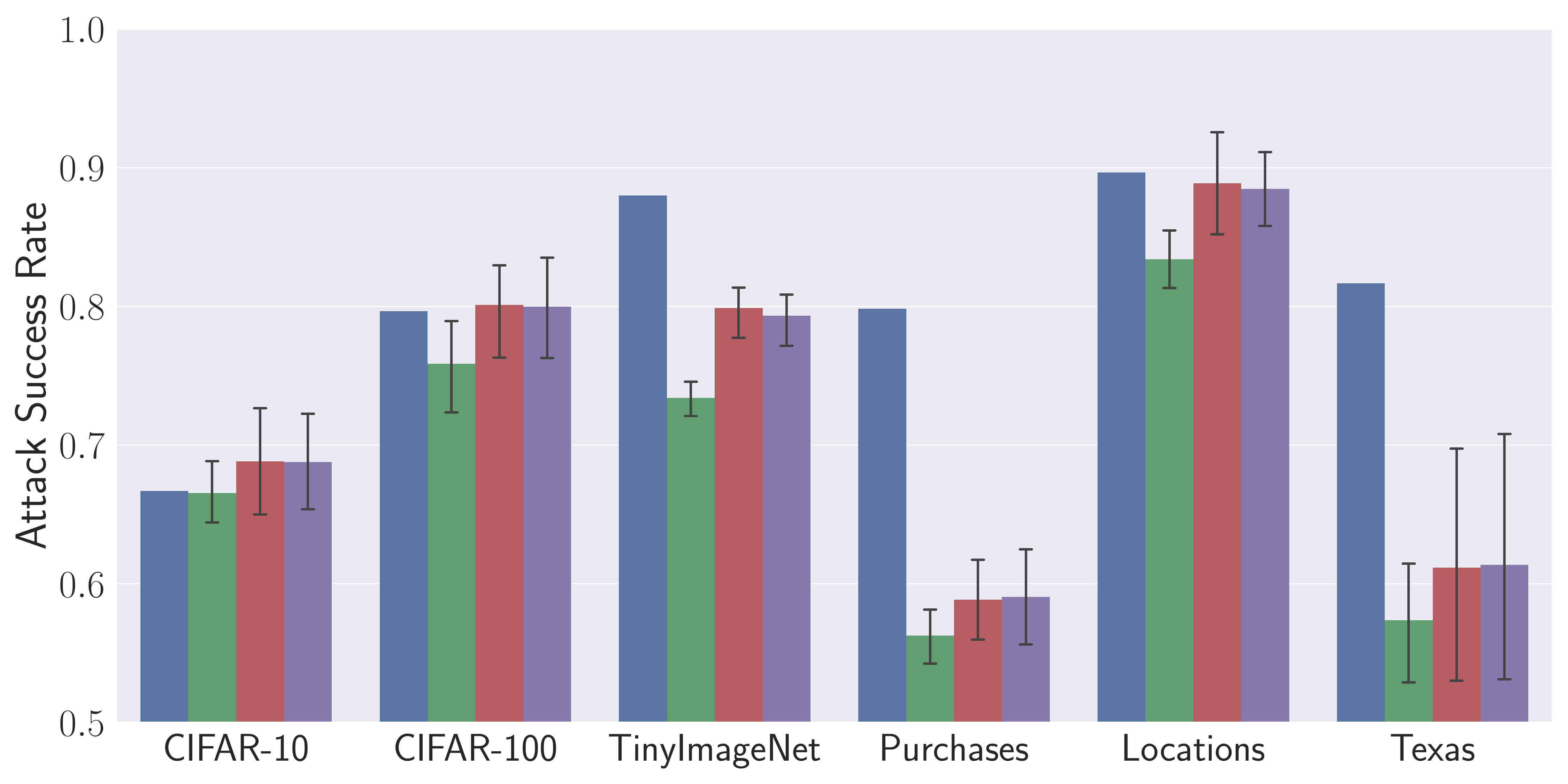}}
    
    \caption{The attack performance of different membership inference attacks on all datasets. 
    The blue and green bars indicate the original attack on the vanilla and multi-exit models, while the red and purple bars indicate our hybrid attack on the multi-exit model. 
    From left to right (first/second/third columns), computer vision tasks on VGG-16/ResNet-56/WideResNet-32 and non-computer vision tasks on FCN-18-2/FCN-18-3/FCN-18-4.}
\label{fig:appendix_attackASR}
\end{figure*}

\newpage

\begin{figure*}[!htbp]
    \centering
    \subfloat[Score-based]{\includegraphics[width=0.25\linewidth]{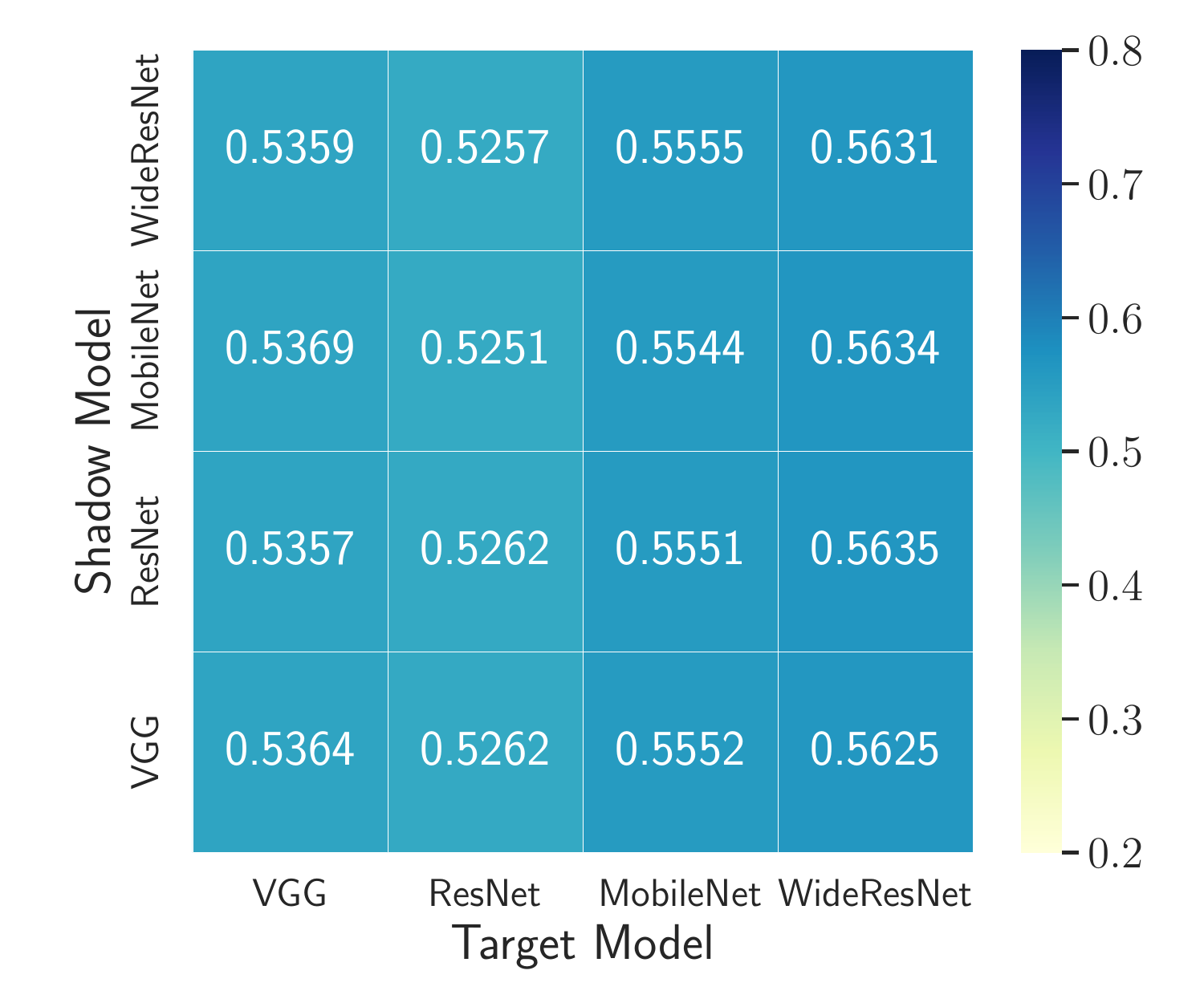}}
    \subfloat[Score-based (hybrid attack)]{
    \includegraphics[width=0.25\linewidth]{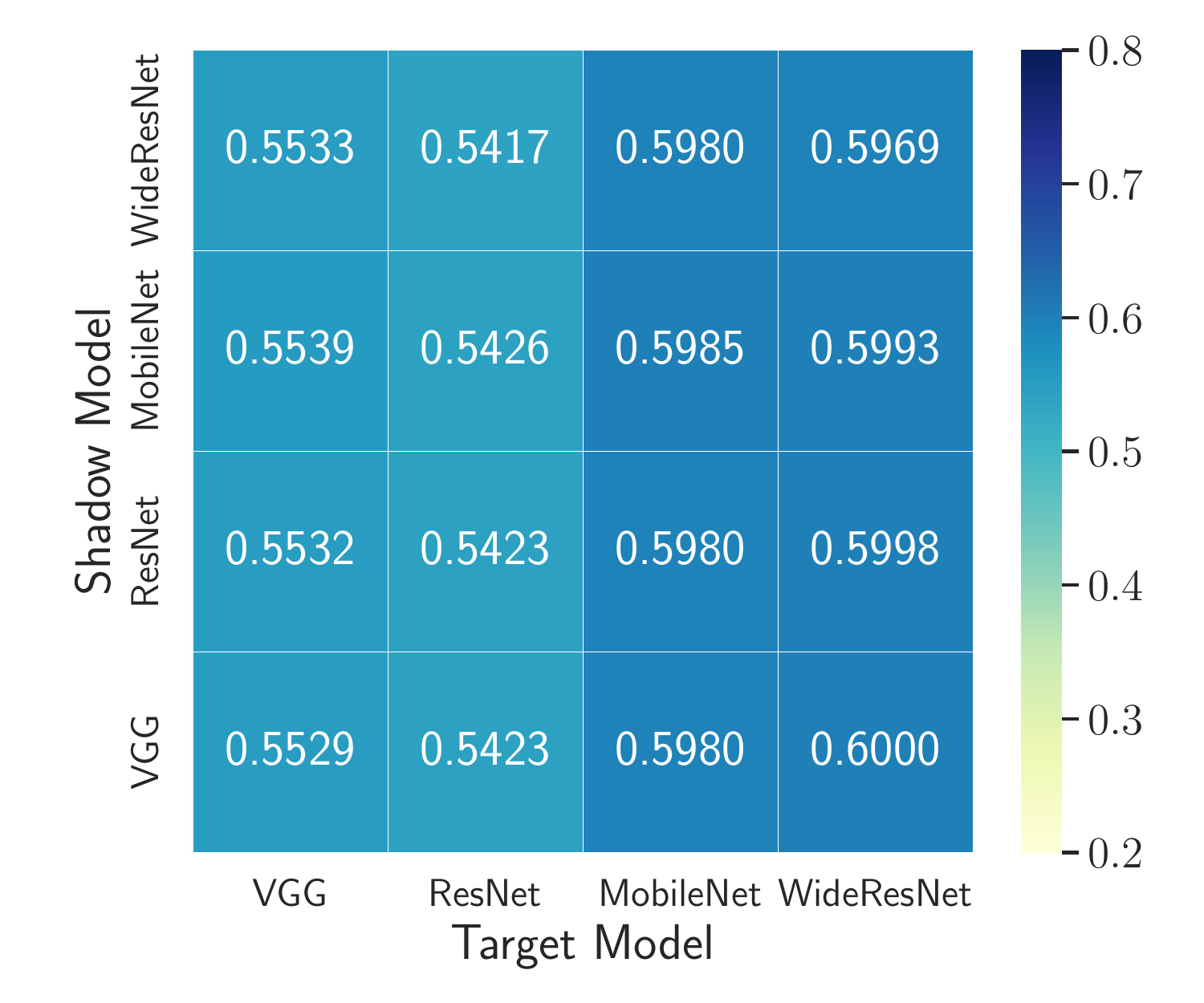}}
    \subfloat[Score-based]{\includegraphics[width=0.25\linewidth]{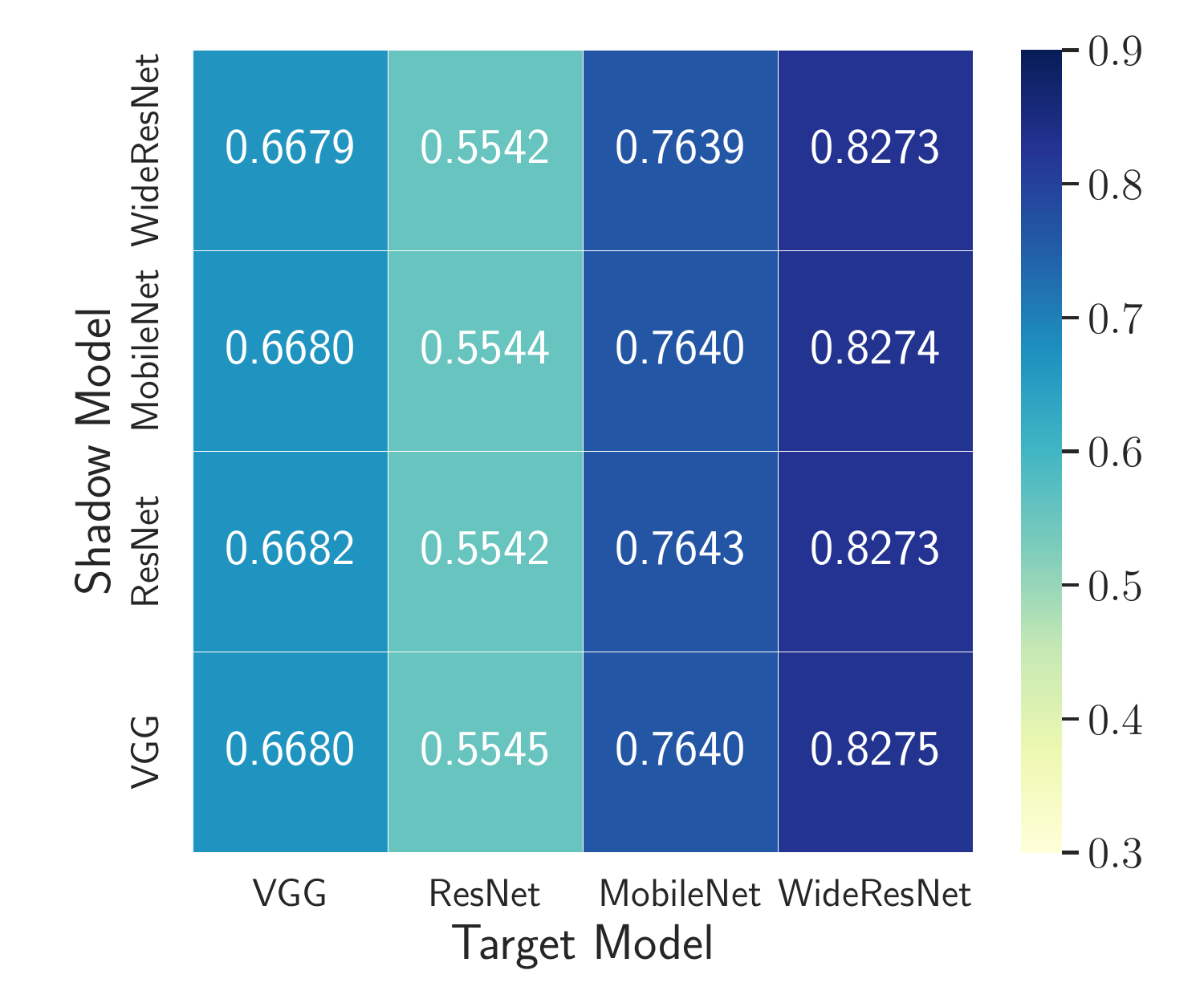}}
    \subfloat[Score-based (hybrid attack)]{
    \includegraphics[width=0.25\linewidth]{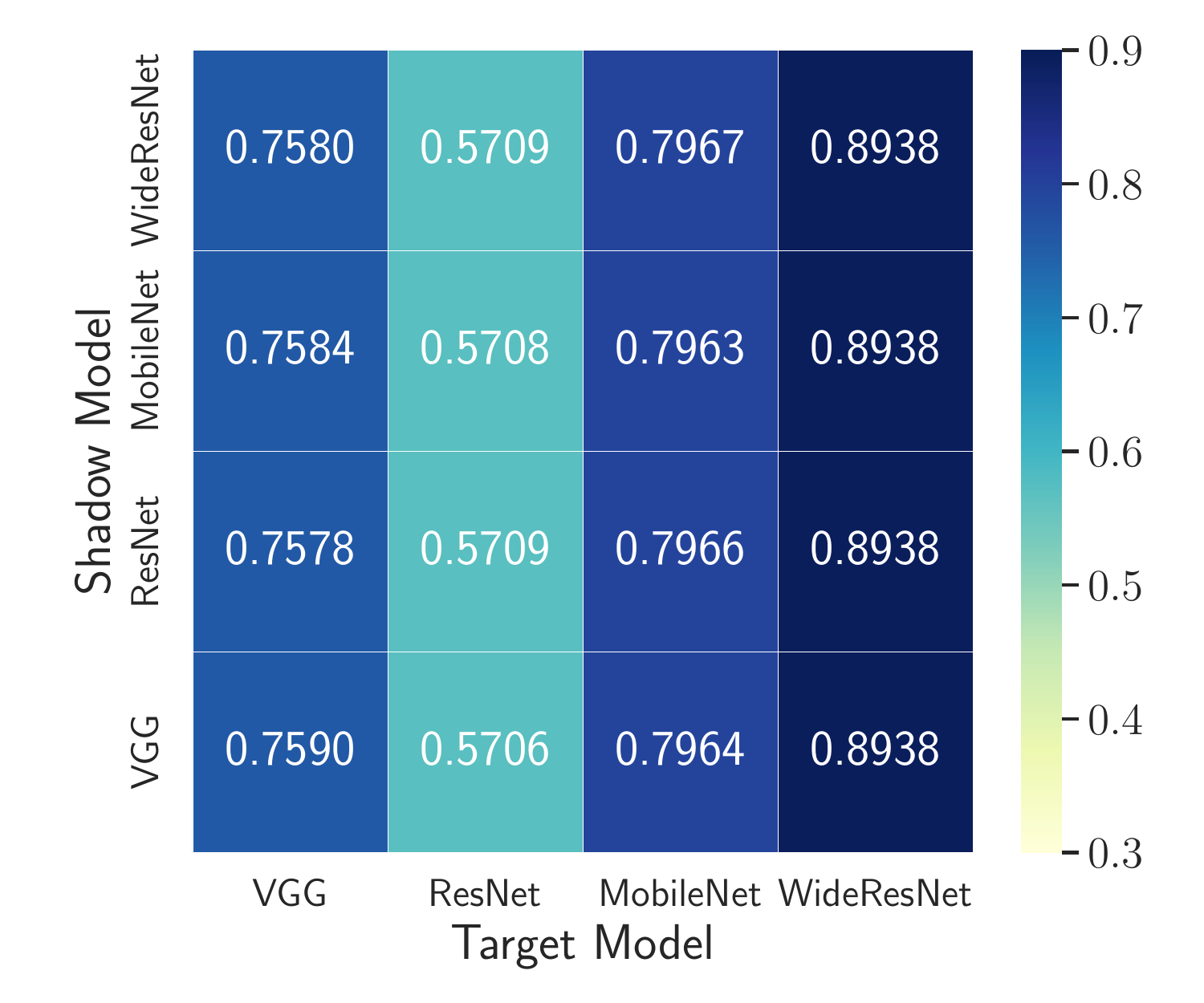}}
    \caption{The attack performance when the shadow model has different architecture compared to the target model. 
    The former two plots are on CIFAR-10, and the latter two plots are on TinyImageNet.}
    \label{fig:model_transfer_cifar10_tinyimagenet}
\end{figure*}

\begin{figure*}[!htbp]
    \centering
    \subfloat[Score-based]{\includegraphics[width=0.25\linewidth]{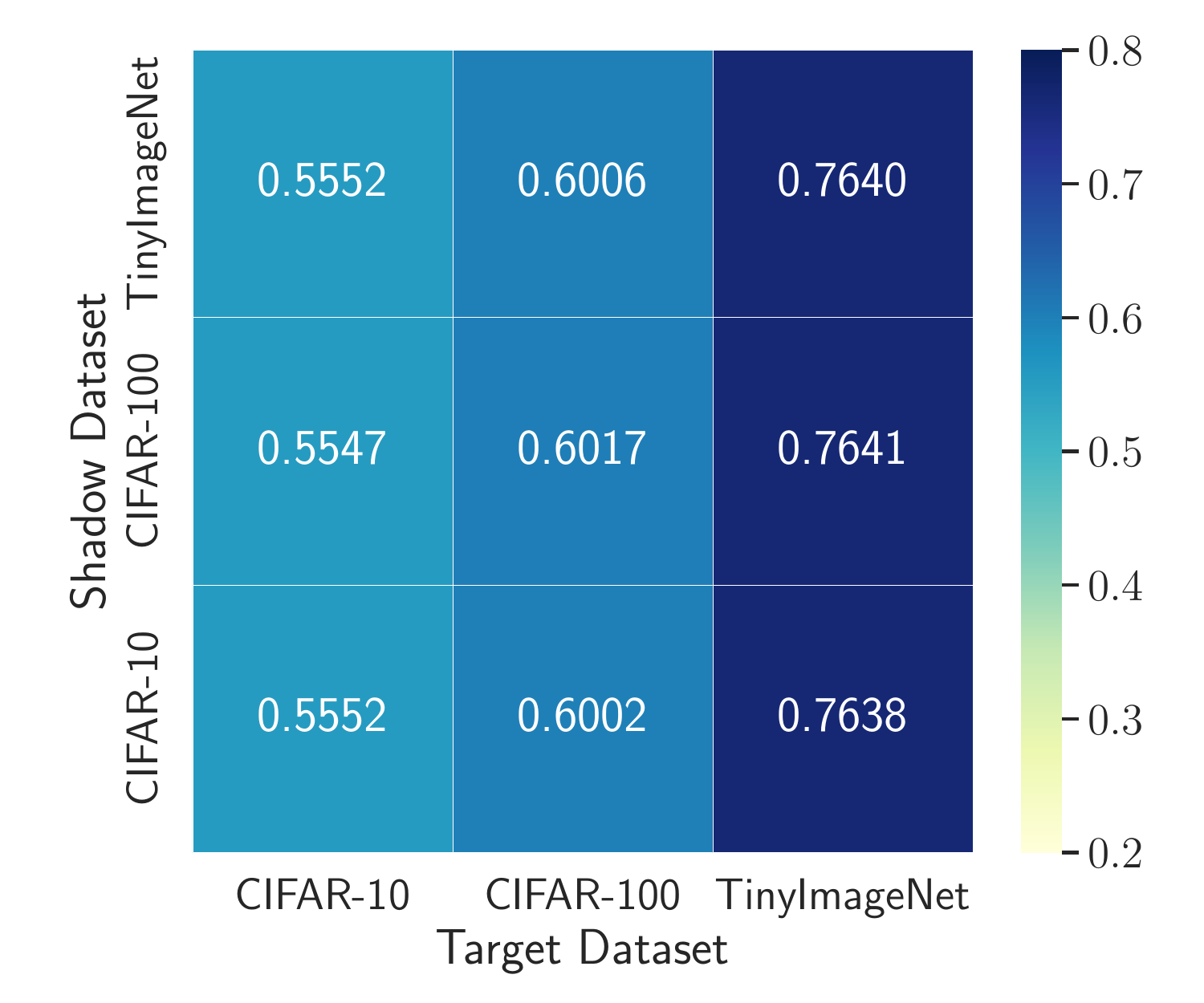}}
    \subfloat[Score-based (hybrid attack)]{
    \includegraphics[width=0.25\linewidth]{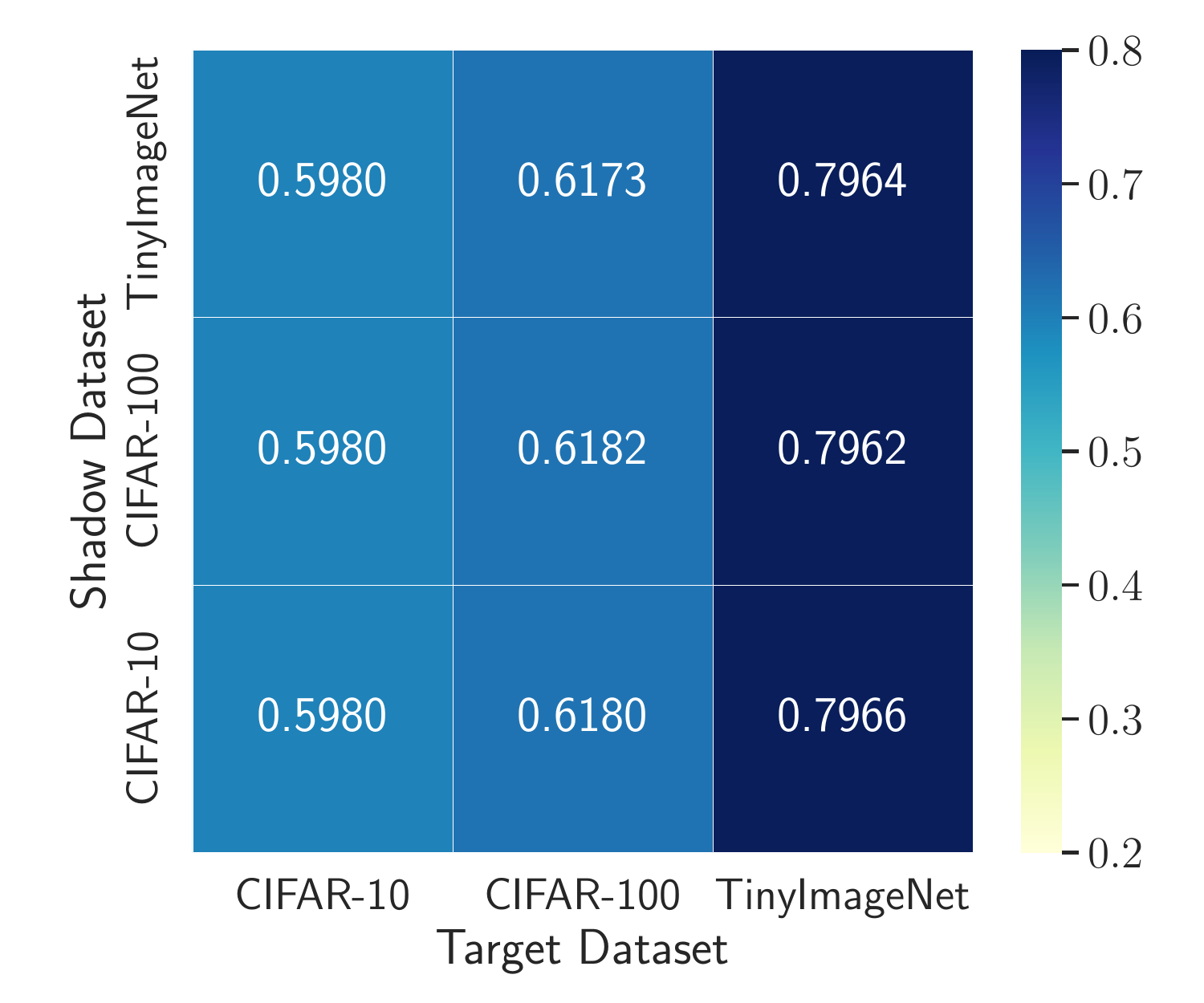}}
    \subfloat[Score-based]{\includegraphics[width=0.25\linewidth]{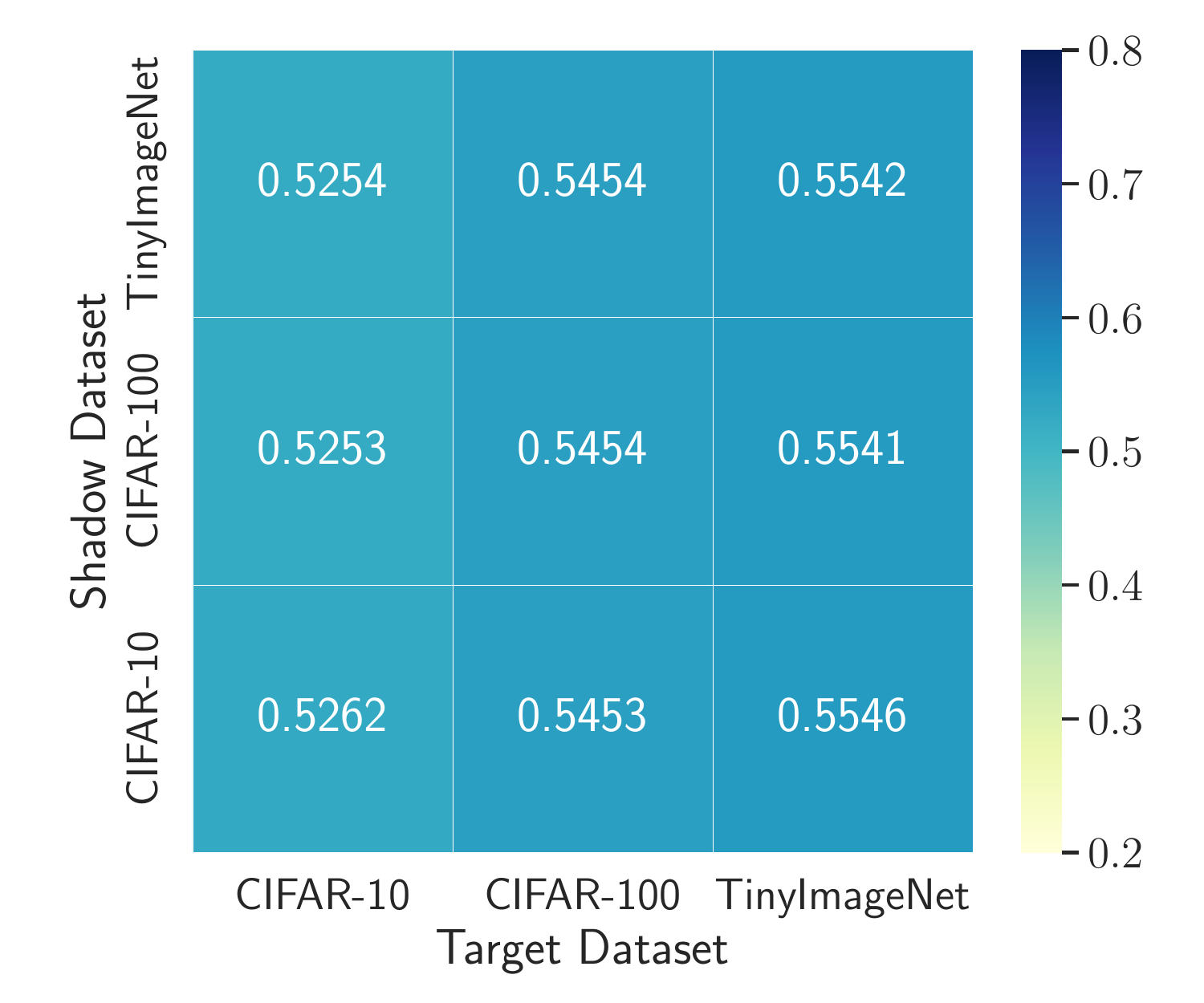}}
    \subfloat[Score-based (hybrid attack)]{
    \includegraphics[width=0.25\linewidth]{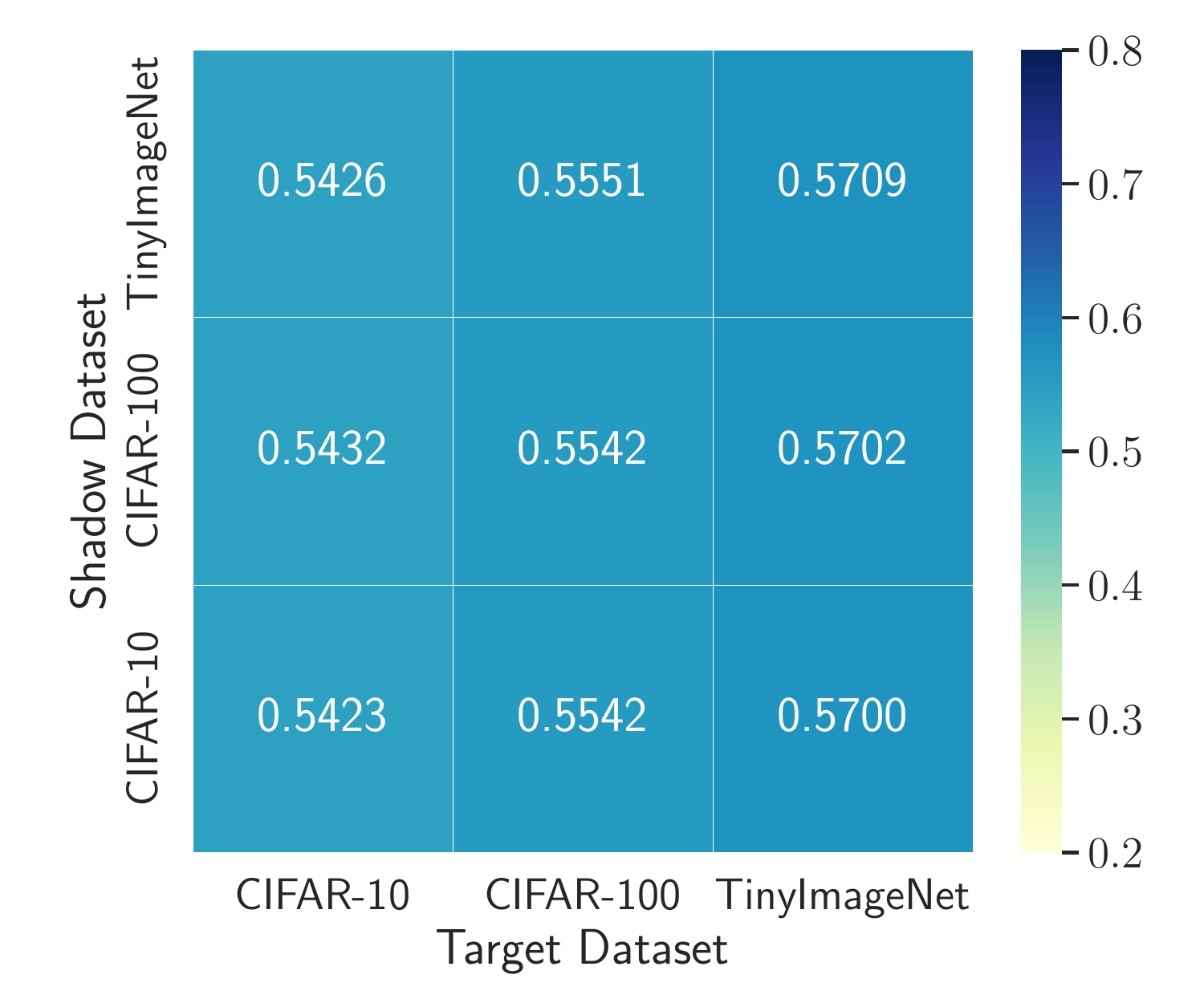}}
    \caption{The attack performance when the shadow dataset comes from different distributions of the target dataset. 
    The former two plots are on VGG-16, the latter two plots are on ResNet-56.}
    \label{fig:dataset_transfer_vgg_resnet}
\end{figure*}
\subsection{Possible Defense}
\label{MaxTimeGuard}

To empirically evaluate \textit{TimeGuard} that delays giving prediction to max inference time, i.e., it is impossible for the adversary to determine the exit depth.
Thus, we consider an adaptive adversary who attempts to steal the exit information by observing the prediction (score) only. 
The intuition of this adaptive adversary is that different exit points actually represent different capability models, which would present a statistical difference in their predictions.
The adversary first feeds a set of data samples to the shadow model, records each sample's predicted score and exit depth (index), then constructs a dataset.
The adversary then trains a classifier (4-layer MLP) using the dataset and leverages the classifier to predict the exit depth of the predicted scores returned from the target model.
Finally, the adversary mounts the hybrid attack.

We build such a classifier to predict the exit information of VGG-16 and MobileNet.
The prediction accuracy of exit depth varies from 45\% (5 or 6 exits) to 70\% (2 exits), and the adaptive attack performance against the 2-exit model is still a little better than the original attack, while the attack performance against the model with more exits drops sharply, which lead to much lower averaged \texttt{ASR} score, as shown in \autoref{fig:adaptive_adversary}.
We can observe that the attack performance of an adaptive adversary is worse than the original attack in most cases.
The results show that excessive false exit depths exploited by the adversary can even lead to performance degradation.

In conclusion, our \textit{TimeGuard} that delays giving prediction to max inference time indeed can mitigate membership leakages stemming from the hybrid attack. Besides, we acknowledge that \textit{TimeGuard} can only be applied to some energy-constrained applications and not to the real-time constrained applications, e.g., self-driving cars. 
Therefore, we propose a novel \textit{TimeGuard} with high efficiency in \autoref{sec:defense}.

\end{document}